\documentclass[amsmath,amssymb,aps,superscriptaddress,
]{revtex4-2}

\usepackage{graphicx}
\usepackage{bm}
\usepackage{braket}
\usepackage{upgreek}
\usepackage{float}
\usepackage[dvipsnames,svgnames,x11names]{xcolor}
\usepackage{ulem}
\usepackage{gensymb} 

\newcommand*\arucl{{$\alpha$-RuCl$_3$}}
\newcommand*\etal{{\textit{et al.}}}

\begin{document}

\title{Lessons from \arucl\ for pursuing quantum spin liquid physics in atomically thin materials}

\author{Claudia Ojeda-Aristizabal}
\affiliation{Department of Physics and Astronomy, California State University Long Beach, Long Beach California, 90840}
\author{Xiaohu Zheng}
\affiliation{Beijing Academy of Quantum Information Sciences, Beijing 100193, China}
\affiliation{State Key Laboratory of Materials for Integrated Circuits, Shanghai Institute of
Microsystem and Information Technology, Chinese Academy of Sciences, 865 Changning Road, Shanghai 200050 China}

\author{Changsong Xu}
\affiliation{Key Laboratory of Computational Physical Sciences (Ministry of Education), Institute of
Computational Physical Sciences, State Key Laboratory of Surface Physics, and Department of
Physics, Fudan University, Shanghai 200433, China}

\author{Zohar Nussinov}
\affiliation{Department of Physics, Washington University in St. Louis, 1 Brookings Dr., St. Louis MO 63130, USA}
\affiliation{Institute of Materials Science \& Engineering, Washington University in St. Louis, 1 Brookings Dr., St. Louis MO 63130, USA}
\affiliation{Department of Physics and Quantum Centre of Excellence for Diamond and Emergent Materials (QuCenDiEM),
Indian Institute of Technology Madras, Chennai 600036, India}

\author{Yukitoshi Motome}
\affiliation{Department of Applied Physics, The University of Tokyo, Tokyo 113-8656, Japan}

\author{Arnab Banerjee}
\affiliation{Department of Physics and Astronomy, Purdue University, West Lafayette IN 47907, USA}
\affiliation{Quantum Science Center, Oak Ridge National Laboratory, Oak Ridge TN 37831, USA}

\author{Adam W.\ Tsen}
\affiliation{Institute for Quantum Computing, University of Waterloo, Waterloo, ON, Canada}
\affiliation{Department of Chemistry, University of Waterloo, Waterloo, ON, Canada}

\author{Michael Knap}
\affiliation{Technical University of Munich, TUM School of Natural Sciences, Physics Department, 85748 Garching, Germany}
\affiliation{Munich Center for Quantum Science and Technology (MCQST), Schellingstr. 4, 80799 M{\"u}nchen, Germany}

\author{Rui-Rui Du}
\affiliation{International Center for Quantum Materials, Peking University, Beijing 100871, China}

\author{Gajadhar Joshi}
\affiliation{Center for Integrated Nanotechnologies, Sandia National Laboratories, Albuquerque, New
Mexico 87123, USA}

\author{Andy Mounce}
\affiliation{Center for Integrated Nanotechnologies, Sandia National Laboratories, Albuquerque, New
Mexico 87123, USA}

\author{Youngwook Kim}
\affiliation{Department of Physics and Chemistry, Daegu Gyeongbuk Institute of Science and Technology (DGIST), Daegu 42988, Republic of Korea}

\author{Benjamin M.\ Hunt}
\affiliation{Department of Physics, Carnegie Mellon University, Pittsburgh, Pennsylvania 15213, USA}

\author{Dmitry Shcherbakov}
\affiliation{Department of Physics, Washington University in St. Louis, 1 Brookings Dr., St. Louis MO 63130, USA}

\author{Boyi Zhou}
\affiliation{Department of Physics and Astronomy, Stony Brook University; Stony Brook, New York 11794, USA}
\affiliation{Center for Integrated Science and Engineering, Columbia University; New York, New York 10027, USA}

\author{Ran Jing}
\affiliation{Department of Physics and Astronomy, Stony Brook University; Stony Brook, New York 11794, USA}

\author{Mengkun Liu}
\affiliation{Department of Physics and Astronomy, Stony Brook University; Stony Brook, New York 11794, USA}
\affiliation{National Synchrotron Light Source II, Brookhaven National Laboratory; Upton, New York 11973, USA}

\author{Hui Zhao}
\affiliation{Department of Physics and Astronomy, The University of Kansas, Lawrence, Kansas 66045, USA}

\author{Bolin Liao}
\affiliation{Department of Mechanical Engineering, University of California, Santa Barbara, CA 93106, USA}

\author{Martin Claassen}
\affiliation{Department of Physics and Astronomy, University of Pennsylvania, Philadelphia, PA 19104, USA}

\author{Onur Erten}
\affiliation{Department of Physics, Arizona State University, Tempe, AZ 85287, USA }

\author{Yong P.\ Chen}
\affiliation{1.	Department of Physics and Astronomy, Aarhus University, Aarhus-C, Denmark}
\affiliation{2.	WPI-AIMR Advanced Institute for Materials Research, Tohoku University, Sendai, Japan}
\author{Erik A.\ Henriksen}
\affiliation{Department of Physics, Washington University in St. Louis, 1 Brookings Dr., St. Louis MO 63130, USA}
\affiliation{Institute of Materials Science \& Engineering, Washington University in St. Louis, 1 Brookings Dr., St. Louis MO 63130, USA}
\email[Corresponding author~]{henriksen@wustl.edu}

\date{\today}

\begin{abstract}

Quantum spin liquids can arise from Kitaev magnetic interactions, and exhibit fractionalized excitations with the potential for a topological form of quantum computation. This review surveys recent experimental and theoretical progress on the pursuit of phenomena related to Kitaev magnetism in layered and exfoliatable materials, which offer numerous opportunities to apply powerful techniques from the field of atomically thin materials. We primarily focus on the antiferromagnetic Mott insulator \arucl, which exhibits Kitaev couplings and is readily exfoliated to single- or few-layer sheets, and thus serves as a test bed for developing probes of Kitaev phenomena in atomically thin materials and devices. We introduce the Kitaev model and how it is realized in \arucl\ and other material candidates; and cover \arucl\ synthesis and fabrication into van der Waals heterostructure devices. A key discovery is a work-function-mediated charge transfer that heavily dopes both the \arucl\ and proximate materials, and can enhance Kitaev interactions by up to 50\%. We further discuss a wide range of recent results in electronic transport and optical and tunneling spectroscopies of \arucl\ devices. The experimental techniques and theoretical insights developed for \arucl\ establish a framework for discovering and engineering superior two-dimensional Kitaev materials that may ultimately realize elusive quantum spin liquid phases.

\end{abstract}

\maketitle

\tableofcontents

\section*{Introduction}

Quantum spin liquids are fascinating systems in which spins, frustrated from ordering due to geometric constraints or many-particle interactions, form a highly quantum entangled ground state \cite{Anderson1973,Savary2016,Knolle2019,Broholm2020}. Numerous models for quantum spin liquids exist, but the variants arising from Kitaev magnetic interactions---where frustration is driven by a bond-direction-dependent spin coupling on the honeycomb lattice---is of particular interest \cite{Kitaev2006,Matsuda2025}. Not only are versions of the Kitaev model exactly solvable in two dimensions, but the basic excitations can be non-Abelian anyons with unusual fractional statistics, such that moving one anyon around another generates a non-trivial unitary transformation that is in principle useful for fault-tolerant quantum computation \cite{Kitaev2003,Kitaev2006,Nayak2008,KL,NussinovBrink,Simonbook}. From the fundamental physics of quantum spin liquids to potentially practical quantum technologies, experimentally realizing the Kitaev model is highly desirable.



In a seminal work, Jackeli and Khaliullin demonstrated how these Kitaev magnetic couplings, and so in principle the resulting Kitaev quantum spin liquid (KQSL), can be realized in octahedrally coordinated systems having strong spin-orbit coupling (SOC) \cite{Jackeli2009,Chaloupka2010}. The Kitaev-type bond-direction-dependent interactions appear in layered iridates of the form A$_2$IrO$_3$ (A=Li, Na), as a consequence of the microscopic exchange of the magnetic moments inside the edge-sharing IrO$_6$ octahedra which causes the isotropic part of the exchange Hamiltonian to vanish and an anisotropic interaction to appear. Many 5$d^5$ iridum compounds were identified as potential hosts of Kitaev physics, including Li$_2$IrO$_3$ and many polytypes A$_3$LiIr$_2$O$_6$ (A=H, Cu, Ag) \cite{Modic2014-kl,Takayama2015,Takagi2019-ke, Kitagawa2018-lg}, and eventually the 4$d^5$ \arucl\ system as well \cite{plumb_2014}, where edge-sharing RuCl$_6$ octahedra mediate the anisotropic exchange of the magnetic moment of the Ru$^{3+}$ ions at the center of the octahedra \cite{banerjee_proximate_2016,TREBST20221}. Despite small non-cubic distortions of the RuCl$_6$---also present for IrO$_6$ in the iridium compounds---to date it is samples of \arucl\ that have shown the clearest signatures of expected KQSL phenomena.

\arucl\ is so far the most important \textit{layered} and \textit{exfoliatable} material exhibiting Kitaev-type magnetic couplings. This offers a unique opportunity to pursue quantum spin liquid phenomena from the vantage point of the field of atomically thin materials, where enormous strides have been made over the past two decades in developing techniques to fabricate and measure high quality devices based on monolayer or thicker flakes of layered materials. The still-recent discovery of single layer magnetic materials has delivered numerous important techniques for exploring magnetism in the typically very small ($\sim10$s of $\mu$m in size) and very thin ($\sim$nm thick) flakes \cite{Lee2016-ig,huang2017layer,gong2017discovery,burch_magnetism_2018}. Advances in the ability to transfer and stack disparate atomically thin samples into van der Waals heterostructures allows a degree of control over the physical structure and environment of exfoliated flakes that offer novel methods for controlling magnetism via strain, charge doping, and proximity effects \cite{Geim2013}. And since many methods suited to measuring bulk samples---neutron and x-ray scattering, bulk probes of magnetization and thermal properties, \&c.---do not carry over well to vanishingly tiny samples, new methods are being developed to probe unconventional magnetic effects such as the Kitaev couplings. 

Here we review the state of investigations into exfoliated \arucl\ flakes and devices that have been carried out, with the overarching goal of pursuing the fascinating physics of quantum spin liquids. Of course, \arucl\ is a real material that embodies magnetic interactions beyond just the Kitaev couplings, which act to impede clear observation of KQSL phenomena \cite{Winter2016,Yadav2016,Winter2017,Hou2017}. Nonetheless, being the foremost example of a layered material with Kitaev magnetism, it is a good laboratory for studying this unusual magnetism and developing the probes that will enable study of more ideal materials that we hope will be discovered in the near future. 

The paper is organized as follows:\ In Section \ref{kitaev} we introduce the Kitaev model of magnetism and present several known and possible material realizations, with a focus on \arucl. Section 
\ref{synthesis} covers the synthesis of \arucl, and Section \ref{devices} discusses the isolation of monolayer films and incorporation into van der Waals heterostructures. In Section \ref{transport} numerous studies on the electronic properties of \arucl\ flakes in isolation, or in close proximity to other 2D materials---primarily graphenes---are reviewed. Section \ref{tunneling} reviews numerous recent advances in tunneling spectroscopy of \arucl\ flakes, while Section \ref{lightmatter} discusses optical probes of \arucl, and Section \ref{neutrons} presents results on the 2D nature of \arucl\ that can be extracted from bulk neutron probes. Finally we close with a brief discussion of the prospects for future discoveries in ultrathin Kitaev materials.

\section{The Kitaev model:\ material realizations and new candidates}\label{kitaev}

\begin{figure*} 
    \centering 
    \includegraphics[width = 0.4\textwidth]{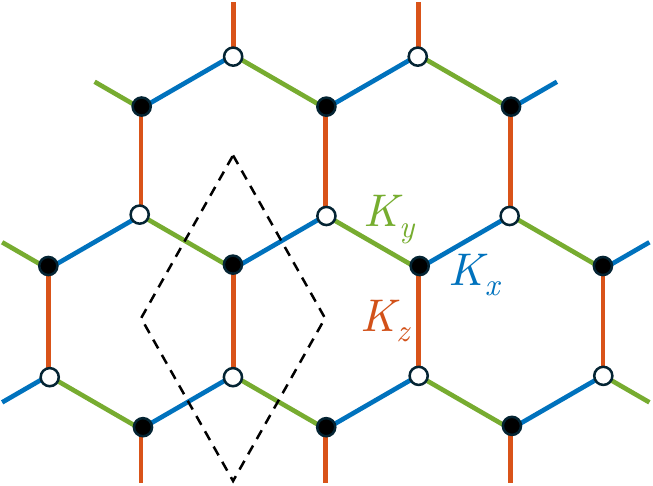}
    \caption{(color online) Honeycomb lattice in Kitaev's model with bond-direction-dependent couplings $K_i$, $i=x,y,z$. Open and closed circles identify the two triangular sublattices. The dashed outline marks red line the unit cell of the lattice.}
    \label{honeycomb}
\end{figure*}

Searching for suitable candidate systems began shortly after the Kitaev  model was proposed \cite{Kitaev2006}. The model describes a spin-1/2 system on a honeycomb lattice, hosts both gapped and gapless spin liquid ground states, and can be solved exactly by a variety of methods \cite{Kitaev2006,Feng2007,Vidal2008,ChenNussinov,NussinovOrtiz,Kells2009,NussinovBrink}. In the original Kitaev model, the spin-1/2 degrees of freedom interact via Ising-like couplings with a magnetic easy axis that depends on the spatial orientation of the bond. As the honeycomb lattice offers three different bonds $x$, $y$, and $z$ (each rotated at 120$\degree$), the Hamiltonian for this model can be  written as
\begin{equation}
\label{eq:Kit0}
H=-K_x\sum_{x-bonds}\sigma_i^x\sigma_j^x-K_y\sum_{y-bonds}\sigma_i^y\sigma_j^y-K_z\sum_{z-bonds}\sigma_i^z\sigma_j^z \end{equation}
where $K_x$, $K_y$, and $K_z$ are coupling constants, $\sigma^x$, $\sigma^y$, and $\sigma^z$ are Pauli operators and indices $i$, $j$ refer to the two sublattices of the honeycomb lattice. A unit cell of the lattice contains one vertex belonging to each sublattice (see Fig.\ \ref{honeycomb}). Interestingly, this Kitaev honeycomb model is exactly equivalent to a $p$-wave BCS type Hamiltonian on a square lattice \cite{ChenNussinov,NussinovBrink}.

Because the interactions of neighboring bonds cannot be satisfied simultaneously, frustration hampers magnetic ordering and may trigger a KQSL phase. Importantly, this frustration results from the innate ``compass'' like character of Eq.\ \ref{eq:Kit0}, where the internal spin components are associated with interactions along different lattice directions. Numerous systems display such frustration \cite{compass}. In general, the spin operators in Eq.\ \ref{eq:Kit0} need not reflect the bare spin, but are generally pseudospins that may further involve the orbital state.

When one of the couplings in Eq.\ \ref{eq:Kit0} is sufficiently larger than the others (so that the moduli of all three $|K_{x,y,z}|$ do not satisfy the triangle inequalities and no longer form the sides of a triangle), the system enters a gapped QSL phase, the ``A phase'' of the model. When all of the couplings are equal, the system is maximally frustrated and becomes a QSL featuring gapless Majorana excitations \cite{Kitaev2006,pachos,ChenNussinov,NussinovBrink,annurev:/content/journals/10.1146/annurev-conmatphys-033117-053934,TREBST20221}. This gapless QSL---the ``B phase'' of the model---persists so long as the couplings $|K_{x,y,z}|$  satisfy the triangle inequalities, and so occupies a finite region centered about the maximally frustrated symmetric $K_{x}=K_{y}={K_{z}}$ point. 

The A phase of Eq.\ \ref{eq:Kit0} hosts Abelian Z$_2$ topological order that is adiabatically connected to the one  present in Kitaev's toric code model \cite{Kitaev2003,Kitaev2006,pachos,ChenNussinov,NussinovBrink,annurev:/content/journals/10.1146/annurev-conmatphys-033117-053934,TREBST20221}; but the more ``interesting'' region with non-Abelian statistics is the gapless region centered on the maximally frustrated point. In order to 
open a gap and render the non-Abelian anyon excitations stable, one may augment the Hamiltonian of Eq.\ \ref{eq:Kit0} by an additional Zeeman-like term, which may be emulated by an effective next-nearest-neighbor three-spin coupling \cite{Kitaev2006,NussinovBrink,TREBST20221}.

In real Kitaev material candidates, the ions that form the honeycomb lattice also experience additional interactions beyond the Kitaev model. These ``Kitaev-Heisenberg'' models can be described by additional isotropic Heisenberg term amending the Kitaev interactions in Eq. \ref{eq:Kit0} \cite{Jackeli2009,plumb_2014,NussinovBrink}. A more complex and much studied Hamiltonian, the $JK\Gamma$ model \cite{Rau2014,Winter2017}, includes further off-diagonal antiferromagnetic terms:
\begin{equation}
\label{Ham:JKG}
    H=\sum_{\langle ij \rangle|| \gamma-bonds,\ \alpha \neq \beta \neq \gamma}(J\boldsymbol{S}_i \cdot \boldsymbol{ S}_j- 4 K_\gamma
S_i^{\gamma}S_j^{\gamma}+\Gamma(S_i^{\alpha}S_j^{\beta}+S_i^{\beta}S_j^{\alpha})).
\end{equation}
The nearest neighbor sites $\langle i j \rangle$ define the bond direction $\gamma$. 
Here $J$ is the Heisenberg exchange, $K$ the Kitaev exchange, and the $\Gamma$ terms capture off-diagonal exchange. The operator $\boldsymbol{S}_i$ is a (pseudo)spin located at the $i$-th lattice site. In the simplest case of Eq.\ \ref{Ham:JKG}, whenever symmetry considerations are applicable in viable material realizations, the Kitaev couplings and off-diagonal couplings are uniform along all lattice directions; strain and other effects can yield anisotropic couplings. The Hamiltonian of Eq.\ \ref{Ham:JKG} can be further generalized by allowing for (pseudo)spins $S$ of arbitrary size beyond the spin $S=1/2$ system of Eq.\ \ref{eq:Kit0}.

The first two terms of this model are well studied and reviewed in the literature \cite{Chaloupka2010, Chaloupka2013, Gohlke2017, NussinovBrink,PhysRevB.90.195102,PhysRevB.84.100406}. When the Heisenberg coupling is large compared to the Kitaev one, $|J|\gg |K|$, the ground state has a N\'{e}el antiferromagnetic (AFM) order for $J>0$ and ferromagnetic (FM) order for $J<0$. When the two couplings have similar magnitude, the ground state will depend on their signs with viable 
zig-zag 
(alternating zigzaging 
ferromagnetic chains) 
and ``stripy'' orders 
\cite{NussinovBrink,Chaloupka2013, Janssen_2019}. 

When the $\Gamma$ term is included, new magnetic phases further appear such as incommensurate spiral order or 120$\degree$ order, where spins are in a coplanar spiral that is either incommensurate to the lattice or at relative angles of $0$ or $\pm2\pi/3$ on the same sublattice \cite{Rau2014,Winter2017}. Additional refinements of the Hamiltonian of Eq.\ \ref{Ham:JKG} include the incorporation of not only the nearest neighbor Kitaev $K_1 \equiv K$, Heisenberg exchange $J_1 \equiv J$, and off-diagonal $\Gamma_1 \equiv \Gamma$ couplings 
but also couplings ($K_n,J_{n},\Gamma_{n}$) between spins that are $n^{th}$-nearest neighbors.


We reiterate that, fundamentally, in the $S=1/2$ Kitaev-Heisenberg type models and their descendants, the $S=1/2$ degrees of freedom generally do not solely represent bona fide spins. Instead, 
the low-energy single-ion states are spanned by effective $S=1/2$ pseudo-spin Kramers doublets that reflect spin and orbital mixing due to strong spin-orbit coupling. In the respective up and down pseudospin states, states of definite orbital angular momentum and spin polarization are superposed. Many low energy effective Kitaev-Heisenberg models are primarily derived by considering multiple \textit{superexchange} processes that may contribute to coupling between the pseudospin degrees of freedom \cite{Jackeli2009,plumb_2014}.

In the rest of this Section we review numerous candidate materials to host Kitaev interactions, with an emphasis on layered/two-dimensional systems. Not all materials are necessarily exfoliatable, but by casting a wide net we hope to spur discovery of novel systems that can be exfoliated or grown as thin films.

\subsection{Na$_2$IrO$_3$ and \arucl}
\label{sec:NR}

Several years after Kitaev introduced and solved the model of Eq.\ \ref{eq:Kit0} in order to illustrate concepts in topological quantum computing \cite{Kitaev2006,Nayak2008,KL,NussinovBrink,Simonbook}, numerous works suggested that when augmented with additional interactions, such as those in Eq.\ \ref{Ham:JKG}, the basic model Hamiltonian can be directly realized in various candidate materials. Historically the first compound suggested to host dominant Kitaev interactions was Na$_2$IrO$_3$, proposed by Jackeli and Khaliullin in 2009 \cite{Jackeli2009}. This was soon followed by a series of studies on other candidate materials such as Li$_2$IrO$_3$ and \arucl\  \cite{Modic2014-kl,Takayama2015,plumb_2014}. These materials are characterized by a honeycomb lattice, edge-sharing octahedra, and heavy $4d/5d$ transition metals of  spin-$1/2$. 

For Na$_2$IrO$_3$ \cite{hou2018first}, a minimal $J_1$-$K_1$-$\Gamma_1$-$J_3$ model has been proposed, with dominant ferromagnetic Kitaev interaction $K_1 = -10.00$ meV, antiferromagnetic Heisenberg couplings $J_1 = 1.63$ meV and $J_3 = 0.83$ meV, and an off-diagonal interaction $\Gamma_1 = 0.90$ meV. This model yields a zigzag antiferromagnetic ground state, consistent with experimental observations. Due to structural distortions, the three nearest-neighbor bonds in both Na$_2$IrO$_3$ and \arucl\ are not equivalent; they split into two (X and Y) and one (Z) bond types. The exchange parameters cited are typically averaged over these inequivalent bonds. Importantly, both Na$_2$IrO$_3$ and \arucl\ exhibit ferromagnetic Kitaev interactions ($K<0$) \cite{Das2019,ran_spin-wave_2017,koitzsch_low-temperature_2020,Sears2020-om}, although the pure Kitaev model itself does not impose constraints on the sign of $K$.

At low temperatures, both Na$_2$IrO$_3$ and \arucl\ adopt in-plane zigzag AFM orders. This state can arise from the combination of FM Kitaev interaction and sizable off-diagonal exchange $\Gamma_1$ \cite{Janssen_2019}. For \arucl, DFT calculations \cite{Hou2017} have determined that: (i) the first-nearest-neighbor (1NN) interactions include $J_1 = -1.8$ meV (FM), $K_1 = -10.6$ meV (FM), and $\Gamma_1 = 3.8$ meV (from IrO$_6$ distortions); (ii) second-nearest-neighbor interactions are negligible; and (iii) third-nearest-neighbor couplings include $J_3 = 1.25$ meV (AFM) and $K_3 = 0.65$ meV. These results suggest that a $J_1$-$K_1$-$\Gamma_1$-$J_3$-$K_3$ model is needed to describe the magnetism of \arucl. The competition between $J_1$ and $J_3$ leads to a zigzag AFM ground state rather than a KQSL.

Further insights from ab initio quantum chemistry calculations \cite{Bhattacharyya2024} suggest that in \arucl, the direct Coulomb exchange contribution for the Kitaev coupling is approximately 1.75 meV. By comparison, the ligand-mediated superexchange and dynamical correlations contribute $\sim$ 2 meV. Thus direct exchange cannot be neglected by comparison to the indirect superexchange pathways originally proposed by Jackeli and Khaliullin \cite{Jackeli2009}. This observation may require revisiting the microscopic origin of bond-directional interactions in such 4$d$ systems.

While Na$_2$IrO$_3$ and \arucl\ laid the groundwork for Kitaev materials, they are constrained by ferromagnetic $K$ and fixed pseudospin-1/2. Recent progress has expanded the scope to 3$d$ transition metal systems with antiferromagnetic Kitaev interactions and larger spins. For instance, $d^7$ cobaltates were proposed to host pseudospin-1/2 states \cite{liu2018pseudospin,liu2020kitaev}, though strong Kitaev terms were not initially found. Later, CrI$_3$ and CrGeTe$_3$ were predicted to exhibit sizable Kitaev interactions via ligand SOC \cite{xu2018interplay,xu2020possible}, and NiI$_2$ was identified as an $S=1$ Kitaev candidate with triangular geometry \cite{stavropoulosMicroscopicMechanismHigherSpin2019}. In parallel, rare-earth magnets with spin-orbit-entangled Kramers doublets have emerged as promising platforms, offering a complementary route to Kitaev physics~\cite{Motome2020}. All these developments have opened new directions for exploring Kitaev physics, as discussed in the next sections.

Nonetheless, \arucl\ remains the most-explored Kitaev layered system, driven largely by measurements in bulk materials that have identified phenomena related to the Kitaev magnetism. Most prominently these include claims of a half-integer quantized thermal Hall conductance \cite{kasahara_majorana_2018}, which though still under discussion, has captured the imagination of the quantum magnetism community.

\subsubsection{\arucl\ in bulk}\label{bulk}

\begin{figure*} 
    \centering 
    \includegraphics[width = 0.6\textwidth]{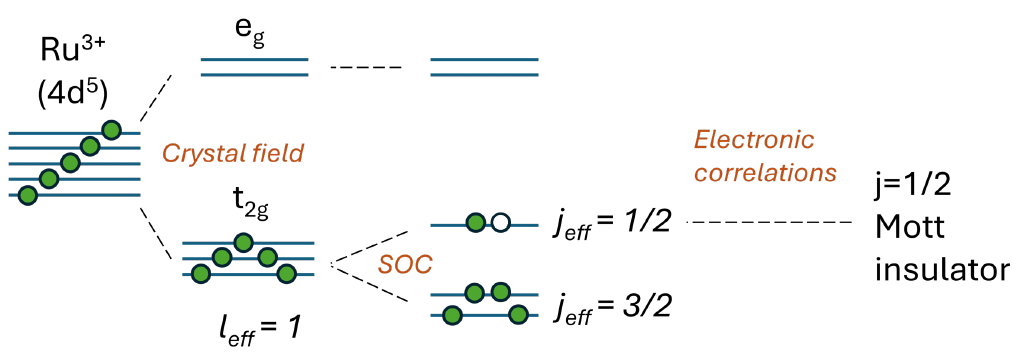}
    \caption{(color online) Crystal field effects, spin-orbit coupling, and electronic correlations lead to the spin-orbit entangled moments $j=1/2$.}
    \label{EnergyLevels}
\end{figure*}

\arucl\ was originally identified as a magnetic semiconductor \cite{Hill1950,FLETCHER1963,Hyde1965,Fletcher1967,binotto_optical_1971,guizzetti_fundamental_1979,rojas_hall_1983}. Although the atomic SOC in Ru is relatively weak compared to Ir, electronic correlations in \arucl\ lead to a reduction in the band width near the Fermi level, where spin and orbital degrees of freedom are entangled. Per Fig.\ \ref{EnergyLevels}, electrons in the Ru ions having a 4$d^5$ configuration are in a low-spin state with $S = 1/2$ and under crystal fields effects of the RuCl$_6$ octahedra, with $l_{eff}$=1. Spin orbit coupling further splits a $j_{eff} = 1/2$ from a $j_{eff} = 3/2$ band, and finally  electronic correlations open an energy gap in the j$_{eff}$ = 1/2 band, creating a correlation-induced insulating state \cite{kim_kitaev_2015,koitzsch_j_eff_2016,Motome2020}. This, together with nearly 90$\degree$ superexchange paths in an almost ideal octahedral crystal, realizes the Kitaev interaction that attempts to align the ${x,y,z}$ components of spin along each of the three nearest-neighbor bonds.

The magnetic properties of \arucl\ were soon subjected to numerous experimental and theoretical investigations. Magnetic susceptibility, specific heat measurements, and numerous other experiments suggested that below approximately 7 K in high-quality crystals (rising to 14 K in samples with structural disorder), antiferromagetic order appears in \arucl\ due to several non-Kitaev interactions \cite{johnson_monoclinic_2015,sears_magnetic_2015,majumder_anisotropic_2015,cao2016low,leahy_anomalous_2017}. The effect of these less desirable terms can be suppressed by applying an in-plane magnetic field of around 8 T \cite{johnson_monoclinic_2015,Yadav2016,zheng_gapless_2017,little_antiferromagnetic_2017}. Meanwhile, \arucl\ undergoes a structural phase transition near 150 K from a monoclinic structure ($C2/m$) at high temperatures to trigonal ($R\bar{3}$) at low temperatures, accompanied by a sizable magnetic susceptibility hysteresis \cite{PhysRevB.108.144419,PhysRevB.109.L140101,Park2024-fz,kubota_successive_2015,cao2016low,Glamazda2017}. 

Raman scattering studies provided evidence for unconventional magnetic excitations in \arucl\, characterized by a broad continuum over a wide temperature range from 7 K to 120 K, suggesting the presence of fermionic quasiparticles from fractionalization of spins in \arucl, consistent with the Kitaev model  \cite{sandilands_scattering_2015,nasu_fermionic_2016,wulferding_magnon_2020}.  Spectral techniques including nuclear magnetic resonance (NMR) and inelastic neutral scattering (INS) were also employed to explore the fractional quantum excitations. The NMR results provided evidence suggesting that \arucl\ exhibits a magnetic-field-induced QSL state and a field-dependent spin excitation gap \cite{baek_evidence_2017,zheng_gapless_2017,cui_high-pressure_2017,jansa_observation_2018}, as predicted in the Kitaev model \cite{Kitaev2006,yoshitake_fractional_2016}; INS experiments revealed quasielastic excitations at low energies around the Brillouin zone centre and an hour-glass-like magnetic continuum at high energies, which points towards the emergence of Majorana fermions in \arucl\ \cite{banerjee_proximate_2016, banerjee_neutron_2017,do_majorana_2017,ran_spin-wave_2017}. These fractional excitations in Kitaev QSL are described as itinerant Majorana fermions and localized Z$_2$ fluxes.

Since 2017, a series of thermal transport experiments on \arucl\ yielded rather unusual behavior, generally interpreted in terms of Majorana excitations. An early heat transport experiment reported a striking enhancement of the in-plane thermal conductivity and an anomalous torque response after the magnetic order was suppressed by high magnetic fields, which was interpreted in terms of the Kitaev model \cite{leahy_anomalous_2017}. Shortly thereafter, thermal Hall transport measurements in tilted magnetic fields revealed a half-quantization of thermal Hall conductance, considered a smoking gun for the presence of Majorana fermions \cite{kasahara_majorana_2018,yokoi_half-integer_2021,bruin_robustness_2022,imamura_majorana-fermion_2024,Matsuda2025}. These findings are the subject of ongoing debate, with potential contributions from other thermal transport mechanisms including phonons and magnons \cite{yu_ultralow-temperature_2018,vinkler-aviv_approximately_2018,PhysRevLett.121.147201,czajka_oscillations_2021,li_giant_2021,Lefranois2022,Lefranois2023,czajka_planar_2023,Dhakal2024-np}. For instance, a recent theoretical study suggests that the unusual thermal Hall signals may be attributed to topological Majorana fermions over a wide range of temperatures and magnetic fields \cite{okuboThermalHallTransport2025}. Sample dependence surely plays a role, and has been explored in recent studies \cite{Kasahara2022,PhysRevMaterials.7.114403,Namba2024,PhysRevMaterials.8.014402}. Indeed, the potential for stacking faults in this layered material is a strong motivation for pursuing studies in exfoliated single- and/or few-layered samples.

\subsection{CrI$_3$, Cr$_2$Ge$_3$Te$_6$, and related materials}

The experimental discovery of ferromagnetism in monolayer CrI$_3$ and Cr$_2$Ge$_4$Te$_6$ in 2017 helped to make two-dimensional magnets a major research focus \cite{huang2017layer,gong2017discovery}. Both materials exhibit van der Waals layered structures, with Cr$^{3+}$ ions ($3d^3$, $S=3/2$) located at the centers of edge-sharing CrI$_6$ or CrTe$_6$ octahedra that form a honeycomb lattice. The high crystalline symmetry ensures that all Cr-Cr bonds are equivalent, resembling the geometry of \arucl. However, a key distinction is the effective spin-orbit coupling in these Cr-based compounds originates primarily from the heavy ligands (I or Te) \cite{lado2017origin,xu2018interplay}, in contrast to \arucl\ where the SOC is hosted by the $4d$ transition metal. This mechanism is captured by a tight-binding model developed on a Cr$_2$L$_2$ cluster (L = I, Te), showing that the Kitaev interaction scales as $K \propto \lambda_L^2 V_{pd\pi}^4$, where $\lambda_L$ is the SOC strength of the ligand and $V_{pd\pi}$ is the $pd$ hopping integral \cite{xu2018interplay}. This result establishes a new route for engineering Kitaev interactions in 3$d$ magnets using heavy ligands, and opens the possibility of tuning such interactions through ligand choice or chemical substitution.

Another key distinction is the sign of the Kitaev interaction. In Cr-based systems, the Kitaev term is antiferromagnetic ($K>0$), in contrast to the ferromagnetic Kitaev interactions typically found in $d^5$ systems like \arucl. From the cluster-level analysis, this sign can be traced to the SOC-induced anisotropic hopping paths between ligand $p$ orbitals, which introduce an AFM contribution when the spins are oriented perpendicular to the Cr$_2$L$_2$ plane. 

Although both CrI$_3$ and Cr$_2$Ge$_2$Te$_6$ are ferromagnetic in their monolayer form, they exhibit markedly different magnetic anisotropies:\ CrI$_3$ shows strong Ising-like behavior, whereas Cr$_2$Ge$_2$Te$_6$ is nearly isotropic. First-principles calculations reveal these differences arise from the interplay between single-ion anisotropy (SIA) and anisotropic exchange interactions---particularly the Kitaev interaction \cite{xu2018interplay}. Both materials host sizable Kitaev interactions, even though they are based on 3$d$ transition metal ions; here the pseudospin in the effective Kitaev model description the real spin, $S = 3/2$. The ratio $|K/J|$ is estimated to be $\sim 0.35$ in CrI$_3$ and $\sim 0.054$ in Cr$_2$Ge$_2$Te$_6$, indicating the Kitaev term plays a non-negligible role in the spin dynamics and magnon dispersion of these systems.


Epitaxially strained Cr-based monolayers like Cr$_2$Ge$_2$Te$_6$ (and especially CrSiTe$_3$) may host KQSL phases in $S=3/2$ systems, making them promising platforms to explore beyond the traditional $S=1/2$ models \cite{xu2020possible}; a phase diagram analysis shows that antiferromagnetic Kitaev interactions with $S=3/2$ can stabilize rich nontrivial magnetic states, especially under the application of external fields. This highlights the importance of exploring higher-spin Kitaev models and paves the way for future studies in 3$d$-based materials.

These findings significantly broaden the landscape of Kitaev materials, suggesting that non-negligible Kitaev interactions may be widespread across Cr-based and more generally 3$d$ systems. In particular, transition metal trihalides (MX$_3$) and trichalcogenides (MGY$_3$), where X or Y are heavy elements (I, Te, Se), represent promising material families. Due to the sensitivity of Heisenberg interactions to lattice constants, some of these systems may lie near a magnetic instability between FM and AFM states, where Kitaev terms can dominate and quantum spin liquid physics may emerge.

\subsection{NiI$_2$ and related materials}

Nickel diiodide (NiI$_2$) has recently emerged as a promising platform for hosting Kitaev interactions and 2D multiferroicity \cite{ju2021possible,song2022evidence}. Bulk NiI$_2$ is a van der Waals layered material that crystallizes in a rhombohedral structure (space group $R\bar{3}m$), where each layer comprises edge-sharing NiI$_6$ octahedra forming a triangular magnetic lattice.  Ni$^{2+}$ ions adopt a high spin $3d^8$ configuration, resulting in spin $S=1$ and a local moment of $\sim 2\ \mu_B$ \cite{kuindersma1981magnetic}. Below 75 K, bulk NiI$_2$ exhibits a proper screw (PS) magnetic ground state with an in-plane propagation along the $\langle 1\bar{1}0 \rangle$ direction and a 55$^{\circ}$ canting of the spin rotation plane, giving rise to a type-II multiferroic spin-induced polarization \cite{kurumaji2013magnetoelectric}. 

Strong Kitaev interactions can arise in high-spin $S=1$ systems like NiI$_2$, due to strong spin-orbit coupling on iodine and large Hund's coupling on the transition metal as shown in a tight-binding model \cite{stavropoulosMicroscopicMechanismHigherSpin2019}. This model further predicts an antiferromagnetic Kitaev term ($K>0$) in monolayer NiI$_2$. Here, the $S=1$ Kitaev based model directly uses the triplet manifold of the high-spin 3d$^8$Ni$^{2+}$.

This analysis has been extended using density functional theory combined with the four-state energy-mapping method \cite{amoroso2020spontaneous}. These results suggest that the nearest-neighbour exchange tensor in monolayer NiI$_2$ is strongly anisotropic, with a nearest-neighbour coupling $J_1$ tensor having principal eigenvalues $\lambda_\alpha, \lambda_\beta, \lambda_\gamma \approx -8.1,\,-8.0,\,-4.8$ meV. This corresponds to an antiferromagnetic Kitaev term $K \approx 3.2$ meV, giving a ratio $|K/J_{1}| \approx 0.4$; the nearest neighbor $J_1 \approx -7.0$ meV employed in the latter ratio is the isotropic value of the exchange (this isotropic nearest neighbor  $J_1$ value is the average of the above principal  eigenvalues).

In another recent study, a realistic first-principles-based spin Hamiltonian for NiI$_2$ was constructed using a symmetry-adapted cluster expansion method \cite{li2023realistic}. Crucially, the presence of a sizable AFM Kitaev interaction was confirmed along with identification of a significant biquadratic term. The model includes Heisenberg, Kitaev, biquadratic, single-ion anisotropy, and interlayer couplings. Monte Carlo and conjugate-gradient simulations reproduce the proper screw ground state with $\langle 1\bar{1}0 \rangle$ propagation, long period ($\sim 7.3a$), and $\sim 46^\circ$ canting (consistent with experimental values). The analysis reveals that the Kitaev interaction, not $J$-$J_3$ frustration, stabilizes both the propagation and canting. Additionally, the biquadratic term favors collinearity and helps refine the ground state, while the Kitaev-induced anisotropy underpins the multiferroicity. Topological spin textures such as merons and antimerons are also predicted.

Taken together, these studies not only confirm the viability of achieving Kitaev physics with $3d$ transition metals and heavy ligands but also extend Kitaev candidate materials to triangular lattices. A broader family of compounds with the general formula MX$_2$ and MY$_2$ (M = $3d$ transition metal; X = I, Br; Y = Te, Se) may host similar physics and deserves further exploration.

\subsection{Na$_3$Co$_2$SbO$_6$ and related materials}

Na$_3$Co$_2$SbO$_6$ was proposed as a possible Kitaev candidate by Refs.\ \cite{sanoKitaevHeisenbergHamiltonianHighspin2018,liuPseudospinExchangeInteractions2018,liu2020kitaev}. This system also possesses 3$d$ electrons; the CoO$_6$ octahedra share edges and form honeycomb lattice. The Sb ions locate at the hollow sites of honeycomb hexagons. In the octahedral environment, Co$^{2+}$ ion with $t^5_{2g}e^2_g$ configuration possesses spin $S = 3/2$ and unquenched orbital angular momentum, resulting in a pseudospin of $S=1/2$ in the Kitaev interactions. A tight-binding analysis  finds that Na$_3$Co$_2$SbO$_6$ is proximate to a KQSL phase \cite{liu2020kitaev}. Recently, inelastic neutron scattering experiments and spin wave fitting were carried out \cite{kim2021antiferromagnetic};  a model with 1NN and 3NN Heisenberg terms ($J_1$ and $J_3$), Kitaev interactions ($K$), and off-diagonal terms ($\Gamma$ and $\Gamma'$ ), was fit to the measured magnon spectrums. The conclusions are that (i) Na$_3$Co$_2$SbO$_6$ exhibits AFM Kitaev coefficient with $K=3.60$ meV; (ii) the Heisenberg terms yield $J_1$=-4.70 meV and $J_3$=0.95 meV; and (iii) off-diagonal terms are non-negligible with $\Gamma$ = 1.30 meV and $\Gamma'$ = -1.40 meV.

Other potential Na$_3$Co$_2$SbO$_6$-type Kitaev systems include cobalt based systems Na$_2$Co$_2$TeO$_6$, BaCo$_2$(AsO$_4$)$_2$ and another nickel based system Na$_3$Ni$_2$BiO$_6$ \cite{stavropoulosMicroscopicMechanismHigherSpin2019,liu2020kitaev,Zhong2020,Zhang2022}. For cobalt based systems,  the results are rather controversial. Some studies are uncertain about the sign of the Kitaev interaction \cite{songvilayKitaevInteractionsCo2020, samarakoonStaticDynamicMagnetic2021, kim2021antiferromagnetic, sandersDominantKitaevInteractions2022, linFieldinducedQuantumSpin2021}, while others even question its existence \cite{guInplaneMultiqMagnetic2024, yaoExcitationsOrderedParamagnetic2022}; clearly further experimental and computational works are in demand. Unlike Co$^{2+}$, Ni$^{2+}$-based systems exhibit $3d^8$ configuration with $S=1$ and $L=0$. It was believed that Na$_3 $Ni$_2$BiO$_6$ may host strong Kitaev interaction, as the substantial SOC of Bi could potentially induce it through proximity. However, recent work obtained a realistic model for such nickel system and excludes this proximity mechanism, as orbital hybridization between Ni-$d$ orbitals and Bi-$p$ orbitals is absent \cite{chenstrength2025}.

\subsection{Rare-earth materials}
Rare earth magnets, with spin-orbit-entangled Kramers doublets, offer a complementary route to realizing Kitaev physics. Microscopic analyses of edge sharing rare earth networks---exemplified by ytterbium-based systems---have shown how super-exchange in this geometry can yield strongly anisotropic interactions on honeycomb and related lattices, motivating a search for $f$-electron Kitaev platforms beyond the $d$-electron cases discussed above \cite{Rau2018}. Subsequently, ab initio-based studies predicted that the honeycomb oxides $A_2$PrO$_3$ ($A$= alkali metal) host antiferromagnetic Kitaev couplings for the Pr$^{4+}$ $f^{1}$ $\Gamma_7$ doublet, in contrast to the typically ferromagnetic Kitaev couplings in 4$d$ and 5$d$ materials, and mapped systematic trends across both honeycomb and hyperhoneycomb frameworks in this family \cite{Jang2019, Jang2020}. A large scale computational survey further expanded the design space to all 4$f$ configurations with Kramers doublets, identifying regimes ---particularly $f^3$ (Nd$^{3+}$) and $f^{11}$ (Er$^{3+}$)--- where the Kitaev coupling $\lvert K\rvert$ can rival or exceed the Heisenberg coupling $\lvert J\rvert$, providing broad guidance for rare earth material discovery \cite{Jang2024a}. For a comprehensive overview of these design principles and their relation to the Jackeli-Khaliullin mechanism, see the review in Ref.\ \cite{Motome2020}.

Experimentally, the pioneering work of Hinatsu established synthesis strategies for layered rare earth oxides with honeycomb arrangements \cite{Hinatsu2006}, laying the foundation for later targeted studies of Kitaev candidates. Building upon and extending this approach, recent efforts achieved selective synthesis and structural characterization of honeycomb and hyperhoneycomb Na$_2$PrO$_3$, providing a platform to test theoretical predictions of anisotropic exchange \cite{Okuma2024a}. Inelastic neutron scattering on honeycomb $\alpha$-Na$_2$PrO$_3$ reveals low energy magnetism best described by a dominant Heisenberg term ($J\approx 1.1$~meV) with at most a subdominant AFM Kitaev component ($\lvert K\rvert\lesssim0.2 J$), indicating that this member lies on the $J$-dominated side of the rare earth Kitaev landscape \cite{Daum2021}. 

By contrast, single crystal studies on hyperhoneycomb $\beta$-Na$_2$PrO$_3$ uncover a strongly non-collinear order with highly dispersive gapped excitations consistent with compass model physics dominated by bond dependent off diagonal ($\Gamma$) anisotropy, demonstrating that tri-coordinated rare earth lattices can realize generalized compass limits distinct from a pure Kitaev regime \cite{Okuma2024b}. These experimental findings prompted a re-evaluation of the exchange interactions in $A_2$PrO$_3$, which showed that realistic octahedral crystal field splittings can suppress $K$ below $J$, reconciling the original AFM Kitaev prediction with the observed $J$- or $\Gamma$-dominated behavior \cite{Jang2024b}.

Beyond Pr$^{4+}$, rare earth honeycomb halides illustrate additional pathways. For instance, SmI$_3$ hosts a $\Gamma_7$ Kramers doublet on the edge sharing network with antiferromagnetic correlations and no long range order down to $0.1$~K, satisfying key local criteria for AFM Kitaev interactions while suggesting proximity to a disordered regime \cite{Ishikawa2022}. Additionally, YbCl$_3$ realizes a $J_{\rm eff}=1/2$ Kramers doublet on a honeycomb lattice with strong in-plane anisotropy, a reduced ordered moment, and field tunable phase boundaries, consistent with a proximate Kitaev regime despite the emergence of low temperature NÃ©el order \cite{Xing2020}.

Taken together, rare-earth honeycomb and hyperhoneycomb materials have emerged as promising platforms for Kitaev and generalized compass physics. They combine robust ingredients---spin-orbit entangled Kramers doublets and edge sharing geometries---with a remarkable tunability of exchange anisotropy. The balance among Heisenberg, Kitaev, and off-diagonal interactions depends sensitively on ion choice, crystal field scale, and lattice topology, enabling regimes that range from $J$-dominated to compass model limits. This tunability, together with the diversity of rare earth configurations, positions these systems as fertile ground for realizing proximate Kitaev spin liquids and exploring field induced quantum phases.

\section{\arucl\ synthesis}\label{synthesis}

\subsection{Crystal growth}

\begin{figure*}[t!]
    \centering 
    \includegraphics[width = 0.5\textwidth]{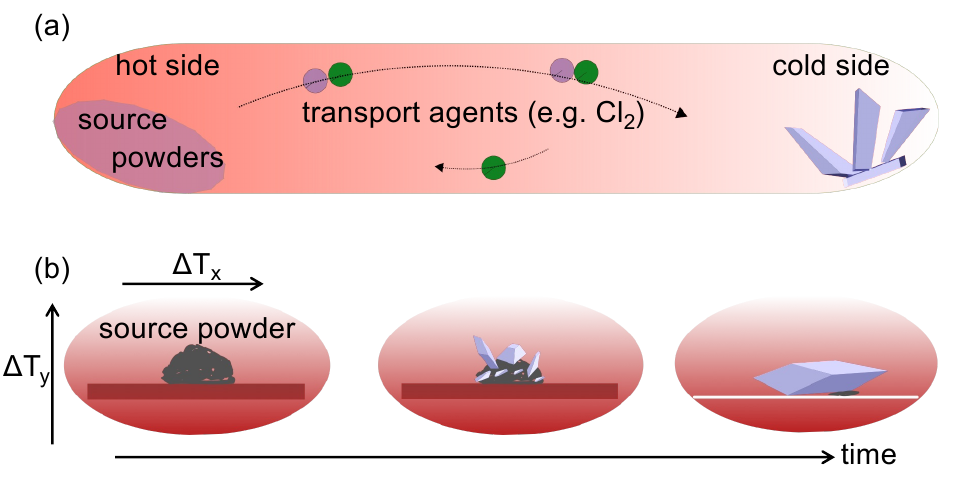}
    \caption{(color online) Top: Illustration of (a) typical vapor transport growth with a large horizontal temperature gradient, and (b) self-selecting vapor transport growth performed with a small temperature gradient around the starting powder. Weak temperature gradients applied in the horizontal and vertical directions help the grain selection. Over time, part of the starting powder is converted to small crystals growing on top of the powder; at the end of an ideal growth, one large single crystal forms consuming all starting powder as well as the intermediate smaller crystals \cite{yan2023self}. American Physical Society (APS).}
    \label{crystals-1}
\end{figure*}

\arucl\ does not melt but instead sublimes at high temperatures. Consequently, all crystal growth processes are conducted via vapor transport, leveraging its high vapor pressure effectively. While often referred to by different names such as chemical vapor transport, sublimation, Bridgman, and self-selecting vapor growth, all growth methods in the literature fall under the category of vapor transport growth (see Ref.\ \cite{kim2022alpha} for a summary of the vapor transport growth conditions). These methods can be broadly classified into two types based on the temperature gradient applied during crystal growth. In the first, vapor transport in horizontal tube furnaces utilizes a large temperature gradient ranging from, for example, 50$\degree$C to 250$\degree$C,  illustrated in Fig.\ \ref{crystals-1}(a). Synthesis can be performed with the aid of a transport agent such as Cl$_2$ (at 730-660 $\degree$C in Ref.\ \cite{majumder_anisotropic_2015} or 750-650$\degree$C in Ref.\ \cite{hentrich2018unusual}) or TeCl$_4$ (700-650$\degree$C in Ref.\ \cite{may2020practical}), or even without any agent since \arucl\ itself has a reasonable vapor pressure at elevated temperatures. The starting \arucl\ powder is placed at the hotter end of a growth ampoule, and plate-like \arucl\ crystals form at the cooler end within just several hours. \arucl\ crystals grown in this manner normally appear as thin plates. This approach has been widely used in the community to produce single crystals suitable for various measurements such as magnetic, thermodynamic, thermal transport studies. $\beta$-RuCl$_3$ forms when the growth temperature is lower than 500$\degree$C \cite{Hyde1965}, while at higher temperatures it becomes unstable and transforms to \arucl.

\begin{figure*} 
    \centering 
    \includegraphics[width = 0.7\textwidth]{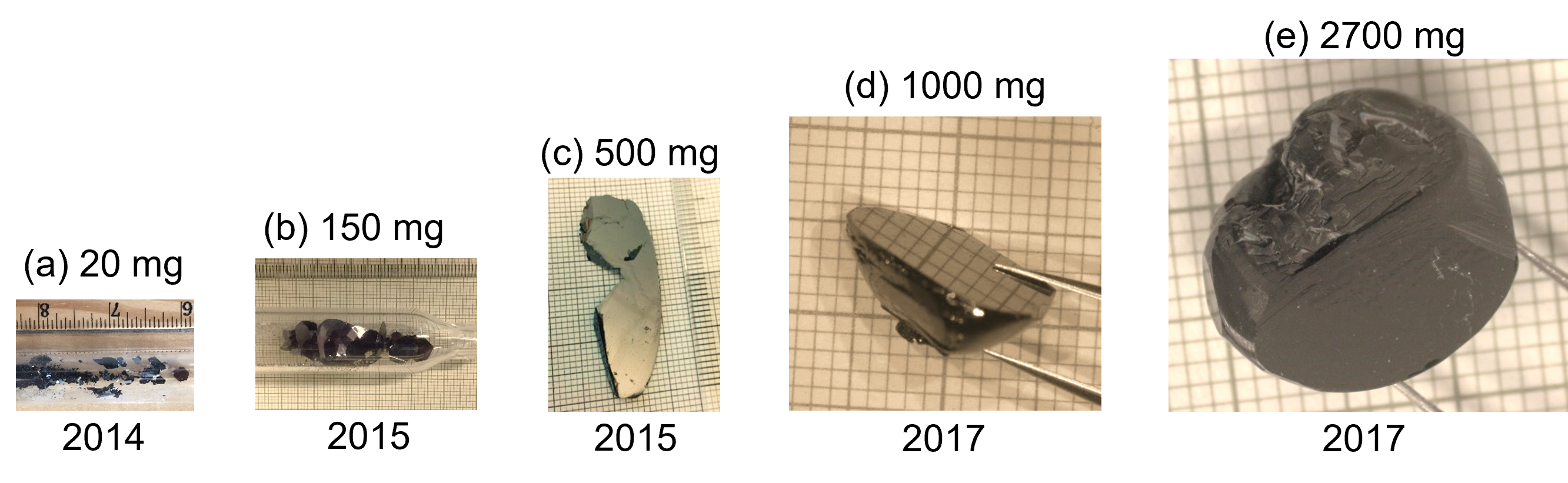}
    \caption{\arucl\ crystals grown at different stages by J-Q Yan, \etal\ \cite{yan2023self}. (a) A conventional vapor transport growth performed in a horizontal tube furnace with a temperature gradient of about 45 $^{\circ}$C between the starting powder at the hot end and crystals at the cold end. (b)-(e) Self-selecting vapor transport growths performed in a box furnace. The increasing crystal dimension results from a better control of the temperature gradient around the starting powder. The number beneath each panel shows in what year the growth was performed. Reproduced with permission from \cite{yan2023self}, APS.}
    \label{YanFig2}
\end{figure*}

The second approach, the self selecting vapor growth (SSVG) technique, employs an extremely small (less than 2$\degree$C) vertical temperature gradient \cite{yan2023self}. SSVG is normally performed in a box furnace with a well defined temperature gradient around the powder as illustrated in Fig.\ \ref{crystals-1} (bottom). In this case, the starting powder is sealed under vacuum in a fused quartz ampoule. The sealed ampoule is then put inside of a box furnace with the desired temperature gradient around the starting powder. After dwelling at a furnace set point temperature of 1060$\degree$C for 6 h, the furnace is cooled to 800$\degree$C at 2-10$\degree$C/h. When the cooling rate is in the range 2-4$\degree$C/h, all of the powder inside of the ampoule converts into one single crystal. The SSVG grown crystals are rather thick with a small mosaic that allows for inelastic neutron scattering on a single crystal, without the need to coalign many smaller pieces. Plate-like crystals can be obtained by increasing the cooling rate or the temperature gradient near the starting powder. The size and geometry of the crystals can also be tailored by adjusting the amount of starting powder and the inner diameter of the quartz ampoule. Crystals grown in this manner have minimal amount of stacking faults and order antiferromagnetically at approximately 7.5 K. See Fig.\ \ref{YanFig2} for pictures of crystals using both vapor transport and SSVG methods. Multiple stages of vapor transport can yield very pure crystals \cite{Namba2024}.


Different growth processes produce crystals with distinct thermal transport properties \cite{lee2021}. For example, crystals grown by the Bridgman technique  have higher thermal conductivity but weaker oscillatory features compared to those grown by chemical vapor transport \cite{kubota_successive_2015}. Since \arucl\ does not melt and involves only solid and vapor phases during crystal growth, syntheses performed using a Bridgman furnace are still a type of vapor transport growth. Specifically, they can be categorized as either chemical vapor transport or self-selecting vapor growth depending on the temperature gradient employed and the location of the nucleation and growth \cite{yan2023self}. As discussed latter, the nature and degree of stacking disorder can vary in crystals depending on the growth conditions and techniques employed and can result in the variation of physical properties.  

\subsection{Effects of stacking disorder}

\arucl\ is prone to stacking disorder due to the weak vdW bonding between the honeycomb layers. The importance of the layer stacking on the magnetic properties was realized at an  early stage \cite{cao2016low,banerjee_proximate_2016}.  A few magnetic anomalies are typically observed in the temperature range 7-14 K. Studies on crystals with different amount of stacking disorder suggest that the 14 K magnetic order has a different wavevector with a two-fold out-of-plane periodicity. The mixing of different types of stackings gives rise to the multiple magnetic anomalies at differing temperatures, which has been employed as an effective  indicator of the crystal quality. One single magnetic transition around T$_N$=7 K is generally taken as a simple and convenient criteria for crystal selection. 

\begin{figure*} 
    \centering 
    \includegraphics[width = 0.8\textwidth]{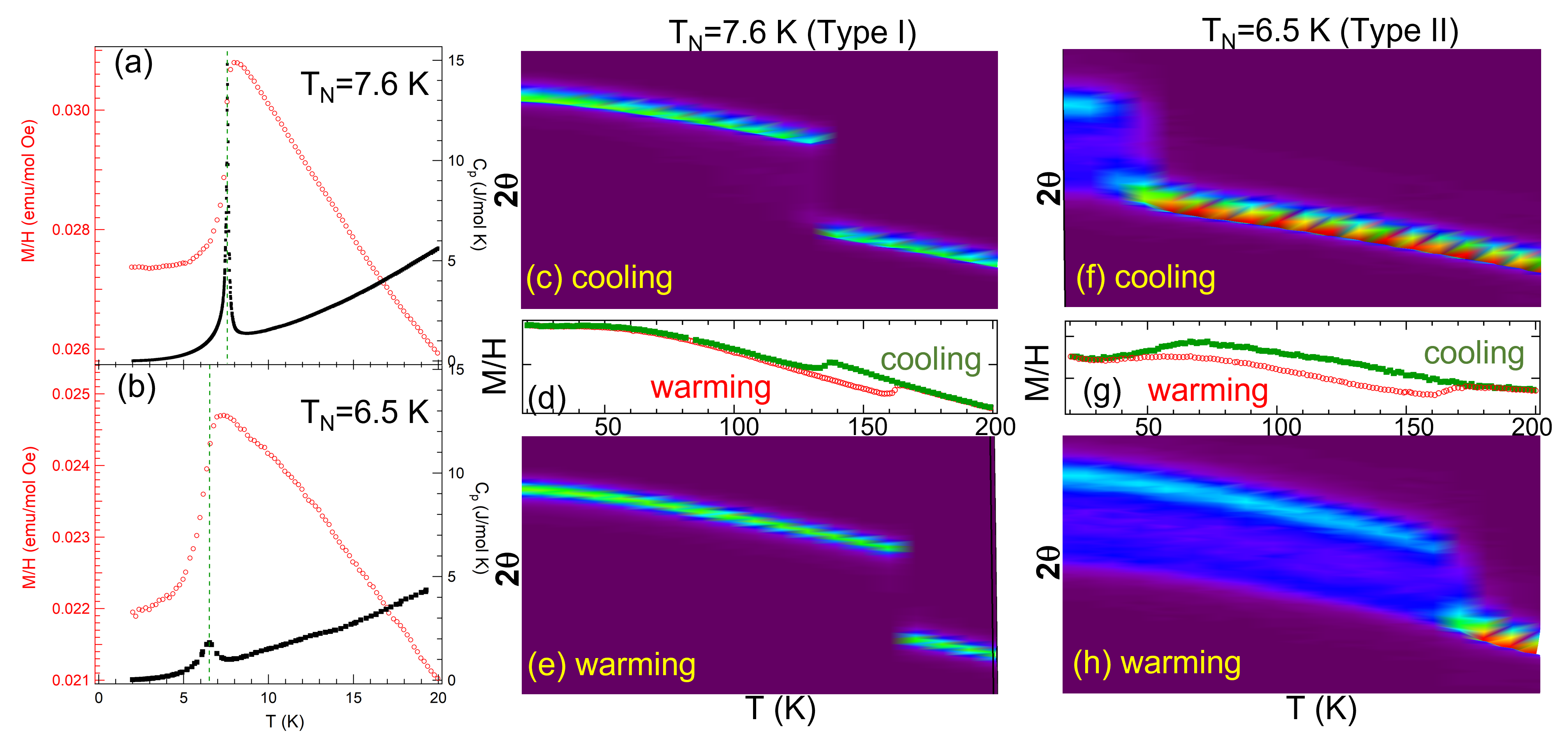}
    \caption{(color online) Magnetic order and structure transition in two typical types of \arucl\ single crystals. (a), (b) Temperature dependence of magnetization and specific heat (C$_p$) below 20 K. The magnetization was measured in a magnetic field of 1 kOe applied along the zigzag direction (perpendicular to the Ru-Ru bond). The vertical dashed lines highlight the NÃ©el temperature, T$_N$, defined as the temperature where C$_p$ peaks. T$_N$ = 7.6 K in (a) is 1.1 K higher than that in (b). (c)-(e) Temperature dependence of magnetization and 005 reflection in the temperature range 20-200 K for the crystal with T$_N$ = 7.6 K. (f)-(h) Temperature dependence of magnetization and 005 reflection in the temperature range 20-200 K for the crystal with T$_N$ = 6.5 K. The temperature dependence of magnetization in (d),(g) was measured in a magnetic field of 10 kOe applied perpendicular to the honeycomb plane. All data in [(c)-(h)] were collected in a temperature sweep at a rate of 2 K/min. Reproduced with permission from \cite{PhysRevMaterials.8.014402}, APS.}
    \label{MagCpXRD-1}
\end{figure*}

The N\'eel temperature has been found to vary from 6 to 8 K even for those crystals with one single magnetic transition around 7 K \cite{kim2022alpha,sears2017phase,do_majorana_2017}. Recently a careful investigation of the correlation between the layer stacking, structure transition, magnetic and thermal transport properties of different \arucl\ crystals with T$_N$ varying from 6 K to 7.6 K found this variation results from a small amount of stacking disorder in \arucl\ crystals \cite{PhysRevMaterials.8.014402}. As shown in Fig.\ \ref{MagCpXRD-1}, crystals with minimal stacking disorder show sharp anomalies at T$_N$=7.6 K in the temperature dependence of both magnetization and specific heat, and also a structure transition at 140 K upon cooling and 170 K upon heating. Meanwhile in crystals with a T$_N$ lower than 7 K, the structural transition occurs at a much lower temperature and also is not complete, as evidenced by the coexistence of reflections from both high-temperature and low-temperature (down to 20 K) phases in x-ray diffraction measurements. Neutron diffraction studies suggest that crystals with a lower T$_N$ has more stacking disorder. It should be noted that the amount of stacking disorder in these crystals is too small to generate observable magnetic anomalies in the  10-14 K range in either magnetization or specific heat measurements. However, these isolated stacking faults can suppress T$_N$ by disturbing the interlayer magnetic interactions. This suggests a strong correlation between the layer stacking, structure transition, magnetic, and thermal transport properties, and highlights the importance of interlayer coupling in \arucl. Importantly, it also suggests well-defined criteria for selecting \arucl\ crystals with minimal amount of stacking disorder:\ a structure transition around 140 K when being measured upon cooling and one single magnetic anomaly around 7.6 K in specific heat. 

Stacking faults are the major factor leading to sample dependent issues in  studies of \arucl. This becomes critical in attempts to reproduce the observation of a half integer quantized thermal Hall effect, and to understand the origin of oscillatory features in the thermal conductivity of \arucl\cite{kasahara_majorana_2018,czajka_oscillations_2021}. While the magnitude of thermal conductivity and thermal Hall effect can be sample dependent, the critical magnetic fields at which the oscillatory features show up and the sign structure of the thermal Hall effect are sample independent \cite{PhysRevMaterials.7.114403,Zhang2024a}. Stacking faults can form during crystal growth, from strain fields induced by the sample environment and handling, or as the result of the structural phase transition. The thermal and mechanical history also affects the type, population, and distribution of stacking disorder. These issues are expected to be quite important in studies exploring \arucl\ flakes and mesoscopic devices, due to the stresses imparted during the exfoliation, transfer, and re-stacking processes \cite{Geim2013}. 

\section{Atomically thin films and heterostructures}\label{devices}


The investigation of \arucl\ in the two-dimensional limit is of growing interest due to its van der Waals layered nature and the ease of exfoliating to single and few-layer flakes. As noted above, such samples are typically only a few 10s of $\mu$m in size and 1-100 nm thick, and so bring numerous opportunities and challenges. Many standard techniques for exploring quantum magnetism---neutron and x-ray scattering and bulk thermodynamic probes---cannot be used due to the vanishingly small sample volumes and therefore signal sizes. In their place, many experimental techniques developed over the past two decades for probing atomically thin materials can be brought to bear. These include electron tunneling, electronic transport, and spectroscopic methods re-tooled for small samples (e.g.\ nano-ARPES); direct thin film growth \cite{Grnke2018,Park2020,Wang2022,Asaba2023}; and most importantly the ability to transfer atomically thin flakes to be re-stacked with a variety of other materials into novel quasi-3D heterostructures whose overall properties are an amalgamation of those inherited from the constituent flakes \cite{Geim2013,Yankowitz2019}.

\subsection{Monolayers of \arucl:\ experiment}


The first steps toward exploring exfoliated \arucl\  were undertaken to explore magnetism of single layers in turbostratically disordered stacks made by restacking flakes that were first exfoliated via sonication. These materials revealed a magnetic transition temperature near 7 K similar to that observed in bulk samples \cite{Weber2016}. 

Mechanical exfoliation was first pursued for studies of the Raman response vs sample thickness from monolayers up to $\sim40$ nm thick, and from room temperature to 10 K, which revealed a change in symmetry in the thinnest samples and broadening of a mode coupled to continuum magnetic excitations \cite{Zhou2018,Du2018}.  The appearance of additional Raman modes in the thinnest samples is taken as evidence of a symmetry reduction in thin films, while the Fano lineshape of the strongest Raman modes is suggestive of spin-phonon coupling to a continuum of magnetic excitations. In this context, this scattering is presumably due to fractionalized Majorana spinons \cite{sandilands_scattering_2015,Perreault2015,nasu_fermionic_2016}. Both Raman spectroscopy and electronic transport were studied in nanoflakes, that identified features in the resistivity correlated to a structural transition near 180 K, and leading to a variable range hopping regime at lower temperatures \cite{Mashhadi2018}.

 \begin{figure}[tb]
 	\centering
 	\includegraphics[width=0.7\columnwidth]{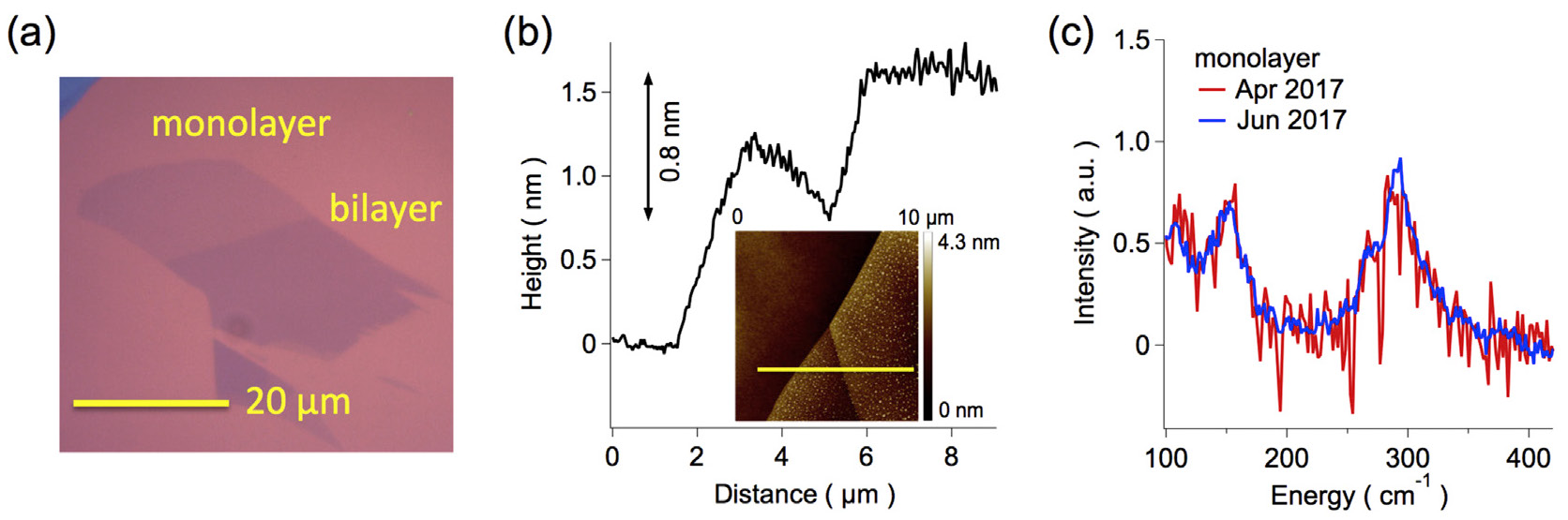}
 	\linespread{1}
 	\caption{(a) Optical image of \arucl\ mono/bilayer flake. (b) AFM scan of flake. (c) Raman spectra of monolayer acquired several weeks apart show no significant change \cite{Zhou2018}, \textcopyright\ 2019 Elsevier.}
    \label{fig:exfolrucl}
\end{figure}

As mentioned, \arucl\ is found to be easily mechanically exfoliated by sticky tape methods, and readily yields a range of thicknesses from monolayers to 100s of nm thick flakes and ranging in size up to 100s of microns in extent, see Fig.\ \ref{fig:exfolrucl}. Fortuitously, such thin \arucl\ samples are chemically and thermally stable in air, though elevated temperatures lead to degradation above 150 $\degree$C \cite{Breitner2023}. \arucl\ has long been reported as soluble in acetone in its hydrated form \cite{Griffith1975}, and \arucl\ flakes have been reported to develop contaminated regions after acetone exposure \cite{Cai2025}. For best results, care should be taken to avoid exposure to both water and acetone during device fabrication. 

Despite the ease of isolating single or few-layer exfoliated samples, exploring the unusual Kitaev magnetism in typical flakes is challenging. As a Mott insulator with a $\sim1$ eV gap \cite{binotto_optical_1971,sandilands_optical_2016,zheng_tunneling_2023}, the conductivity rapidly decreases below room temperature \cite{Mashhadi2018,Barfield2023}. Typical optical probes of magnetism such as magneto-optic Kerr effect are not sensitive to the Kitaev couplings, and bulk probes of magnetic effects such as neutron scattering are insensitive to vanishingly small exfoliated flakes. In this light, numerous researchers use van der Waals heterostructures to study exfoliated \arucl\ samples as we review in Section \ref{transport}.


\subsection{Monolayers of \arucl:\ theory}

As a vdW material, \arucl\ can be exfoliated down to a monolayer, akin to graphene. Ab initio calculations suggest that the cleavage energy of \arucl\ is smaller than or comparable to that of graphene, such that exfoliation should be relatively straightforward \cite{Sarikurt2018}. This prediction aligns with experimental evidence elaborated in the subsequent section. The ease of exfoliation also extends to its related compounds, RuBr$_3$ and RuI$_3$ \cite{Ersan2019}. These materials have attracted considerable attention due to their unique electronic and magnetic states due to the potential for metal-insulator transitions and the quantum anomalous Hall effect through ligand substitution \cite{Huang2017, Kaib2022, Liu2023}. The successful exfoliation of \arucl\ and its relatives into monolayers opens up new avenues for exploring their unique properties and potential applications.

Predicting the stable magnetic state of the monolayer \arucl\ is challenging. While the zigzag antiferromagnetic state is stable in bulk, it competes with the ferromagnetic states in the monolayer. For instance, several $ab$ initio studies have shown the FM ground state \cite{Huang2017, Sarikurt2018}, while others have predicted the zigzag-AFM ground state \cite{Iyikanat2018,Tian2019-mv}. Additionally, it has been suggested that the FM ground state turns into the zigzag-AFM state due to Coulomb interactions \cite{Ersan2019}. These variations are presumably due to the lattice structure and the balance between SOC and electron correlation in this monolayer system.

In the monolayer form of \arucl, lattice strain can significantly affect the ground state. It has been shown that the zigzag-AFM state exhibits a transition to the FM state when tensile strain greater than 2\% is applied, and the magnetocrystalline anisotropy can be controlled by the applied strain \cite{Vatansever2019,Kaib2021}. The strain-stress relationship and elastic modulus have also been studied, revealing that the electronic and magnetic properties of \arucl\ can be finely tuned by varying the biaxial in-plane tensile strain \cite{Salavati2019}. These strain engineering approaches allow for the manipulation of the electronic and magnetic properties of this system.

The Janus monolayer of \arucl, which involves asymmetric substitution of ligands on either side of the monolayer, exhibits AFM Kitaev interactions \cite{Sugita2020}. The Janus-type polar structure is known to introduce a unique asymmetry that can lead to novel electronic and magnetic properties in transition-metal vdW materials. The AFM Kitaev interactions in the Janus monolayer of \arucl\ are particularly intriguing, since the Kitaev interactions are usually FM according to the Jackeli-Khaliullin mechanism discussed in the preceding section \cite{Jackeli2009}. This suggests the potential to tune Kitaev interactions in the polarity perpendicular to the monolayer through Janus substitution and even proximity to other materials.

Exploring these properties in the monolayer of \arucl\ and its analogs opens up exciting possibilities for future research and applications. Additionally, making heterostructures by combining the monolayer \arucl\ with other vdW materials, such as graphene, provides a new platform for creating and controlling the fractional excitations inherent to the KQSL potentially realized in \arucl, as detailed in the subsequent sections. Thus, the ability to manipulate their electronic and magnetic states through lattice strain, chemical substitutions, and the proximity effects in the heterostructures could pave the way for advancements in quantum computing and other cutting-edge technologies.

The charge-doping effect found when \arucl\ is placed against graphene or other materials has been explored in detail by DFT and related calculations \cite{Biswas2019,Gerber2020}. The charge transfer is calculated to be approximately 0.05-0.06e/Ru atom for (-0.01e per C atom in graphene), with the charge clustered largely in Cl orbitals closest to the graphene, and the overall charge transfer essentially limited to the two layers that are directly in contact. Similar charge-doping effects have been predicted for the allotrope $\beta$-RuCl$_3$, which also has a sizable work function $\sim6$ eV \cite{Razpopov2024}.

Placing any two van der Waals materials together leads to new strain fields as the lattices relax against each other. These effects have been studied by DFT for the case of \arucl/graphene and \arucl/hexagonal boron nitride heterostructures, where the magnetic couplings in relaxed structures are found to be altered from intrinsic \arucl. In particular, the Kitaev coupling $K$ increases by as much as 50\%, while the $J$ and $\Gamma$ terms of Eq.\ \ref{Ham:JKG} slightly decrease, overall shifting the phase diagram closer to the ideal Kitaev system \cite{Biswas2019}. Strain and doping effects in \arucl\ monolayers can choose between competing ferromagnetic or zigzag antiferromagnetic ground states (which arise from the competing magnetic couplings) \cite{Iyikanat2018,Sarikurt2018,Tian2019-mv}.

\subsection{Heterostructures with graphene, hBN, and other vdW materials} \label{stack}

The layered nature, ease of exfoliation, and stability in air of \arucl\ make it a prime candidate for incorporation into vdW heterostructures along with other atomically thin materials \cite{Geim2013,Yankowitz2019}. The role of materials beyond \arucl\ can include, for instance, using hexagonal boron nitride (hBN) as an inert insulating layer \cite{Wang2020,balgley_ultrasharp_2022}, graphene as a proximate conductor \cite{mashhadi_spin-split_2019,zhou_evidence_2019,PhysRevLett.126.097201,rossi_direct_2023,Kim2023-apl}, and other layered materials  \cite{Yang2023-ke}. Inspired by the remarkable physics in moir\'e graphene and transition metal dichalcogenide systems \cite{Bistritzer2011,Cao2018a,Cao2018b,Andrei2020,Mak2022}, recent studies have explored the potential for moir\'e structures to tailor the Kitaev magnetism as well \cite{Akram_NanoLett2021,Haskell_PRB2022,Nica_npjQM2023,Keskiner_NanoLett2024,Akram_NanoLett2024} (see Sec.\ \ref{moire}).

The most striking result in heterostructures of \arucl\ so far is the discovery of a significant charge transfer from neighboring graphene layers to \arucl\ \cite{zhou_evidence_2019}. The charge transfer is driven by the large work function, $W=6.1$ eV, of \arucl\ \cite{Pollini1996}, which lies well below ${\approx}4.5$ eV typical of graphene \cite{Yu2009}, so when these materials are brought close to each other electrons from graphene fall into the potential well of \arucl. This leaves the graphene significantly hole-doped, at approximately $p=3\times10^{12}$ cm$^{-2}$ \cite{mashhadi_spin-split_2019,zhou_evidence_2019}. The charge transfer effect has been used to increase the transparency of electronic contacts to transition metal dichalcogenide WSe$_2$, highlighting the generic nature of the charge transfer \cite{Wang2020,Pack2024}. Strong interfacial dipoles appear between graphene and \arucl\ that effectively behave as a ferroelectric below a critical temperature and induce a strong hysteresis in the graphene gating response, and may enable control over the magnetic states in graphene as well \cite{Shi2023,liu2025,Kim2025}. These topics are discussed further in Sec.\ \ref{transport}.

\begin{figure}[tb]
 	\centering
 	\includegraphics[width=0.6\columnwidth]{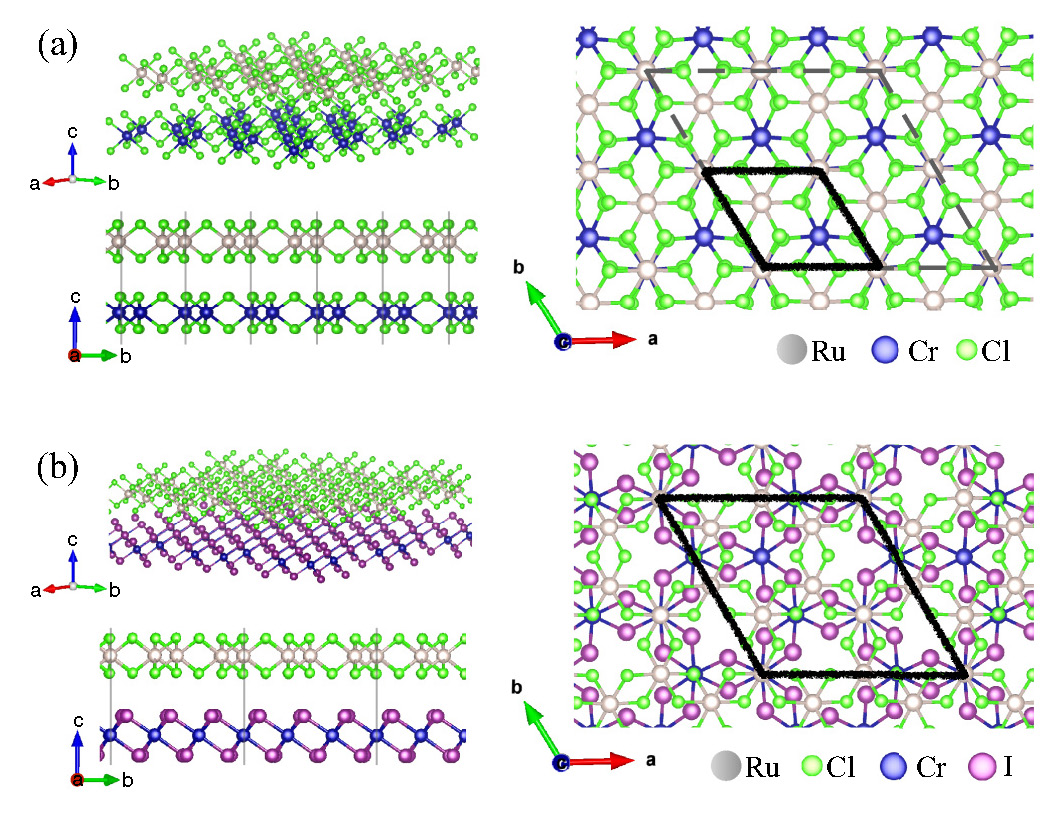}
 	\linespread{1}
 	\caption{Structures of the heterostructures of \arucl\ and CrX$_3$ with (a) X=Cl and (b) X=I. The gray, blue, green, and purple spheres represent Ru, Cr, Cl, and I atoms, respectively. The black rhombus denotes the unit cell, while the gray dashed rhombus in (a) is a $2{\times}2$ supercell used in the ab initio calculations. Reprinted with permission from\ \cite{zhangPossibleRealizationKitaev2024}, APS.}
    \label{fig:heteroRCr}
\end{figure}

Recently, heterostructures of a monolayer of \arucl\ with vdW magnets CrX$_3$ (X=Cl, I) were theoretically studied (Fig.\ \ref{fig:heteroRCr}) \cite{zhangPossibleRealizationKitaev2024}. Both CrCl$_3$ and CrI$_3$ are known to exhibit FM order even in their monolayer forms, with in-plane XY and out-of-plane Ising type magnetic anisotropy, respectively \cite{Huang2017,Bedoya-Pinto2021}. Hence, in these heterostructures, the possible KQSL state in the \arucl\ layer is expected to be modulated by the internal magnetic field from the ferromagnetic state in CrX$_3$ layer through the proximity effect. The ab initio-based study suggests that the heterostructure of \arucl/CrCl$_3$ could facilitate the realization of a KQSL state in the \arucl\ layer at zero magnetic field, suppressing the zigzag-AFM order by an internal magnetic field from the ferromagnetic CrCl$_3$ layer through the proximity effect. 

Meanwhile, \arucl/CrI$_3$ was found to come close to an insulator-to-metal transition by interlayer hybridization between Ru 4$d$ and I 5$p$ orbitals, providing a platform for exploring exotic metallic and superconducting states through the effect of carrier doping to the Kitaev magnet. This study opens up possibilities for tailoring the electronic and magnetic properties in Kitaev magnets by leveraging the diverse range of vdW materials and their flexible stacking control.

\subsection{Moir\'e structures}\label{moire}

Thanks to the weak interlayer interaction, vdW materials can be stacked with arbitrary translations and rotations between each successive sheet. If two neighboring layers have a lattice mismatch or are twisted relative to each other, a new long-range pattern called a moir\'e emerges. In the recent years, moir\'e superlattices of vdW materials have been shown to be tunable quantum platforms that gave rise to an array of emergent phases including unconventional superconductivity, Wigner crystals and fractional anomalous and spin Hall effects among others \cite{Cao2018b, Xie_Nature2021, Regan_Nature2020}. While engineering new electronic phases in twisted TMDs or graphene is and has been widely explored, research on moir\'e superlattices of magnetic materials is at an early stage. A wide range of novel spin textures have been proposed theoretically \cite{Tong_ACS2018, Hejazi_PNAS2020, Hejazi_PRB2021, Akram_PRB2021, Akram_NanoLett2021, Tong_PRR2021, Tong_PRR2021, Ghader_CommPhys2022, Zheng_AFM2023, Kim_NanoLett2023, Fumega_2DMAtt2023, Keskiner_NanoLett2024, Akram_NanoLett2024, Das_arXiv2024, Antao_arXiv2024, Kim_arXiv2024} including moir\'e superlattices of QSLs \cite{May-Mann_PRB2020,Haskell_PRB2022, Luo_PRB2022, Nica_npjQM2023}. Experimentally, non-coplanar spin textures have been realized in twisted bilayer \cite{Xu2022NatureNano, Song_Science2021}, double bilayer \cite{Hongchao2023NaturePhysics, Cheng_NatElect2023} and double trilayer \cite{Huang_NatComm2023} CrI$_3$. 

\begin{figure}[t!]
 	\centering
 	\includegraphics[width=0.6\columnwidth]{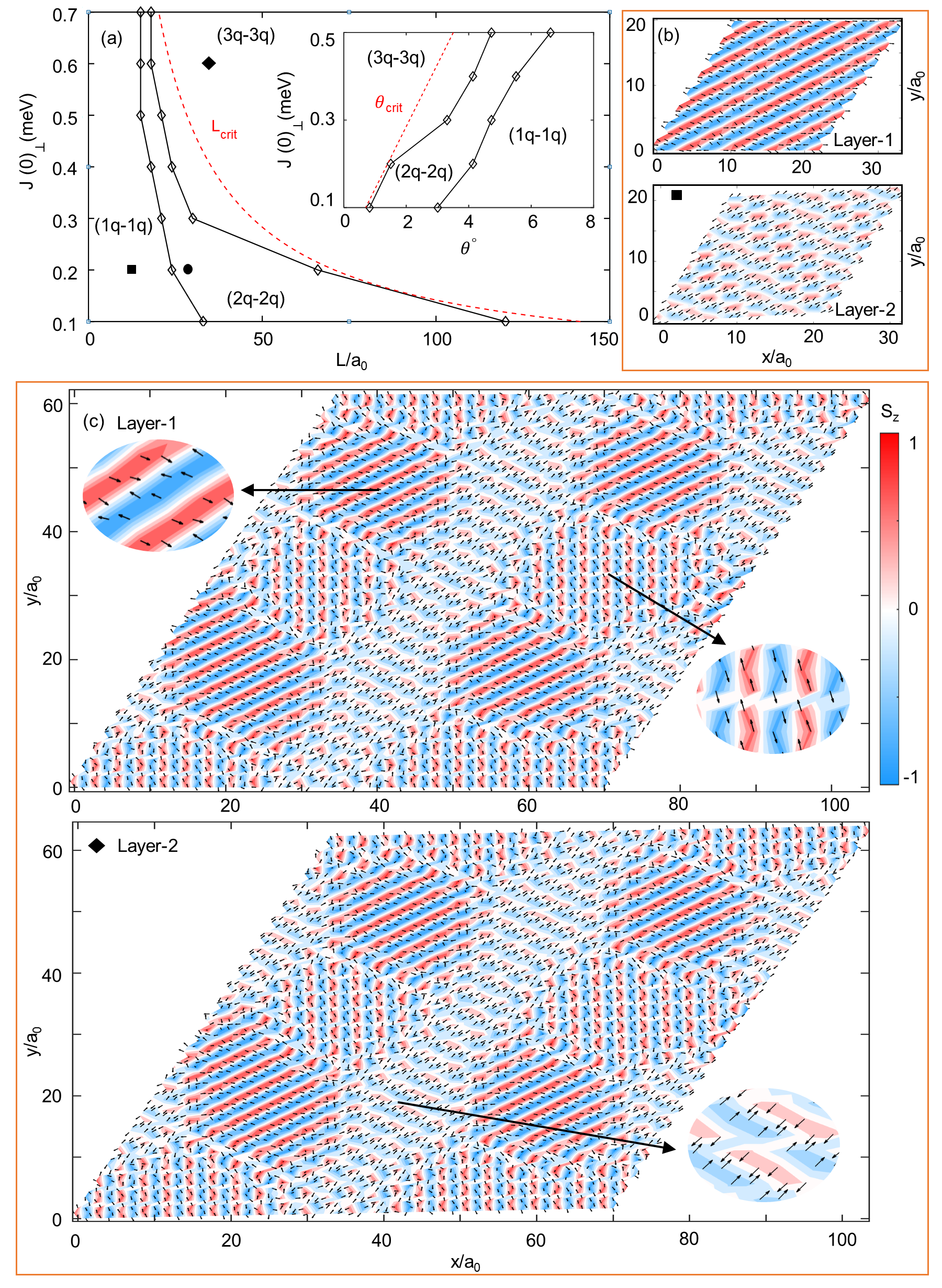}
 	\caption{(a) Theoretical phase diagram of twisted bilayer \arucl\ as a function Moir\'e period $L$ and interlayer exchange $J_\perp$. The inset shows the same phase diagram as a function twisted angle $\theta$ and $J_\perp$. Magnetization texture of (b) $1q - 1q$ for $L = 12a_0$ and $J_\perp (0)= 0.2$ meV and (c) $3q - 3q$ for $L = 36a_0$ and $J_\perp (0) = 0.6$ meV. Magnetization textures are shown for 2 Ã— 2 Moir\'e unit cells. Adapted from~\cite{Akram_NanoLett2024}.}
    \label{fig:moire}
\end{figure}

To date, there are no reports of the experimental realization of \arucl\ moir\'e superlattices. Theoretically, twisted zigzag antiferromagnets were first discussed in Ref.\ \cite{Hejazi_PNAS2020}, using a continuum Ginzburg-Landau theory that is valid for weak interlayer interactions and long moir\'e periodicity. Here, a zigzag order parameter can form diamond-shaped domains within the moir\'e unit cell forming a dice lattice in order to minimize the interlayer energy as depicted. 

More recently, Ref.\ \cite{Akram_NanoLett2024} explored twisted \arucl\ by a combination of first principles calculations and atomistic simulations. The authors first determined the stacking dependent interlayer exchange parameter which does not change sign within the moir\'e unit cell, unlike CrI$_3$. Next, using a lattice model, they simulated the classical magnetic phase diagram as a function of twist angle and interlayer interactions, as shown in Fig.\ \ref{fig:moire}(a). While the ground state remains a single-$q$ spiral for large twist angles (Fig.\ \ref{fig:moire}(b)), they discovered a $3q$ multi-domain order for small twist angles (Fig.~\ref{fig:moire}(c)). The $3q$-phase is similar to the one predicted by Ref.\ \cite{Hejazi_PNAS2020}, yet the shape of the domains are slightly different due to lattice effects. The twist angle dependent phase transition arises due to a competition between the energy gain from interlayer exchange and energy cost of the domain wall. Since the local stacking changes within the moir\'e unit cell, the $3q$-phase is optimized to minimize the interlayer energy while the domains are connected via domain walls that cost energy. As the interlayer exchange scales with the area and the domain walls scales by the circumference of the domains, the $3q$ phase is stabilized at small twist angles (large moir\'e periodicity).   

Current theoretical approaches are limited to classical simulations for twisted \arucl\ due to large system size. Even though it is a challenging task to deduce what the effects of twisting may be on the QSL, the classical simulations show that it introduces an additional magnetic frustration for the classical zigzag order and therefore twisting may help to unveil a more exotic quantum order.

\begin{figure*}[t!]
    \centering 
    \includegraphics[width=0.6\textwidth]{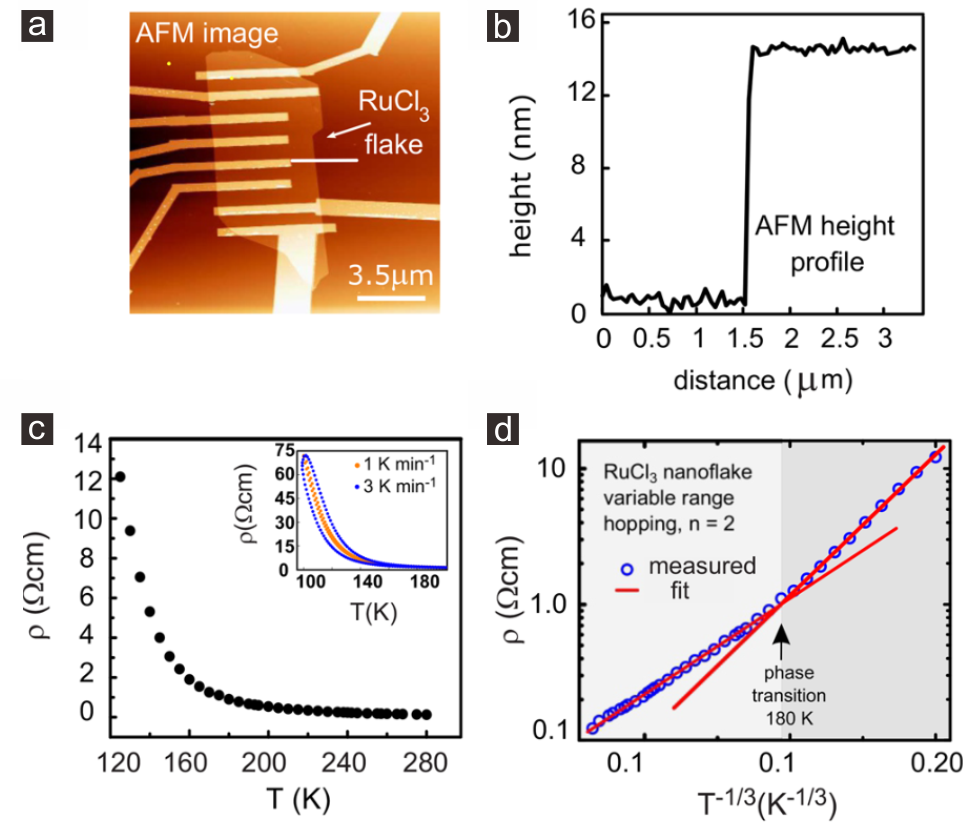}
    \caption{(a) AFM image of \arucl\ device. (b) Height profile of the $\alpha$-RuCl$_3$ sheet within the device. (c) Temperature-dependent resistivity ($\rho$) of a nanoflake with a thickness of 14 nm; the inset presents full temperature cycles at two different rates. (d) Semilogarithmic plot of resistivity ($\rho$) as a function of $T^{-1/3}$, with linear fits (red lines) for a typical heating/cooling cycle, following the VRH model. Reprinted with permission from Ref.\ \cite{Mashhadi2018}. \textcopyright\ 2018 American Chemical Society.}
    \label{mono}
\end{figure*}

\section{Electronic transport}\label{transport}

Electronic transport is among the most common and straightforward methods for studying atomically thin materials. As noted, being a Mott insulator without any low-energy charge excitations, \arucl\ is challenging to explore using electronic probes simply because it rapidly freezes out below room temperature \cite{binotto_optical_1971,sandilands_optical_2016,Mashhadi2018,Barfield2023}. Nonetheless some progress has been made in direct measurements of \arucl, and especially in heterostructures combining \arucl\ with more conducting materials such as graphene.

\subsection{Electronic transport in thin crystals of \arucl}



Electronic transport measurements in \arucl\ nanoflakes are a  sensitive tool for probing the complex interplay among lattice structure, magnetic fluctuations, and the emergence of fractionalized excitations in this prominent KQSL candidate. In nanoflakes with typical thicknesses around 10 nm, the resistivity exhibits behavior consistent with two-dimensional variable range hopping (2D VRH). This is evidenced by a linear dependence in semilogarithmic plots of $\ln(\rho)$ versus $T^{1/3}$, clearly indicating the strongly localized nature of charge carriers within the Mott insulating state of \arucl.

In early transport experiments, a feature was observed in \arucl\ nanoflakes at around 180 K \cite{Mashhadi2018} (see Fig.\ \ref{mono}), where the slope of the $\ln(\rho)$ versus $T^{1/3}$ plot changed distinctly, suggesting a structural transition that roughly coincides with the onset of Kitaev interactions. Detailed analysis revealed that, in a high-temperature regime (approximately 180-284 K), the characteristic temperature $T_0$ extracted from the 2D VRH model is lower (around $5.3\times10^5$ K), while in the lower temperature regime (approximately 125-180 K), $T_0$ increased significantly to about $1.58\times10^6$ K. This abrupt change in $T_0$ was understood as a modified hopping behavior due to a change in lattice structure and enhanced electron correlations. 

The transition temperature observed in the nanoflakes was notably higher than in bulk \arucl, a shift that was tentatively attributed to surface strain effects, enhanced electron correlations, and an increased density of stacking faults introduced during mechanical exfoliation. Furthermore, a modest hysteresis observed in the temperature-dependent resistivity evidenced by differences in the cooling and warming cycles was interpreted as the phase transition being broadened by defect-induced nucleation dynamics. This hysteretic behavior was associated to a coexistence of the high-temperature and low-temperature phases extending over a wider temperature range, likely due to the higher density of defects in the exfoliated nanoflakes compared to bulk crystals.



The energy gap in \arucl\ has been reported to range from 0.2 to 2.2 eV by various experimental techniques \cite{binotto_optical_1971,guizzetti_fundamental_1979,plumb_2014,sandilands_optical_2016,koitzsch_j_eff_2016,sinn_electronic_2016,ziatdinov_atomic-scale_2016,zhou_angle-resolved_2016,warzanowski_multiple_2020,koitzsch_low-temperature_2020,nevola_timescales_2021,annaberdiyev_electronic_2022}, makes direct electronic transport measurements challenging. This insulating character prevents accurate electronic transport measurements in bulk or thin crystals of \arucl, as the sample resistance at low temperatures becomes comparable to the insulation and capacitive impedance of electronic cabling in a transport experiment and may result in leakage currents that corrupt the sample signal. 


\begin{figure*} 
    \centering 
    \includegraphics[width = 0.45\textwidth]{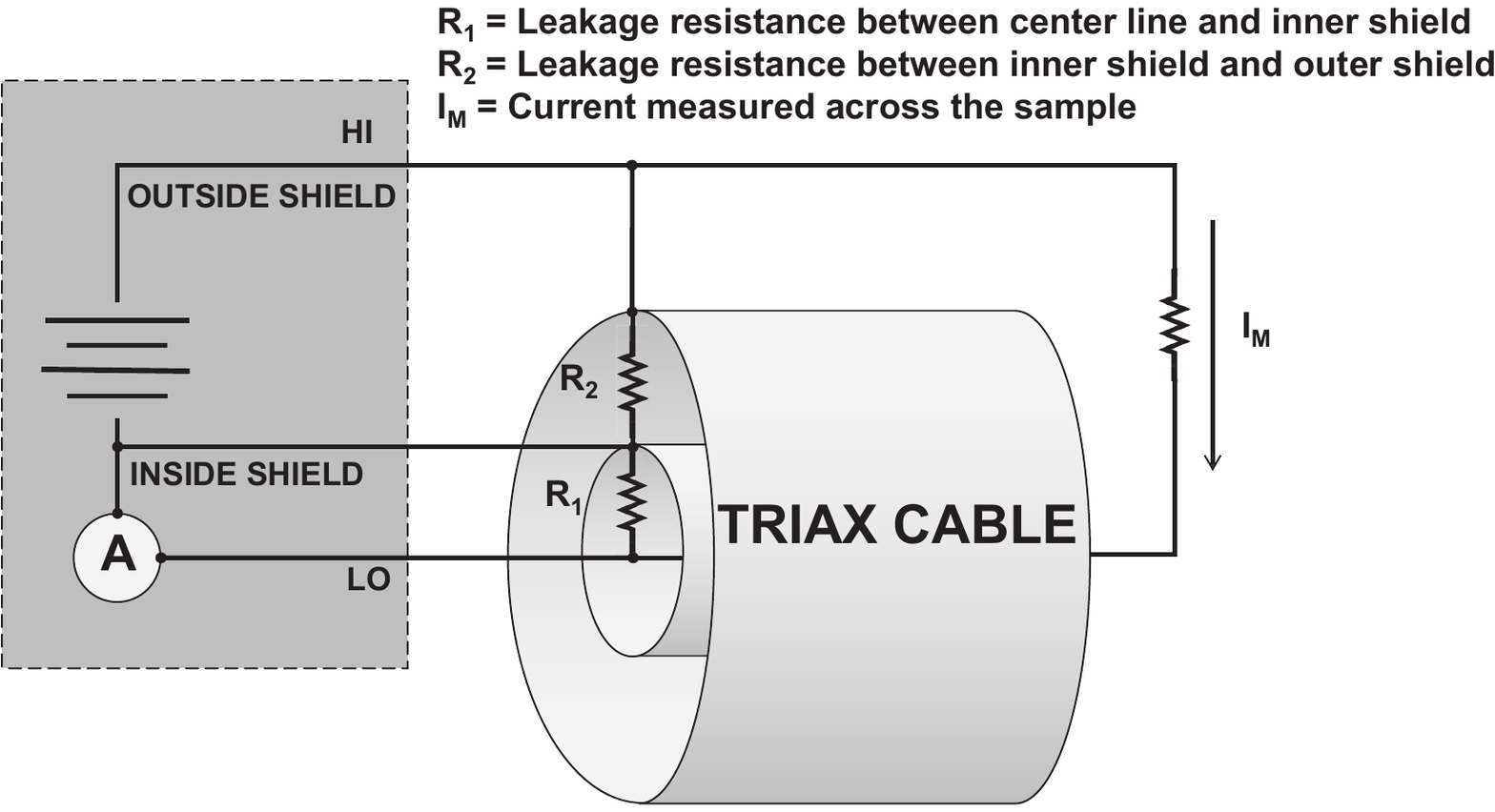}
    \includegraphics [width = 0.45\textwidth]{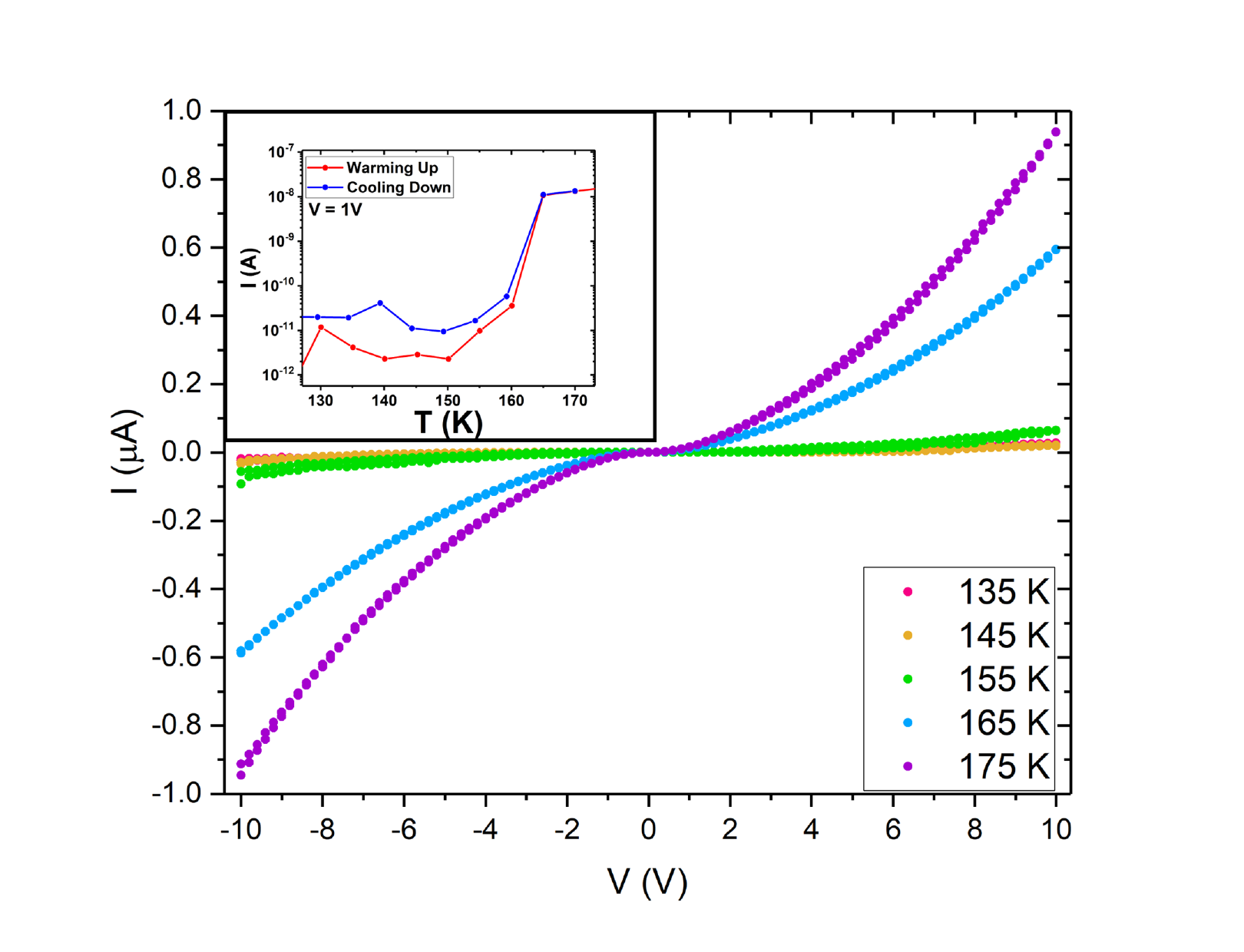}
    \caption{Left: Circuit schematic of the guarded force-voltage-measure-current setup. Right: Signature of the structural phase transition measured with the guarded setup. I-V characteristics of a thin crystal of \arucl, measured up to 175 K, show a striking change in the conductance at 160 K. The inset shows the current at bias voltage 1 V, extracted from I-V curves at different temperatures. Adapted from \cite{Barfield2023}, \textcopyright\ 2023 AIP Publishing.}
    \label{Ojeda-Aristizabal_1}
\end{figure*}

Triaxial cabling---similar to coaxial lines with an additional shield between the inner core and the outside shield---makes electronic transport measurements of high resistance samples possible \cite{Barfield2023}. 
In a guarded configuration that sources a voltage and measures a current (Fig.\ \ref{Ojeda-Aristizabal_1}), the sample is connected to the inner core of the lines and to one of the ammeter terminals. The other ammeter terminal is connected to the inside shield of the triax line, which plays the role of a guard, as the voltage drop across the terminals of the ammeter is very small, resulting in negligible leakage currents between the inner core and the inside shield of the triax cable, while the sample is held at approximately the same potential as the input HI terminal of the voltmeter. Any leakage between the shields is isolated from the ammeter. 

\begin{figure*}[b!] 
    \centering 
    \includegraphics[width = 0.4\textwidth]{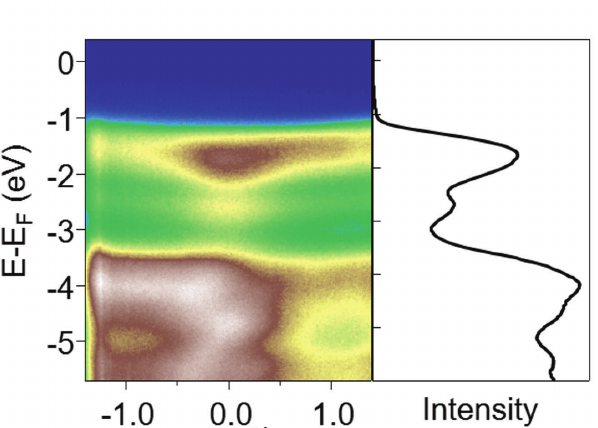}
    \includegraphics[width = 0.4\textwidth]{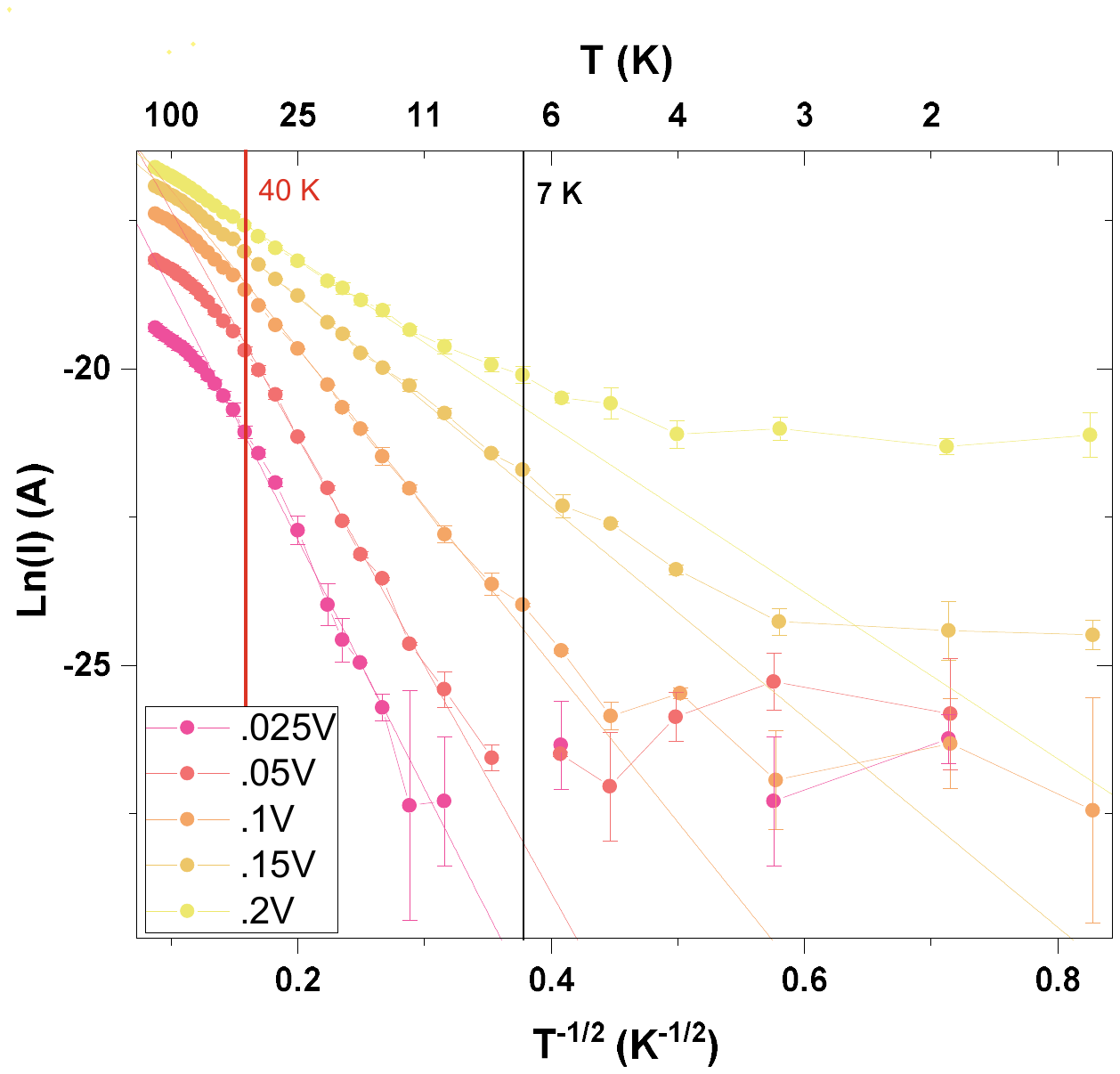}
    \caption{Left: Measured electronic band structure for \arucl\ in a momentum direction across $\Gamma$ within the first Brillouin zone. Momentum integrated (in a range of -1.0\,\AA$^{-1}$ $\leq k_y \leq$ 1.0\,\AA$^{-1}$ ) energy distribution curve (EDC), showing that the density of states vanishes at the Fermi energy. Right: Data at certain bias voltages (indicated in the legend), extracted from IV curves taken at different temperatures. The data fits ES-VRH in a temperature range of 7-40 K. There is a deviation for T$>$40 K, that fits well an Arrhenius law (see \cite{Barfield2023}). Below 7 K, there is a deviation from any known transport mechanism. Adapted from \cite{Barfield2023}, \textcopyright\ 2023 AIP Publishing.}
    \label{Ojeda-Aristizabal_2}
\end{figure*}

Using this approach, the structural phase transition in \arucl\ at $\approx$160 K has a striking signature \cite{Barfield2023}, with a change in resistance of three orders of magnitude (see Fig. \ref{Ojeda-Aristizabal_1}). This structural phase transition was first reported in neutron diffraction experiments \cite{ziatdinov_atomic-scale_2016, PhysRevB.108.144419, Park2024-fz} and earlier transport measurements \cite{Mashhadi2018}. As reported by Barfield, \etal, using this technique, electronic transport experiments are possible in a thin crystal of \arucl\ on Si/SiO$_2$ down to 1.5 K, revealing different transport mechanisms at different temperatures \cite{Barfield2023}. Measurements in a temperature range of $\approx$30 - 130 K obey an Arrhenius law ($I\approx\exp(-E_a/k_BT)$), with a thermal activation energy $E_a\approx$ 9 meV, which is much lower than the energy gap. This conduction regime is attributed to impurity states at the energy gap via charge trapping in the SiO$_2$ substrate, with charge carriers hopping between nearest neighbor electron- and hole-doped regions, as has been found in MoS$_2$ \cite{Ghatak2011-kl}.  In contrast, at low temperatures, the electronic transport is best described by the Efros-Shklovskii variable range hopping (ES-VRH). In general, Mott VRH describes hopping of carriers between remote sites in a narrow band near the Fermi level, as spatially neighboring sites have larger resistances. The hopping length of the carriers increases at lower temperatures, thus the term of variable range hopping. Bulk \arucl\ crystals do not present an energy band near the Fermi level, as reported using ARPES (see Fig.\ \ref{Ojeda-Aristizabal_2}) and in previous experiments \cite{zhou_angle-resolved_2016,sinn_electronic_2016, rossi_direct_2023,Barfield2023}. 

In a thin crystal device, a narrow band of localized states may be present due to the impurities from the substrate. If electronic correlations between the impurities are accounted for, the Miller-Abrahams random resistor network that is behind the theory of VRH is no longer adequate and can be reformulated, leading to the ES-VRH regime, $I\approx\exp\big(-\big(T_0/T\big)^{1/2}\big)$, with $T_0=2.8e^2/4\pi\epsilon\epsilon_0k_Ba$, where $a$ is the localization length of the impurities. This model indeed fits the transport data at temperatures down to 7 K---corresponding to the zigzag antiferromagnetic ordered transition known for the bulk crystals---and up to 40 K. Interestingly, 40 K has been identified as the Curie-Weiss temperature for \arucl\ \cite{sears_magnetic_2015}, that deviates from the critical magnetic ordering transition temperature, suggesting frustration. Thus the ES-VRH behavior occurs in a temperature range spanning the 7 K magnetic ordering transition, and 40 K, at the onset of possible quantum spin liquid fluctuations.

The authors deduce a Coulomb gap of $\Delta\approx$ 4.5 eV, similar to the activation energy deduced in the high temperature regime, and a localization length of $a\approx3$ nm. It is interesting to note that while the electronic transport in a thin crystal of \arucl\ can be understood in a wide range of temperatures, it goes beyond thermal activation or ES-VRH below the magnetic ordering transition at 7 K. (Similar results have been reported in another spin-orbit assisted Mott insulator, Na$_2$IrO$_3$ \cite{PhysRevB.101.235415}). Most importantly, the data is best fitted by ES-VRH in a temperature range between the onset of possible quantum spin liquid fluctuations and the magnetic ordered state.

\subsection{Electronic properties of proximate graphene-\arucl\ devices}

\subsubsection{Charge transfer in \arucl\ heterostructures}

Seeking alternative methods to probe \arucl\ by electronic transport, numerous groups have turned to heterostructures of \arucl\ and graphene or other conducting layered materials \cite{mashhadi_spin-split_2019,zhou_evidence_2019,balgley_ultrasharp_2022, Wang2020,massicotte_giant_2024}. In graphene/\arucl\ devices, deviations from the familiar behavior of graphene can be linked to phenomena originating in the proximate \arucl\ flake. It rapidly became clear that a large charge transfer takes place when graphene and \arucl\ are brought in contact \cite{mashhadi_spin-split_2019,zhou_evidence_2019}. This charge transfer is ultimately driven by the large work function mismatch, with the work function of graphene approximately 4.6 eV vs 6.1 eV for \arucl\ \cite{Wang2020,Pollini1996,Yu2009,Yan2012}. This large work function mismatch heavily $p$-dopes the graphene layer \cite{mashhadi_spin-split_2019,zhou_evidence_2019,Biswas2019,Gerber2020,cRizzo2020-sv,Wang2020,PhysRevLett.126.097201,rossi_direct_2023,Kim2023-apl,Kim2024-cap}, shifting the Dirac point upward by more than 0.5 eV and resulting in an unusually low four-terminal resistance across a wide range of gate voltages, which is an unambiguous indicator of extremely high carrier density (see Fig.\ \ref{g_rucl3}).

The charge transfer properties of \arucl\ are not limited to graphene. The large 6.1 eV work function serves to draw electrons from many proximate materials, including also WSe$_2$ and EuS \cite{Wang2020,Pack2024}. This is a potentially very useful feature in devices, where it has been used to heavily dope contacts for better electronic transport measurements in semiconducting WSe$_2$, and to define sharply defined spatially varying charge doping \cite{balgley_ultrasharp_2022}. 

\begin{figure*} 
    \centering 
    \includegraphics[width=\textwidth]{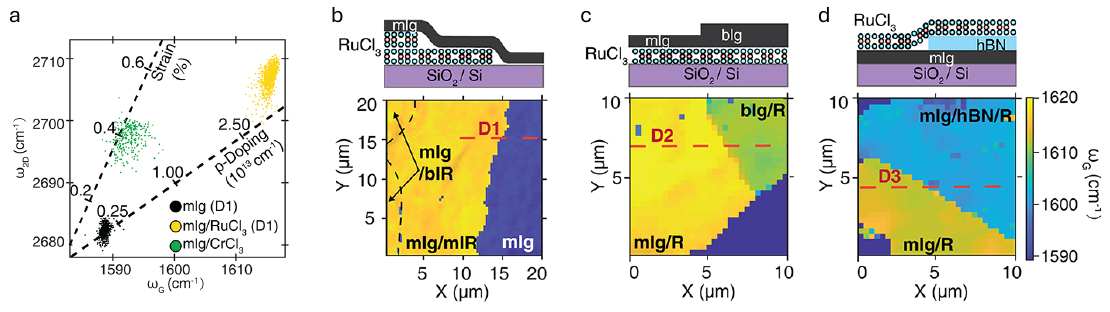}
\caption{(a) Correlation between the graphene G and 2D Raman peaks for different monolayer graphene-based heterostructures, with axes of strain or doping indicated by dashed lines \cite{Wang2020}. (b),(c),(d) Raman maps of the graphene G peak energy for different graphene/\arucl\ heterostructures shown schematically above each map. Adapted under Copyright © 2022 American Chemical Society.}
    \label{wang}
\end{figure*}

The charge-transfer properties of graphene and \arucl\ have been delineated by experiments in multiple heterostructures. With reference to Fig.\ \ref{wang}, the energy of the G and 2D peaks observed in Raman spectroscopy on graphene can be used to investigate its strain and charge-doping. Graphene proximate to \arucl\ layers show significant charge-doping effects but little strain. Since the G peak energy is most strongly correlated with charge-doping, it can be used to create spatial maps in graphene/\arucl\ heterostructures. In Fig.\ \ref{wang}(b)-(d), these maps demonstrate that the charge transfer to monolayer graphene is independent of the number of \arucl\ layers (which is borne out by DFT calculations \cite{balgley_ultrasharp_2022}); but the inverse is not true, namely graphene and bilayer graphene exhibit Raman peak shifts distinct from zero, so the second graphene sheet away from the \arucl\ is also affected; and finally that inserting a thin insulating layer (here, 2 nm of hexagonal boron nitride) can reduce but not eliminate the charge transfer. Indeed, this effect recapitulates ``modulation doping'', the effect where charge donors are spatially separated from the location of the charge that was first used to enhance the mobility of the two-dimensional electron system in GaAs heterostructures \cite{Dingle1978}. Approximately 10 nm of hexagonal boron nitride is needed to eliminate the charge transfer effect.

\begin{figure*} 
    \centering 
    \includegraphics[width=\textwidth]{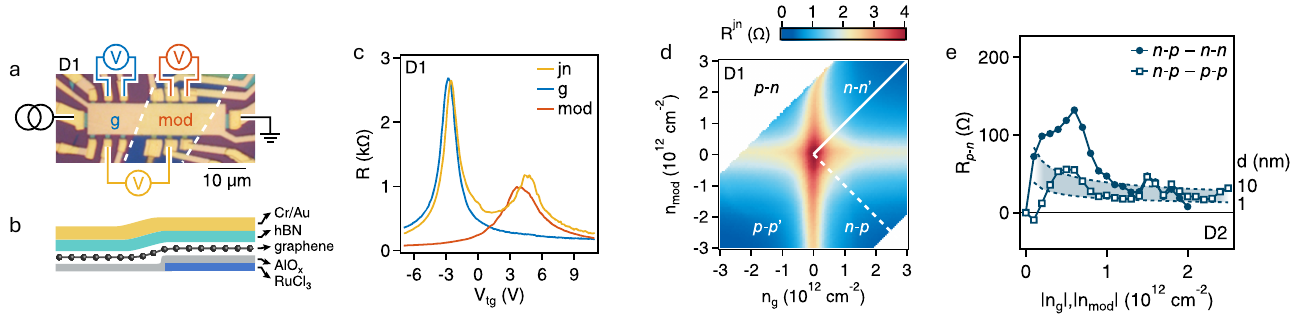}
\caption{\arucl-modulation-doping-defined $p$-$n$ junction. (a) Optical micrograph of the device. The white dashed lines mark the \arucl\ flake boundary separating regions of intrinsic (``g'') and \arucl-modulation-doped graphene (``mod''). (b) Schematic of the device layer profile. (c) Four-terminal resistance measured simultaneously in g (blue) and mod (orange) regions, and also across their interface (yellow), color-coded to the voltage measurement schematics in (a). (d) Four-terminal resistance across the g-mod interface, as a function of the g- and mod-side carrier densities. Labels show the polarity in the four quadrants, either monopolar ($n$-$n$') or bipolar ($n$-$p$). White solid and dashed lines mark where the carrier density on either side of the interface is equal ($n$-$n$) or is of equal magnitude but opposite sign ($p$-$n$). (e) $p$-$n$ junction resistance. The shaded region marks the theoretical resistance for a ballistic device with junction width ranging between 1 and 10 nm.Adapted from \cite{balgley_ultrasharp_2022}, \textcopyright\  2022 American Chemical Society. }
    \label{balgley}
\end{figure*}

The fact that charge-transfer only occurs in the immediate vicinity of \arucl\ enables the realization of a long-sought-for ultra-sharp $p$-$n$ junction in graphene; see Fig.\ \ref{balgley}. Here the atomically sharp crystalline edge of a sheet of \arucl, separated from graphene by 1-2 nm of hexagonal boron nitride, is used to differentially charge-dope only half of a graphene sheet. The presence of both a top and back voltage gate, along with the screening effect of interfacial dipoles that form in response to the charge transfer \cite{liu2025,Kim2025}, enable the charge density on either side of the interface to be independently tuned between uni- and bi-polar regimes. The atomically sharp edge of \arucl\ implies the charge transfer is similarly abrupt, and indeed analysis of the electronic transport across the junction can separate the contribution of the junction resistance to that of the rest of the device. This junction resistance is directly related to the sharpness of the $p$-$n$ junction in the bipolar regime, which is found to vary from $p$ to $n$ type over a lateral distance of less than 10 nm. This is sufficiently sharp to pursue observation of intriguing ballistic electron optic effects in graphene including Klein tunneling, snake states, and Veselago lensing \cite{Young2009,Williams2011,Cheianov2007,Chen2016}. 

\begin{figure*} 
    \centering 
    \includegraphics[width=0.7\textwidth]{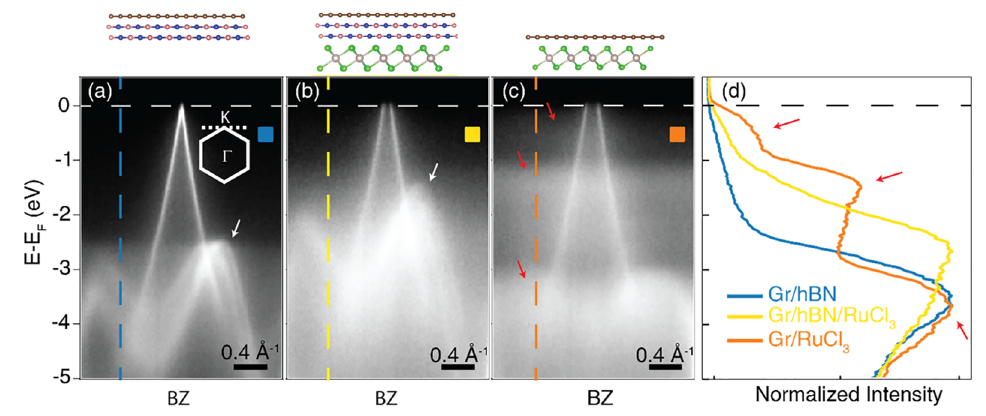}
\caption{(a)-(c) Band structure seen by nano-ARPES in a heterostructure of overlapping graphene, hexagonal boron nitride (hbn), and \arucl\ flakes, collected in the vicinity of the graphene $K$ point as depicted by inset to panel (a). The horizontal dashed line is the chemical potential. White arrows highlight the hbn bands, and red arrows the \arucl\ bands. (d) Energy distribution curves collected along the vertical dashed lines in panels (a-c). The red arrows highlight the corresponding states in panel (c). Adapted from \cite{rossi_direct_2023}.}
    \label{rossi}
\end{figure*}

The charge transfer effect can be strikingly visualized in nano-ARPES experiments conducted on a heterostructure made of overlapping thin sheets of graphene, hexagonal boron nitride, and \arucl, shown in Fig.\ \ref{rossi} \cite{rossi_direct_2023}. Here the band structures near the graphene $K$ point are imaged for graphene on hbn, showing the graphene at charge neutrality (chemical potential crosses the tip of the Dirac cone); graphene modulation-doped by \arucl\ over a thin intervening bilayer flake of hbn; and graphene directly placed on \arucl. A progressive hole-doping of graphene is observed by the upward shift of the Dirac cone relative to the chemical potential. In panels (a) and (b) of Fig.\ \ref{rossi} the hbn bands can be seen to co-travel with the graphene, shifting upward in energy with charge doping. In panel (c), \arucl\ bands are visible along with the graphene Dirac cone. Energy distribution curves (averaging across the momentum space axis) are shown in panel (d) and bring out the presence of the \arucl\ lower Hubbard band near -1.3 eV, and a deeper set of bands identified as Cl $p$ orbitals. The presence of spectral weight around -0.5 eV is unusual and not typically reported in \arucl\ ARPES experiments (see Fig.\ \ref{Ojeda-Aristizabal_2}). In Ref.\ \cite{rossi_direct_2023} this weight is interpreted as states formerly part of the upper Hubbard band which drop below the chemical potential due to electron doping by graphene. Alternatively, it has been suggested that doped \arucl\ exhibits an unusual transition to a charge transfer insulator; this possibility is discussed in more detail below in Sec.\ \ref{stm}.

\subsubsection{Monolayer graphene/\arucl\ heterostructures}

\begin{figure*}[b!]
    \centering 
    \includegraphics[width=0.7\textwidth]{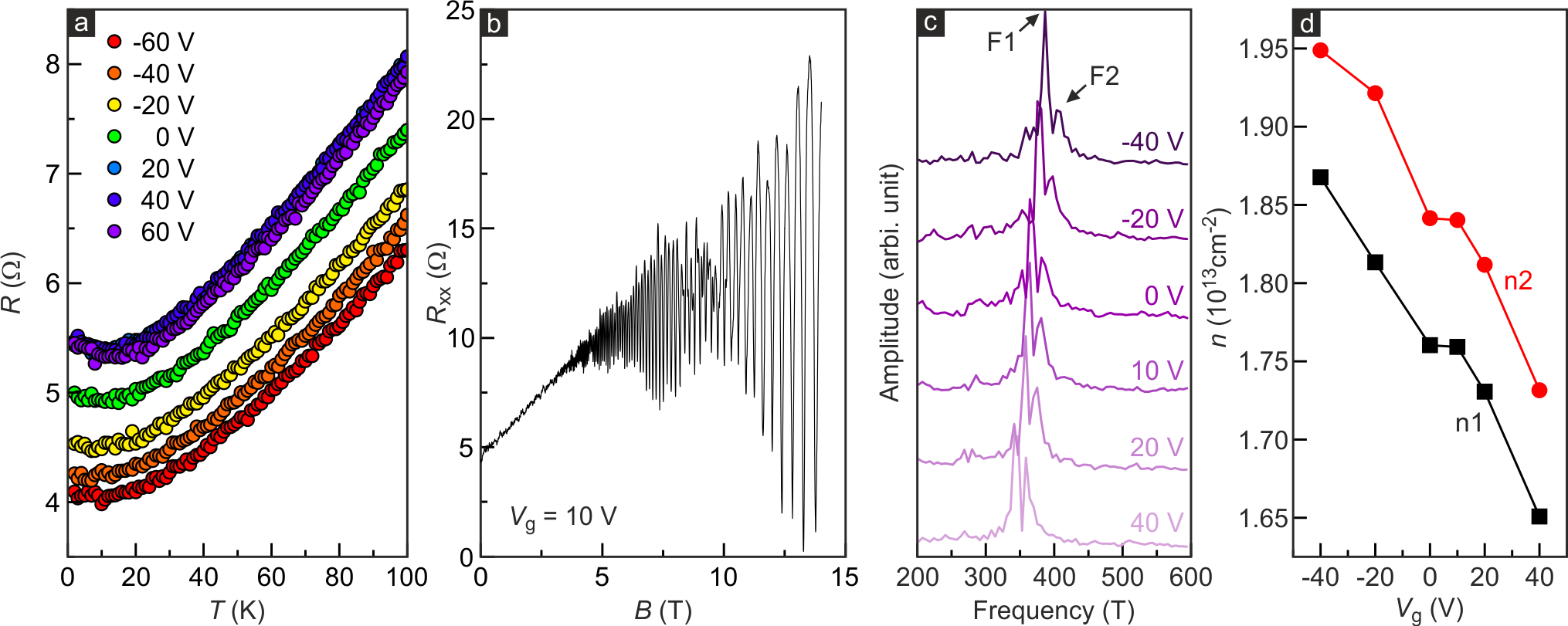}
\caption{(a) Temperature-dependent resistance at gate voltages from $-60$ to $+60$\,V. (b) Longitudinal resistance $R_{xx}$ of the graphene/$\alpha$-RuCl$_3$ heterostructure as a function of magnetic field at  $T=2$\,K and $V_g=10$\,V.  (c) FFT spectra of the SdH oscillations recorded at $V_g = 40$, 20, 10, 0, $-20$, and $-40$\,V. (d) Carrier concentration versus gate voltage extracted from peaks $F_1$ and $F_2$ in (c). Reprinted with permission from Ref.~\cite{mashhadi_spin-split_2019}.  \textcopyright\ 2018 American Chemical Society.}
    \label{g_rucl3}
\end{figure*}

Magnetotransport measurements in monolayer devices uncover a distinctly nonlinear response at low magnetic fields, signifying the presence of multiple conduction channels \cite{mashhadi_spin-split_2019,zhou_evidence_2019,Kim2024-cap}. This nonlinearity implies that while holes are the dominant carriers, a minority electron channel also contributes, most likely arising from states that do not complete closed orbits. Moreover, Shubnikov-de Haas (SdH) oscillations in the longitudinal resistance display a pronounced beating pattern (see Fig.\ \ref{g_rucl3}). Such a beating pattern provides direct evidence that the original, spin-degenerate Dirac cones of graphene are split into two distinct spin-resolved branches. One branch, strongly influenced by spin-selective hybridization with \arucl\ band states, exhibits a distorted dispersion near the Fermi level. In contrast, the other branch retains a more conventional linear dispersion characteristic of pristine graphene.

\begin{figure*}[t!]
    \centering 
    \includegraphics[width=\textwidth]{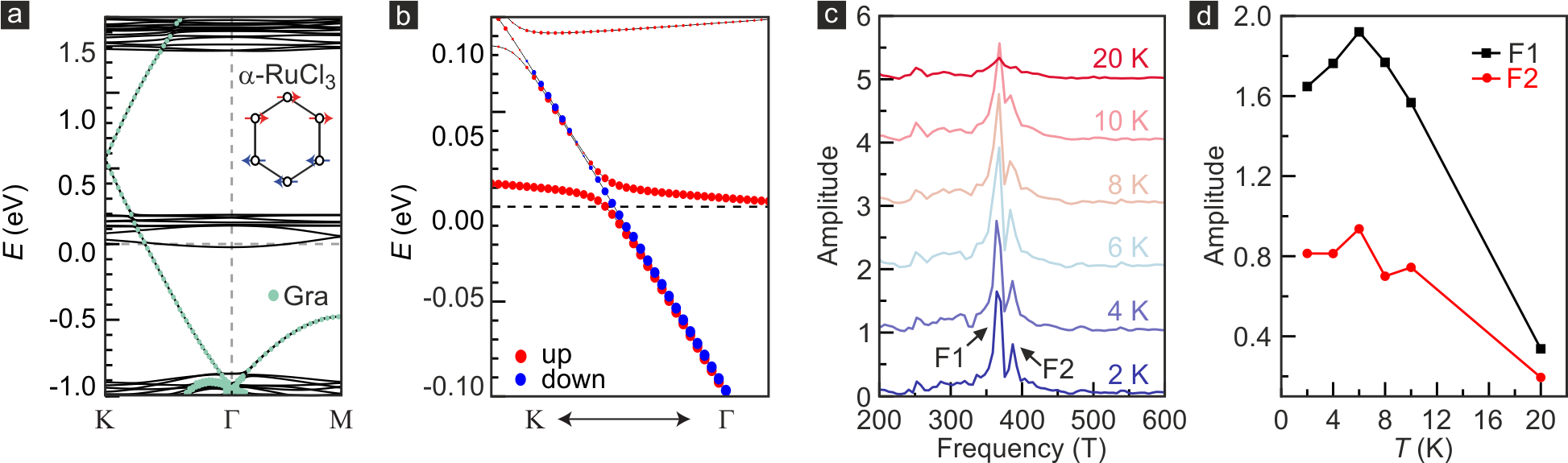}
\caption{(a) DFT band structure of the graphene/$\alpha$-RuCl$_3$ heterointerface; graphene-derived states are shown as green dots. (b) Expanded view from $-0.1$ to $0.1$\,eV highlighting crossings and anticrossings between graphene and $\alpha$-RuCl$_3$ bands. Red and blue dots indicate states with nonzero spin-up or spin-down projections along the $x$-direction, with dot size proportional to $S_x/\sqrt{S_x^{2}+S_y^{2}+S_z^{2}}$. (c) FFT spectra of the magnetoresistance at different temperatures for $V_g = 10$\,V.  
(d) Temperature dependence of the FFT amplitude of the SdH oscillations extracted from (c). Reprinted with permission from Ref.~\cite{mashhadi_spin-split_2019}. \textcopyright\ 2018 American Chemical Society.}
    \label{g_rucl3-2}
\end{figure*}

Density functional theory (DFT) calculations provide support for the experimental observations (see Fig.\ \ref{g_rucl3-2}), indicating that the large work function mismatch leads to an upward shift of the graphene Dirac point by more than 0.5 eV \cite{mashhadi_spin-split_2019,Biswas2019,Gerber2020}. They also reveal the formation of an isolated, weakly dispersive \arucl\ band with dominant spin-up character near the $\Gamma$ point. Hybridization between the graphene $\pi$-bands and this \arucl\ band splits the former into spin-up and spin-down components. This spin-selective hybridization is especially pronounced in the monolayer configuration, where electrons from the graphene Dirac cones are transferred into the \arucl\ states, leaving behind holes that reside in two distinct Fermi pockets.

A collaborative study integrating both theoretical and experimental approaches was used to investigate anomalous quantum oscillations in a heterostructure where graphene is interfaced with a proximate quantum spin liquid, embodied by Mott-insulating \arucl\ \cite{PhysRevLett.126.097201}. Unlike in conventional metals---where Lifshitz-Kosevich (LK) formalism predicts a monotonic decay of oscillation amplitude with increasing temperature---the oscillations in this heterostructure display a pronounced non-LK temperature dependence, with the amplitude reaching a clear maximum around 7 K. This anomalous temperature dependence suggests that additional damping mechanisms, likely linked to enhanced spin scattering as \arucl\ approaches its N\'eel temperature, are at play.

\begin{figure*}[b!] 
    \centering 
    \includegraphics[width=0.7\textwidth]{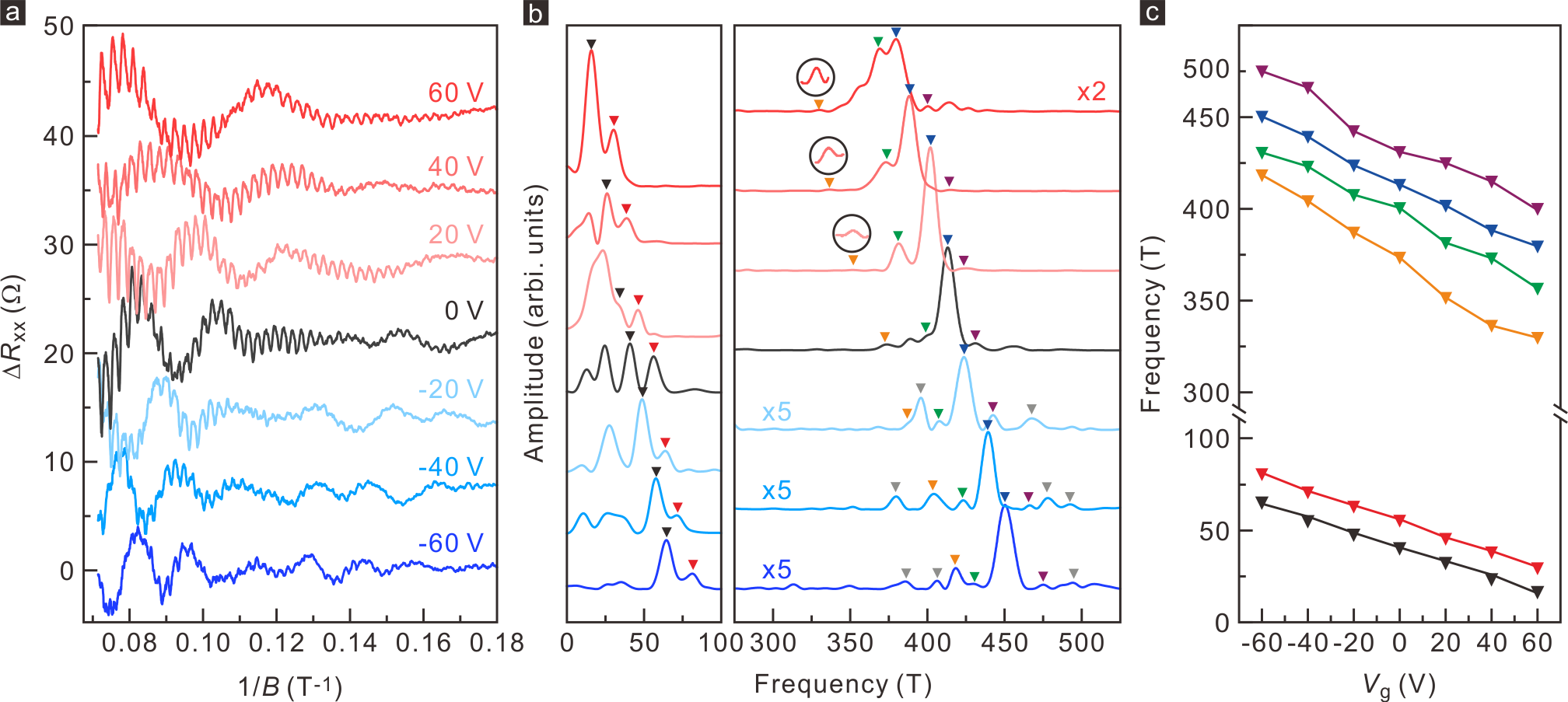}
\caption{(a) $\Delta R_{xx}$ as a function of inverse magnetic field, with different colors representing various gate voltages at $T = 2$\,K. The inset illustrates a schematic of the device.  (b) FFT spectra of the Shubnikov--de Haas oscillations from panel (a), where six frequency peaks are marked by triangles. The inset graphs with circles show magnified views of the peaks marked by yellow triangles, with each dataset scaled by a factor of ten. (c) Gate-voltage-dependent peak sets corresponding to panel (b), with color codes matching the triangle colors in (b). Reprinted from Ref.~\cite{Kim2023-apl} with the permission of AIP Publishing.}
    \label{bi_rucl3}
\end{figure*}

To account for this behavior, a microscopic theory was developed based on a minimal two-layer Kitaev-Kondo lattice model. In this framework, the itinerant electrons of graphene interact via a spin-dependent Kondo coupling with the fractionalized spin excitations of the \arucl\ layer. This coupling enables the normally charge-neutral Majorana excitations predicted by the KQSL model to acquire effective charge, thereby leading to the formation of a heavy Fermi liquid (hFL) state. The reconstructed Landau level structure and the resulting finite-temperature Green's function yield an analytic expression for the quantum oscillation amplitude that departs significantly from the standard LK description. Notably, the model predicts a maximum in the oscillation amplitude at temperatures on the order of the Kondo exchange energy (with $T_{max}\approx J/5$), a prediction that is in quantitative agreement with the experimental data \cite{PhysRevLett.126.097201}.

 \subsubsection{Bilayer graphene/\arucl\ heterostructures}
 
Similar to graphene/\arucl\ heterostructures, \arucl/bilayer graphene (BLG) devices exhibit significant charge transfer driven by the difference in work function between the two materials \cite{Kim2023-apl}(see Fig.\ \ref{bi_rucl3}). This intrinsic charge transfer heavily $p$-dopes the BLG while simultaneously electron-doping the \arucl\ layer, resulting in distinct carrier populations in each component. Consequently, magnetotransport measurements reveal very low resistance, indicative of high carrier density and nonlinear Hall curves that point to multichannel conduction, paralleling observations in graphene/\arucl\ systems.

A closer inspection of the Shubnikov-de Haas (SdH) oscillations uncovers two distinct features. First, rapid $R_{xx}$ oscillations with a pronounced beating pattern are observed. This beating indicates the presence of two or more nearly identical oscillation frequencies, which we attribute to two nearly equal Fermi pockets (large, similarly sized Fermi surfaces) corresponding to higher Fermi wavevectors (see Fig.\ \ref{bi_rucl3}). This pattern suggests that the device hosts at least two sizable Fermi surfaces, likely originating either from the two graphene layers or from spin-split pockets within a single layer \cite{Kim2023-apl}.

In addition to the high-frequency component, a distinct slower oscillatory feature is superimposed on the fast oscillations. This slower oscillation corresponds to a smaller Fermi pocket, implying a lower carrier density or smaller $k_F$ than that associated with high-frequency oscillations (see Fig.\ \ref{bi_rucl3}).

Fast Fourier Transform (FFT) analysis of the SdH data reveals up to six prominent frequency peaks, confirming the presence of multiple Fermi pockets in the BLG heterostructure. In the low-frequency range, two dominant peaks shift toward higher frequencies as the back-gate voltage becomes more negative (as hole doping increases), supporting their assignment to hole-type Fermi pockets. Similarly, in the high-frequency domain, four peaks (comprising two primary and two minor peaks) shift systematically with gate voltage. Additional weak peaks that appear at certain gate voltages may be attributed to magnetic breakdown between Fermi pockets that come into close proximity in $k$-space.

\begin{figure*} 
    \centering 
    \includegraphics[width=0.5\textwidth]{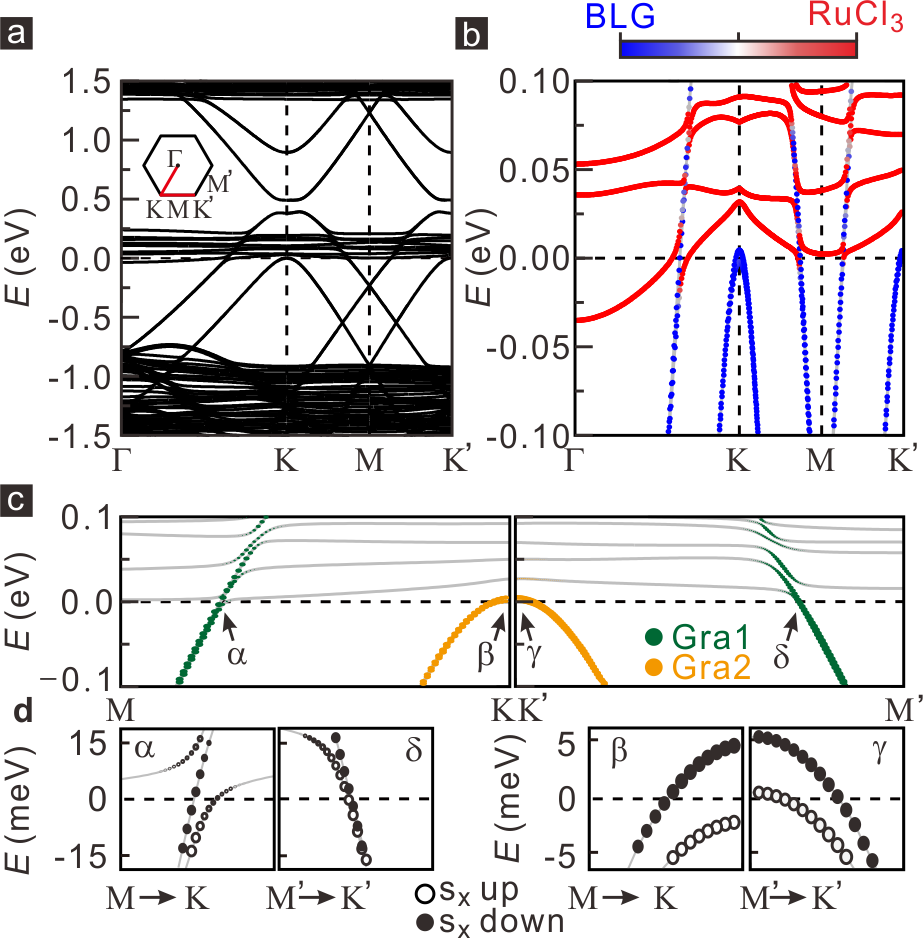}
\caption{(a) Computed band structure of the \arucl/bilayer graphene heterostructure along the high-symmetry path, as indicated in the inset, with the Fermi level set to zero. (b) Atomically resolved band structure near the Fermi level, where blue and red dots denote the contributions from bilayer graphene and \arucl, respectively. (c) Calculated band structures of the heterostructure with layer-specific resolution (upper panel), where green and orange dots indicate contributions from the top graphene layer adjacent to \arucl\ (Gra1) and the bottom graphene layer (Gra2), respectively. Reprinted from Ref.~\cite{Kim2023-apl} with the permission of AIP Publishing.}
    \label{bi_rucl3-2}
\end{figure*}

Band structure calculations indicate that the magnetic proximity effect from \arucl\ strongly modifies the BLG Dirac bands via spin-selective hybridization, similar to the case in monolayer graphene. In the presence of the zigzag antiferromagnetic order of \arucl, a spin-dependent potential is induced in the adjacent graphene layer, splitting the otherwise spin-degenerate bands and leading to the formation of spin-polarized Fermi surfaces. The two low-frequency SdH oscillations, which differ slightly in frequency, can be directly associated with these distinct spin-resolved pockets (see Fig.\ \ref{bi_rucl3-2}).

In bilayer graphene, only the top layer directly contacts \arucl\ and an inherent imbalance arises between the $K$ and $K'$ valleys. Specifically, one valley experiences much stronger hybridization (and hence significant spin splitting), while the other remains relatively unperturbed. This broken valley symmetry manifests itself as valley-polarized Fermi pockets, with subtle size differences that are detectable as additional peaks in the FFT spectrum.

\subsection{\arucl\ devices for spintronics}


Spintronic techniques such as spin-dependent transport can offer powerful tools to probe many quantum materials \cite{Han2020-lk} with interesting spin degrees of freedom and spin excitations, including spin liquids and related materials \cite{doi:10.7566/JPSJ.89.033701,PhysRevLett.125.047204,PhysRevB.104.L060403}.  In addition, spintronic devices such as magnetic tunnel junctions (MTJs) have also been proposed as potential tuning knobs to dynamically control characteristic phases and excitations including anyon quasiparticles in Kitaev materials \cite{PhysRevLett.129.037201}. 

Spin-orbit torques from heavy metals with a high SOC, such as platinum, can be used as sensitive probes of exotic magnetism in insulating systems. In a first-of-its-kind measurement \cite{idzuchi2022spinsensitivetransportspin}, spin-Hall magnetoresistance (SMR) was used to detect spin anisotropy in \arucl\ via electrical transport in a proximate platinum layer. A current passed through Pt generates spin accumulation at the interface through the spin Hall effect. The back-action of the spin configuration, dependent on the relative orientation between accumulated spins and local moments in the proximal \arucl, modulates the resistivity of Pt, yielding angular oscillations that reveal a robust transverse spin anisotropy even in the absence of long-range magnetic order. These SMR signals are diminishing in the isotropic low-field phase below 2 T, but persist across multiple field regimes, from the zigzag ordered phase to the field-induced QSL and partially polarized states. The data are interpreted as showing persistent anisotropic quantum fluctuations transverse to the applied magnetic field and picked up by Pt electrodes. Temperature-dependent measurements reveal multiple anisotropy energy scales, showing an avenue in which a proximate metal can couple to \arucl
. This study demonstrates yet another novel transport-based platform for probing quantum spin liquids and other correlated insulators, opening a pathway to spin-sensitive device architectures for quantum magnets making them more accessible to electrical measurements.



\begin{figure}[tb]
 	\centering
 	\includegraphics[width=0.6\columnwidth]{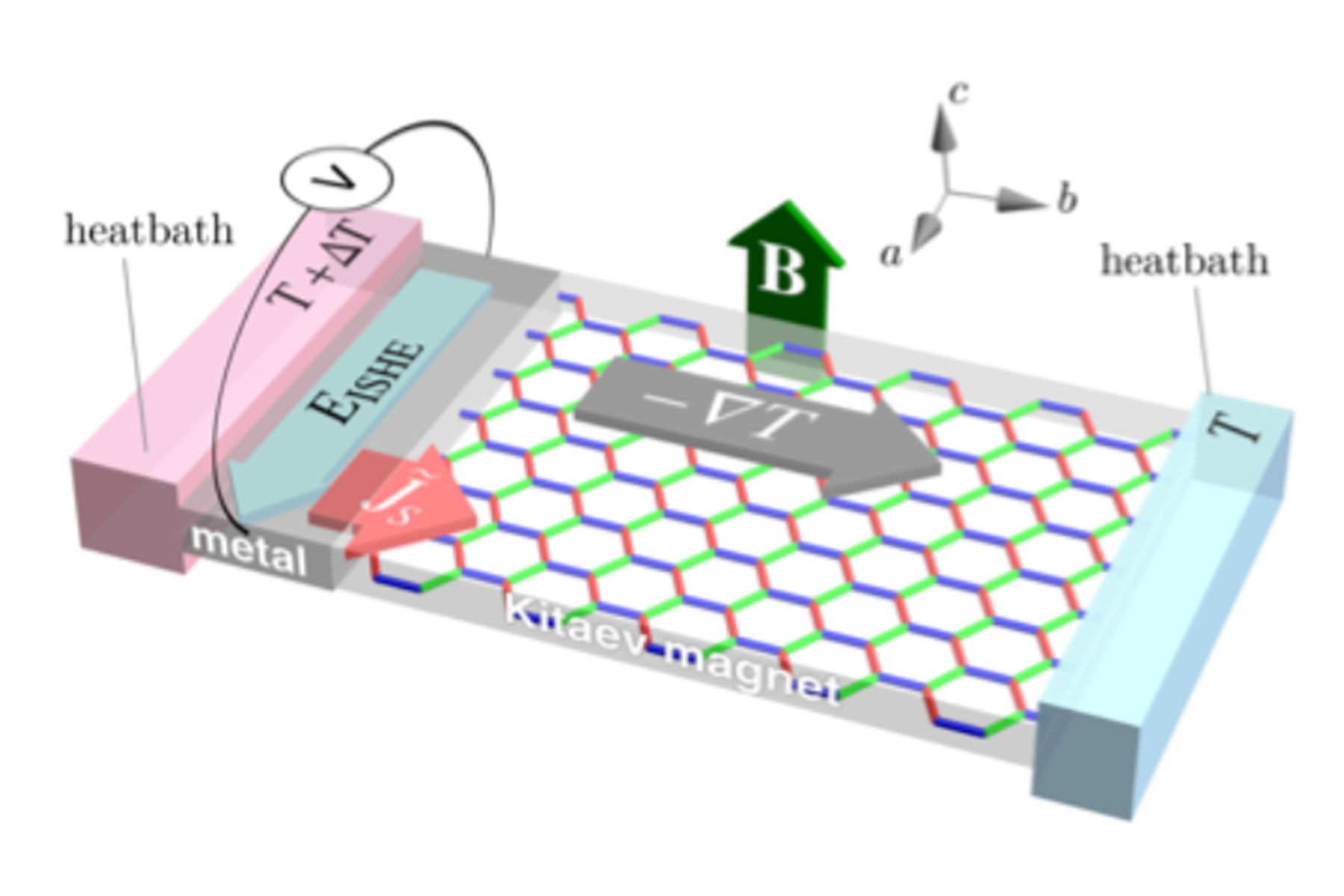}
 	\linespread{1}
 	\caption{Schematic illustration of an experimental setup for measuring the spin Seebeck effect in Kitaev magnets. The Kitaev magnet is attached to heat baths on both sides to establish a temperature gradient (gray arrow), while a magnetic field is applied in the out-of-plane direction (green arrow). A metal with strong spin-orbit coupling, such as Pt, is placed at one end to detect the spin current (red arrow), which is generated by the thermal gradient. This spin current is converted into an electric field (light blue arrow) via the inverse spin Hall effect. Adapted from \cite{Kato2025}.}
    \label{MotomeSeebeck}
\end{figure}

Complementing this approach, the spin Seebeck effect (SSE) was recently studied theoretically for the Kitaev model \cite{Kato2025}. The SSE refers to a thermoelectric response in which a spin current is generated by a thermal gradient, and it is widely used as a sensitive probe for elementary excitations in magnets. While this probe was previously applied to a quantum spin liquid state in a quasi-one-dimensional magnet \cite{Hirobe2017}, no studies had addressed quantum spin liquids in higher dimensions. In the KQSL state in the two-dimensional honeycomb Kitaev model (Fig.\ \ref{MotomeSeebeck}), it was found that low-energy fractional Majorana excitations carry a spin current via the SSE, despite lacking spin angular momentum. Notably, the induced spin current can be either positive or negative depending on the sign of the Kitaev interaction (ferromagnetic or antiferromagnetic), contradicting earlier findings of a consistently negative SSE in one-dimensional systems \cite{Hirobe2017}. Furthermore, the study reveals contrasting field-angle dependencies between the low-field KQSL and the high-field ferromagnetic state, offering a potential experimental signature to distinguish these phases. These findings suggest that the SSE not only serves as a powerful tool for detecting fractional quasiparticles in quantum spin liquids but also holds promise for their generation and control, with implications for topological quantum computing.

\section{Tunneling spectroscopy}\label{tunneling}

Arising from Mott insulators, electrons in KQSL are strongly localized to lattice sites. The absence of itinerant charge carriers severely limits low-energy charge excitations and as mentioned in the last section,  restrains the application of electrically accessible probes for investigating fundamental properties. Consequently, understanding the behavior of the Kitaev model under charge doping and the inter-coupling between charge and fractionalized quantum states becomes a pertinent theoretical question. 

In the specific case of \arucl, its ability to be exfoliated into quasi-2D monolayers or directly grown as 2D films makes electron tunneling spectroscopy in thin films a promising method for detecting quantum excitations. Tunneling of electrons through the Kitaev thin film can provide information on magnetic excitations, such as fractionalized Majorana excitation, through the analysis of the tunneling spectra \cite{Feldmeier2020,Koenig2020,PhysRevB.102.085412,PhysRevLett.125.227202,Klocke_2021,PhysRevLett.126.127201,PhysRevB.104.235118,Bauer_2023,Takahashi2023,Kao2023,PhysRevLett.132.206501,Kao2024,Zhang2024,Zhang2024b}. In this section, we review current research progress on charge doping in KQSL and the tunneling spectroscopy in \arucl\ thin films. 

\subsubsection{Theoretical considerations of hole doping and the charge-spin coupling}


Understanding the properties of the Kitaev model subjected to hole doping, as well as analyzing the dynamical hole spectral functions, provides valuable insights into the structure of fractionalized quantum spin liquids. As mentioned in section I, at charge neutrality, the exact solution of the Kitaev model shows that the pristine KQSL has a gapless ground state with Dirac-type Majorana excitations \cite{Kitaev2006}, while its static $Z_2$ flux excitations remain gapped. To introduce mobile holes into the system, one can consider a mixed hopping-Kitaev model, as introduced Refs.\ \cite{you_doping_2012, Hyart2012, Okamoto2013, Scherer2014, Trousselet2014}, where the local physical Hilbert space is extended from the two-dimensional spin space to a three-dimensional one, which includes the empty site as well. In that description, a kinetic term can be added for nearest-neighbor hole hopping to the Kitaev Hamiltonian:
\begin{equation}
    H = H_t+H_K = -t\sum_{\langle i j\rangle, \sigma}\mathcal{P}_\mathrm{GW}\left(c^\dagger_{i\sigma} c_{j\sigma} + \mathrm{h.c.} \right)\mathcal{P}_\mathrm{GW} - \sum_{a, \langle i j\rangle_a} K_a \sigma^a_i \sigma^a_j, \label{eq::Kitaev_tJ}
\end{equation}
where $c^\dagger_{j\sigma}$ ($c_{j\sigma}$) create (annihilate) fermions with spin $\sigma$ and doubly occupied sites are excluded by the Gutzwiller projector $\mathcal{P}_\mathrm{GW}$. The holes are related to the spin operators by $\sigma^a_i=(c^\dagger_{i\uparrow}, c^\dagger_{i\downarrow}) \sigma^a (c_{i\uparrow}, c_{i\downarrow})^T$, where $\sigma^a$ are the corresponding Pauli matrices. The spin anisotropy in the Kitaev model originates from the strong SOC of electrons to the $d$-orbitals \cite{Jackeli2009}. Therefore, the spins above are not electronic spins but rather hybridized pseudo-spins. Nevertheless, they can directly be related to the electron operators at low energy with isotropic hopping \cite{you_doping_2012,Kadow2024}.

In the slow hole limit ($t\ll{}K$), the KQSL phase is robust and holes only introduce quasi-static vacancies \cite{Willans2010,Willans2011,Halasz2014,Halasz2016}. The model is still analytically solvable in this limit of quasi-static vacancies and the holes are associated with flux, fermion, and plaquette quantum numbers. Strictly speaking, this quasi-static picture only holds in the limit of ultraslow holes, as long as the holes do not couple to any bulk excitations of the undoped Kitaev model. Nevertheless, it is numerically found that the hole spectral function is captured well by the slow-hole description even beyond this perturbative limit \cite{Kadow2024}. 

In realistic systems, the hole kinetic energy $t$ is expected to be much larger than the Kitaev exchange $K$ ($t\gg K$), and initial works based on parton mean-field theories suggested superconducting ground states in the hole-doped regime \cite{you_doping_2012,Hyart2012,Okamoto2013,Scherer2014}. Numerical results based on Density Matrix Renormalization Group (DMRG) could not confirm a superconducting state on cylinder geometries for either ferromagnetic or antiferromagnetic  Kitaev couplings $K$ \cite{White1992,White1993,Peng2021,Jin2024}. This might hint toward very small scales that are involved in stabilizing a putative superconducting state at finite doping. 

In the case of ferromagnetic Kitaev couplings, which are believed to be relevant for \arucl, it is found instead that the KQSL is remarkably fragile already for small and slow hole doping. For hopping strengths $t$ on the order of the KQSL's flux gap, $\Delta_{}E_{\rm v}\sim0.1K$, the system is already partially FM polarized in the [001] direction by the itinerant holes, spontaneously breaking the time-reversal symmetry \cite{Jin2024}. This is understood from kinetic ferromagnetism, where the hole gains kinetic energy by delocalizing through the system while the spins are orienting themselves ferromagnetically to avoid destructive interference in the wave function \cite{Nagaoka1966}. Since the local interactions of the model are ferromagnetic to start with, this state is easily stabilized already in the presence of slow holes. These results can also be explained from a parton theory of fermionic holes and bosonic spinons/magnons. It unveils that the hole kinetic term effectively serves as a FM Heisenberg coupling destabilizing the KQSL \cite{Jin2024}. In addition, the parton theory shows that the resulting FM order along the [001] direction originates from an order-by-disorder mechanism \cite{Shender1982,Henley1989,Jin2024}. 

Spectroscopic features of the KQSL can be obtained by charge tunnel spectroscopy or angle resolved photoemission spectroscopy (ARPES). The electrons are thereby removed from the system and have to overcome the charge gap. Focusing on the gapped KQSL with anisotropic Kitaev couplings,  two  distinct scenarios have been identified~\cite{Kadow2024}. For ferromagnetic Kitaev couplings, the spin liquid is highly susceptible to hole doping: a Nagaoka ferromagnet forms dynamically around the doped hole, even at weak coupling. By contrast, in the case of antiferromagnetic spin couplings, the hole spectrum demonstrates an intricate interplay between charge, spin, and flux degrees of freedom, best described by a parton mean-field ansatz of fractionalized holons and spinons~\cite{Kadow2024}. The dynamical hole spectral functions thus provide rich information on the structure of fractionalized quantum spin liquids. 

The model discussed so far is the pristine Kitaev model subjected to hole doping. Residual zigzag magnetic order appears in materials at low temperatures because of the inevitable presence of additional interactions such as the off-diagonal symmetric ($\varGamma$) and Heisenberg ($J$) exchanges \cite{Chaloupka2013,Rau2014,Yamaji2016}. In particular the $K$-$\varGamma$ (with $J=0$) is believed to remain disordered \cite{Janssen2017, Catuneanu2018, Gohlke2018}. In more elaborate models, hole doping can significantly lower the critical field above which the field-polarized FM state appears in Kitaev candidate materials, which can therefore further challenge the estimation of model parameters \cite{Jin2024}. 


\subsubsection{Inelastic electron tunneling response}\par

Inelastic electron tunneling response allows for measuring collective magnetic excitations in KQSL even without charge doping. In this setting, the insulating sample is mounted between a metallic tip and a metallic substrate. The tunneling conductance $\partial I/\partial V$ is then related to the spin structure factor as~\cite{Rossier2009_stm,Balatsky2010_stm,Bode2003_stm}
\begin{equation}
    \frac{\partial I}{\partial V} = \frac{2e^2}{\hbar}\sum_{i,j,\alpha,\beta}T({r}-{r}_i)T({r}-{r}_j)
\; c_{\alpha\beta} \, \int_0^{eV} d\omega \, \mathcal{S}_{ij}^{\alpha\beta}(\omega),
\end{equation}
where $\mathcal{S}_{ij}^{\alpha\beta}(\omega)$ is the Fourier transform of the spin correlations of spin components $\alpha, \beta$,  $T(r)$ is a local tunneling matrix element that decays exponentially with the distance between the tip and the sample. Due to the exponential sensitivity of tunneling current, the  detection of the local structure factor
with a spatial resolution close to the atomic distance becomes possible. Furthermore,  $c_{\alpha\beta} = \sum_{\sigma,\sigma^\prime} n_\sigma(\varepsilon_F)N_{\sigma^\prime}(\varepsilon_F) \sigma^{\alpha}_{\sigma^\prime \sigma}\sigma^\beta_{\sigma\sigma^\prime}$, where $\sigma^{\alpha}$ are the Pauli matrices and $n_\sigma(\varepsilon_F)/N_\sigma(\varepsilon_F)$ are the spin-dependent densities of states at the Fermi level for both tip/substrate. 

Crucially, the prefactors $c_{\alpha\beta}$ depend on the \textit{relative spin-polarization} of tip and substrate. This allows for a controlled selection of spin excitations that are to be probed~\cite{Kirschner2014_spstm,Rossier2009_stm,Balatsky2010_stm}. Three limiting cases can be distinguished
\textit{(1)} Non-polarized tip and substrate ($n_+=n_-$ and $N_+=N_-$): $c_{\alpha\beta}\sim \delta_{\alpha\beta}$ and independent of $\alpha$.
\textit{(2)} Fully parallel-polarized tip and substrate ($n_-=N_-=0$): $c_{\alpha\beta} \sim \delta_{\alpha,z}\delta_{\beta,z}$, where $z$ was chosen as the common polarization axis.
\textit{(3)} Fully anti-polarized tip and substrate ($n_-=N_+=0$): $c_{\alpha\beta} \sim (1-\delta_{\alpha,z})(1-\delta_{\beta,z})$. Therefore, depending on the spin-polarization of the tip and the sample distinct type of spin excitations of the sample are measured.

\begin{figure}[t!]
     \centering
     \includegraphics[width=0.8\textwidth]{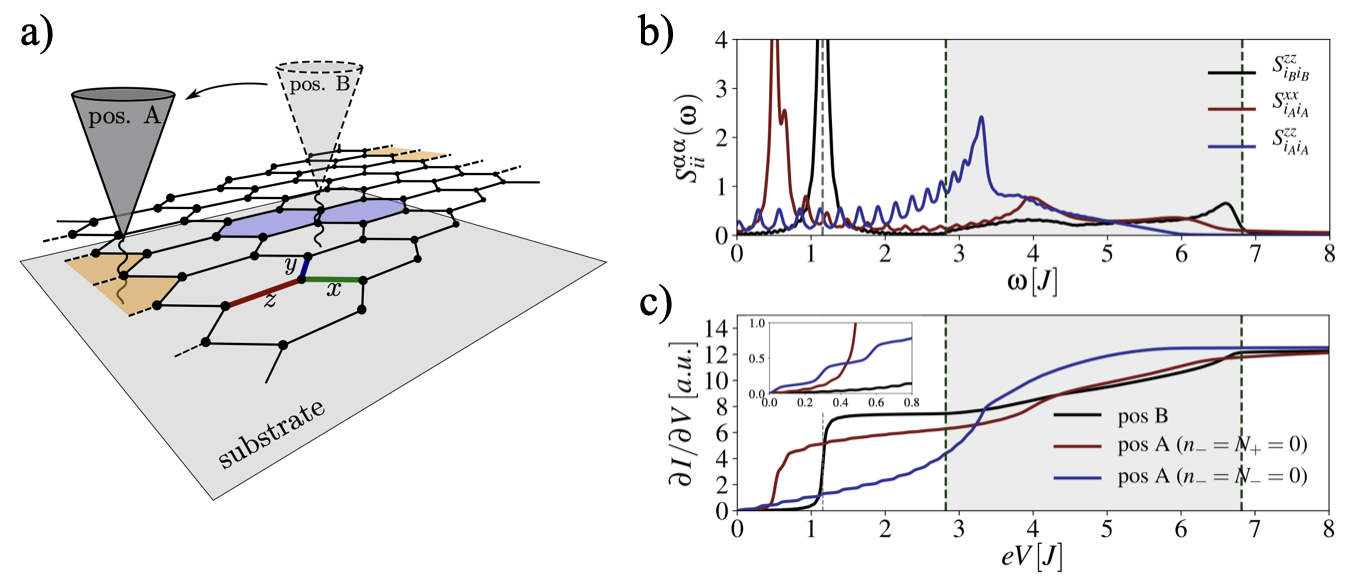}
     \caption{Theoretical prediction for inelastic tunneling spectroscopy of Kitaev systems. (a) Sketch of the tunneling experiment with tips positioned at the edge (Pos. A) and at the bulk (Pos. B). (b) Dynamical structure factor in the bulk of the chiral phase of the extended Kitaev model shows a sharp peak within the gap arising from the fermion bound state (black), whereas at the edge a series of peaks arises from the dispersive edge modes, starting from zero frequency (blue). The spin structure factor $S^{xx}$ evaluated at the edge involves the creation of a boundary and a bulk flux excitation, hence it exhibits a peak at approximately half of the bulk fermion bound state (red).  (c) The conductance shows a step like behavior in the bulk at low bias (black) whereas at the edge it continuously increases with bias due to the presence of the edge state (blue); a signature of chiral edge states and fractionalization. Adapted with permission from \ \cite{Feldmeier2020}, © 2020 American Physical Society.}
     \label{fig:stm}
 \end{figure}

The inelastic tunneling response has been predicted for distinct types of KQSL. In Ref.\ \cite{Feldmeier2020} the extended Kitaev model with a three spin term that is perturbatively generated from a weak field, hosting a non-abelian spin liquid with edge states has been analyzed. The locality of inelastic tunneling experiments thereby allows one to detect the gapless edge excitation near the boundary. The gapless response at the edge is then indicative for fractionalization; see Fig.\ \ref{fig:stm}. Related ideas have been analyzed in Ref.\ \cite{Koenig2020,PhysRevB.102.085412,PhysRevLett.125.227202,Klocke_2021,PhysRevLett.126.127201,PhysRevB.104.235118,Bauer_2023,Takahashi2023,Kao2023,PhysRevLett.132.206501,Kao2024, Zhang2024, Zhang2024b}. Similarly to the spin-polarized inelastric electron response, it is theoretically proposed that the magnetic structure factors can also be measured by NV center magnetometry. Non-Abelian anyons in KQSL can be further characterized by detecting the unique low-energy excitation characteristics of the hybrid modes of dangling Majorana fermions and Majorana zero modes (MZMs) released near the vacancies \cite{Xiao2025}.

\subsubsection{Inelastic tunneling spectroscopy of magnons in 2D \arucl}\label{iets}


\begin{figure}[t!]
     \centering
     \includegraphics[width=0.75\linewidth]{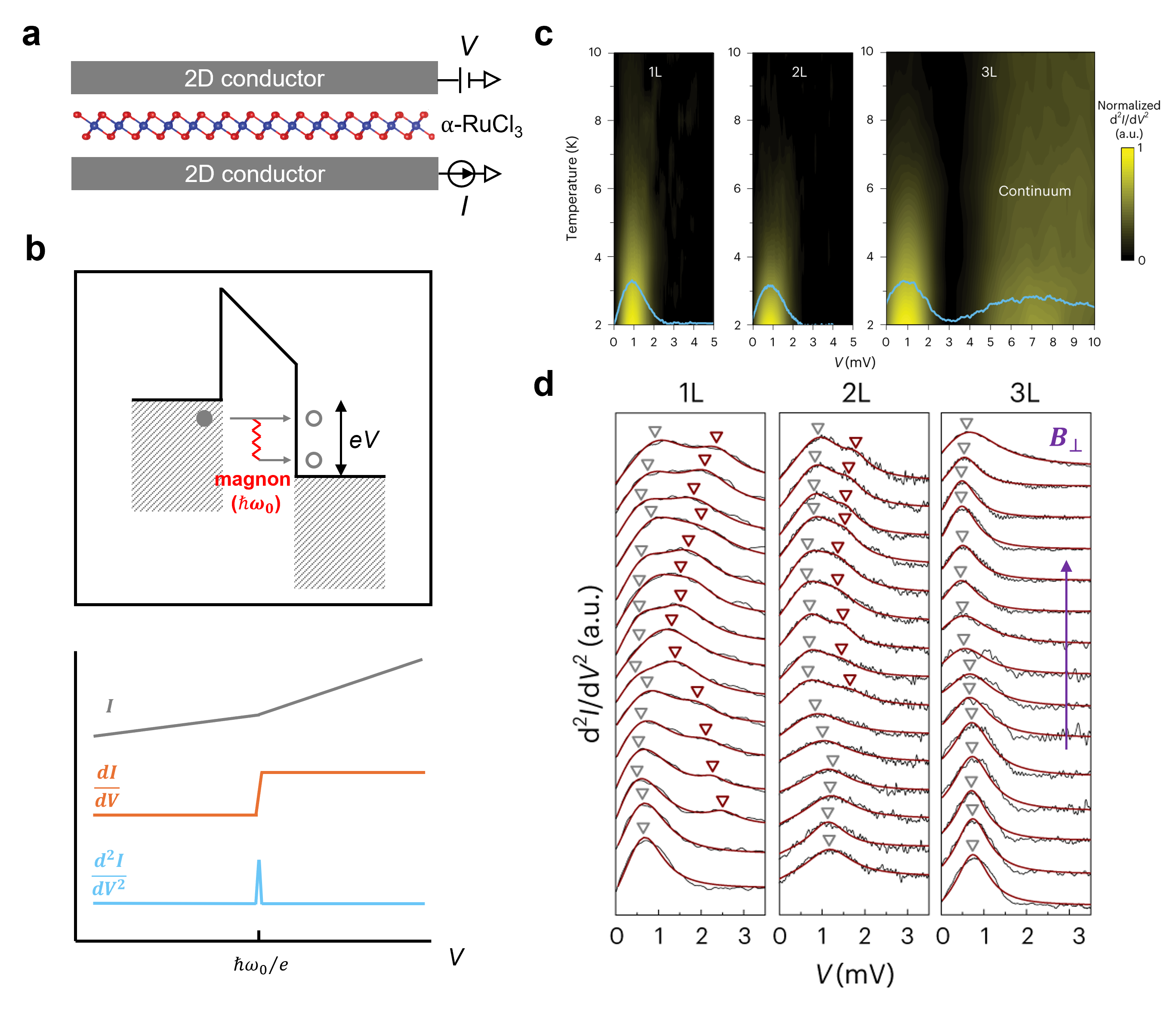}
     \caption{Inelastic tunneling spectroscopy (IETS) of ultrathin \arucl. (a) Device and measurement schematic of metal/2D magnetic insulator (\arucl)/metal tunnel junction. (b) IETS mechanism and transport characteristics. (c) Temperature-dependent IETS spectra of monolayer, bilayer, and trilayer \arucl. The low-energy magnon mode ($\approx$1 meV) for all three samples disappears above the NÃ©el temperature but the broad continuum seen at higher energy ($\approx$5-10 meV) for trilayer persists. (d) Evolution of IETS spectra at 2 K with out-of-plane magnetic field. Shifts in the magnon energies are strongest in the monolayer. A two-magnon feature can be seen (marked by red arrows) that exhibits an energy minimum at $\approx$7 T. This corresponds to the critical out-of-plane field for monolayer \arucl, which undergoes a magnetic anisotropy reversal from easy-plane in the bulk crystal to easy-axis. Data from (c) and (d) are taken from  \cite{Yang2023-ke}, Copyright © 2022, The Authors, under exclusive license to Springer Nature Limited.}
     \label{Tsen_IETS}
 \end{figure} 
 
A variety of experimental schemes based on the theories of electron tunneling and charge doping have been applied to study of the properties of \arucl. After the observation of (anti-)ferromagnetic order in vdW magnetic insulators persisting down to in single atomic layers using optical techniques \cite{Lee2016-ig,gong2017discovery, huang2017layer}, tunneling magnetoresistance soon followed as a convenient transport probe of these 2D systems \cite{Wang2018-ul, doi:10.1126/science.aar4851, doi:10.1126/science.aar3617, Kim2018-wd}. Devices typically consist of a metal/magnetic insulator/metal tunnel junction (see Fig.\ \ref{Tsen_IETS}(a)), whose conductance may depend on the spin configuration of the magnetic barrier. Evaporated metal films can be used for the metal electrodes, but due to the chemical instability of the magnetic compounds, researchers have generally preferred to use another conducting 2D material such as graphene or graphite, so that the entire heterostructure can be assembled in an inert atmosphere using standard co-lamination techniques. The most salient transport signature for such devices is an extremely large negative magnetoresistance seen in layered antiferromagnets (such as ultrathin CrI$_3$ and CrCl$_3$). When an external magnetic field is applied to polarize the spins in the magnet, the barrier is lowered for tunneling electrons with that spin polarization, resulting in a much-enhanced, ``spin-filtered'' conductance.

The same device geometry can also be used to electrically detect the magnetic excitations (magnons or spin waves) and their energies via a technique called inelastic tunneling spectroscopy (IETS) \cite{REED200846}. The voltage bias V applied across the junction directly corresponds to the Fermi energy difference of the two metals (eV), which at low temperature can be regarded as the maximum energy lost by electrons after tunneling. For small bias, the tunneling itself is elastic, as electrons do not interact with excitations within the barrier and energy loss occurs entirely during relaxation at the drain electrode. For larger bias, electrons can instead give their energy to create excitations within the magnetic insulator (such as a magnon or phonon) in an inelastic tunneling process (see Fig. \ref{Tsen_IETS}(b)). If $\hbar\omega_0$ is the energy of the excitation, inelastic tunneling can only occur when eV $\geq\hbar\omega_0$, while the momentum of the excitation should further be given up by electron.

Note that the backwards process (drain-to-source tunneling via annihilation of an excitation in the magnetic insulator) should be less significant as there are fewer magnons or phonons populating excited states compared to the ground or vacuum state. The introduction of an inelastic tunneling pathway thus effectively leads to a stronger increase of the current I with voltage V compared with that for elastic tunneling alone. This enhancement can be observed most strikingly as an abrupt jump in the local conductivity ($dI/dV$) or peaks in its derivative ($d^2 I/dV^2$) at the mode $eV=\hbar\omega_0$ (see Fig. \ref{Tsen_IETS}b, bottom). Magnon scattering can be further distinguished against phonons by the shift of the modes with an applied magnetic field.

The first IETS measurements of magnons in 2D magnets were performed on the chromium trihalides (CrI$_3$, CrBr$_3$, CrCl$_3$) and were used to extract the microscopic exchange parameters for these systems \cite{doi:10.1126/science.aar3617, Ghazaryan2018-lz, doi:10.1073/pnas.1902100116}. The magnon energies generally fell in the range of several to tens of meV; however, modes appeared to vary across studies even for the same material. This could be potentially attributed to different momentum states of the magnons involved in the scattering process. As the Fermi surfaces of graphene or graphite encircle the $K$ points, varying the twist angle between the two graphite electrodes would create an electronic momentum mismatch that thus selects for magnons of certain momenta (and so energy). The twist angle was generally not a controlled parameter for these studies.

Raman and neutron studies have shown bulk \arucl\ hosts magnetic excitations in a similar energy range \cite{sandilands_scattering_2015,banerjee_neutron_2017}, which persist to monolayer thicknesses \cite{Zhou2018,Du2018,Lee2021a}. In principle, due to changes in interlayer coupling, proximate charging, and/or structural distortions, the exchange parameters can be altered in \arucl\ with reduced thickness, which would thus modify the magnetic excitation spectrum and/or its evolution with magnetic field. Yang, \etal, performed IETS measurements on ultrathin \arucl\ using the metal/magnetic insulator/metal tunnel junction geometry \cite{Yang2023-ke}. Instead of graphite, they used 1$T'$/$T_d$ MoTe$_2$ flakes as the metal contacts to avoid the momentum mismatch problem. The Fermi surface of 1$T'$-$T_d$-MoTe$_2$ consists of large and relatively spherical hole pockets centered around the $\Gamma$ point, which can allow for transitions with a large range of momentum changes regardless of twist angle. Assuming magnons across the entire Brillouin zone can be accessed, one may expect IETS features resembling a momentum-integrated magnon spectrum.


The temperature-dependent IETS spectra taken by Yang, \etal,  for monolayer, bilayer, and trilayer \arucl\ are shown in Fig. \ref{Tsen_IETS}(c). At low temperature, a relatively narrow peak is seen centered at $\approx$ 1 meV for all three thicknesses, which corresponds to the lowest energy magnons with high spectral weight at the $Y/M$-points \cite{Yang2023-ke}. These peaks gradually disappear above the N\'{e}el temperature of $\approx$8 K. In contrast, for the trilayer a broad feature is seen at higher energies (between $\approx$5-10 meV) that persists up to 10 K without attenuation. This feature is consistent with the continuum excitations observed in bulk crystals using neutron and Raman scattering and may relate to fractionalized and/or incoherent excitations \cite{sandilands_scattering_2015, banerjee_neutron_2017,Zhou2018,Du2018,Lee2021a}.

The field evolution of the magnons show a departure from that of the bulk crystal. As the out-of-plane direction is the hard axis for bulk \arucl, very little shift in the magnon energies is expected for a perpendicular magnetic field up to $\approx$30 T \cite{Li2021-vk}. As shown in Fig.\ \ref{Tsen_IETS} (d), the low-energy magnon mode observed in IETS is relatively stiff for the trilayer, but shows more change with decreasing thicknesses. For the monolayer, a secondary peak attributed to two-magnon scattering can also been seen which exhibits a clear energy minimum at $\approx$7 T. With additional in-plane magnetic field dependence together with magnetotransport measurements performed on a gated monolayer device, the authors conclude that the magnetic anisotropy is reversed (from easy-plane to easy-axis) in monolayer \arucl. Electron diffraction experiments and ab-initio density function calculations reveal that this change is driven principally by a structural distortion involving an elongation of the vertical Cl atom distance away from the Ru plane. Concomitantly, this results in an enhancement of the Kitaev exchange for the monolayer relative to the bulk crystal.

\subsubsection{Tunneling magnetoresistance}

In a separate work, Ref.\ \cite{massicotte_giant_2024} investigates the magnetic and electronic properties of few-layer \arucl\ by way of tunneling magnetoresistance in graphite/\arucl\ heterostructures. It is widely recognized that a promising approach to realizing the KQSL in \arucl\ is to reduce its dimensionality via mechanical exfoliation, as this enhances order parameter fluctuations and facilitates the emergence of QSL behavior \cite{kim2022alpha}. 

The experiment employs angle-dependent tunneling magnetoresistance (TMR) measurements using magnetic tunnel junctions with ultrathin \arucl\ flakes sandwiched between graphite electrodes. A giant anisotropic magnetoresistance effect is observed, with the tunneling resistance changing as much as 2500\% when a magnetic fields is applied within or perpendicular to the plane of the device. This dramatic response stems from the strong spin-orbit coupling and magnetic anisotropy, which creates highly directional tunneling barriers for electrons with different spin orientations. A two-fold rotational symmetry in the magnetoconductance is observed, contrasting with the six-fold symmetry reported previously for some, but not all, bulk samples \cite{Balz2021,Tanaka2022,LampenKelley2018}. The two-fold symmetry in the magnetoresistance and the absence of any structural phase transition signatures in transport measurements indicate that the exfoliated flakes retain the high-temperature monoclinic C2/m crystal structure down to low temperatures, unlike bulk crystals that typically undergo a phase transition to rhombohedral stacking around 150 K.

Through a systematic study of the temperature and magnetic field dependence, Ref.\ \cite{massicotte_giant_2024} maps out the magnetic phase diagram of few-layer \arucl\ flakes. A zigzag antiferromagnetic ground state is found with the enhanced N\'eel temperature $T_N=14$ K known for structurally disordered samples. These maps reveal a critical magnetic field $B_c \approx 9-10$ T, where the system transitions to a partially polarized quantum disordered state. High-resolution scanning transmission electron microscopy confirms the presence of stacking disorder and faults in the exfoliated flakes. The enhanced magnetic properties arise from the structural disorder introduced during mechanical exfoliation, which effectively pins the crystal in the monoclinic phase and alters the magnetic exchange interactions.

\subsubsection{STM/STS investigations of the electronic properties and Mott transitions in \arucl}\label{stm}


Although current STM/STS-based experimental studies have not yet provided conclusive experimental evidence for the fractionalized excitations in \arucl\ as predicted by theory groups, such experiments have made progress in revealing the intrinsic Mott characteristics of \arucl\ and the evolution of its electronic states in heterostructures. In particular, the magnitude and characteristics of the Mott gap in \arucl\ are currently a subject of ongoing debate. The magnitude of the Mott gap has been found to range widely, from approximately 0.2 to 2.2 eV as determined through different experimental approaches \cite{binotto_optical_1971,guizzetti_fundamental_1979,plumb_2014, sandilands_optical_2016,koitzsch_j_eff_2016,sinn_electronic_2016,ziatdinov_atomic-scale_2016,zhou_angle-resolved_2016,warzanowski_multiple_2020,koitzsch_low-temperature_2020,nevola_timescales_2021,annaberdiyev_electronic_2022}. This substantial ambiguity hampers both the precise estimation of the Coulomb repulsion $U$ in \arucl, a critical parameter for evaluating electron correlations, and the identification of appropriate strategies to electrically access fractionalized excitations in experiments. 

Using angle-resolved photoemission spectroscopy (ARPES), multiple research groups have consistently detected the valence band, also called the lower Hubbard band (LHB), approximately 1.2 eV below the Fermi level (FL) \cite{zhou_angle-resolved_2016,wang_evidence_2021,rossi_direct_2023, Barfield2023}. Nevertheless, ARPES faces challenges in providing information on the conduction bands (upper Hubbard band-UHB). The utilization of photo-emission (PE) and electron-energy-loss spectroscopy (EELS) yields extensive insights into the electron structures within the Mott-Hubbard bands \cite{sandilands_optical_2016,warzanowski_multiple_2020,koitzsch_j_eff_2016,koitzsch_low-temperature_2020,sandilands_spin-orbit_2016}. Four prominent features at around 1.2 eV, 2.1 eV, 3.2 eV and 5 eV have been identified, with the peaks at 1.2 eV and 2.0 eV attributed to the intersite $dd$ transitions involving adjacent Ru sites ($d^5$$d^5$ $\rightarrow$ $d^4$$d^6$). By integrating cluster model calculations, the optical gap of 1.2 eV was attributed to the Mott gap in \arucl \cite{sandilands_optical_2016}. 

However, there is still ongoing debate concerning this 1.2 eV Mott gap.  Sinn, \etal, reported the observation of the conductance band (UHB) using inverse photoemission spectroscopies (IPES) \cite{sinn_electronic_2016}. They found that the UHB locates around 1 eV above FL. Nevola, \etal, recently validated this finding through the use of time-resolved two-photon photoemission spectroscopy (TPPES) \cite{nevola_timescales_2021}. Considering the positions of LHB in the ARPES findings and UHB as identified in the aforementioned studies, it is apparent that the Mott gap could exceed 2 eV, almost \textit{double} the value derived from optical evaluations\cite{sinn_electronic_2016,nevola_timescales_2021}. This substantial Mott gap was additionally bolstered by the calculations employing fixed-node and fixed-phase diffusion Monte Carlo methods \cite{annaberdiyev_electronic_2022}.

 \begin{figure}[htp]
 	\centering
 	\includegraphics[width=1\columnwidth]{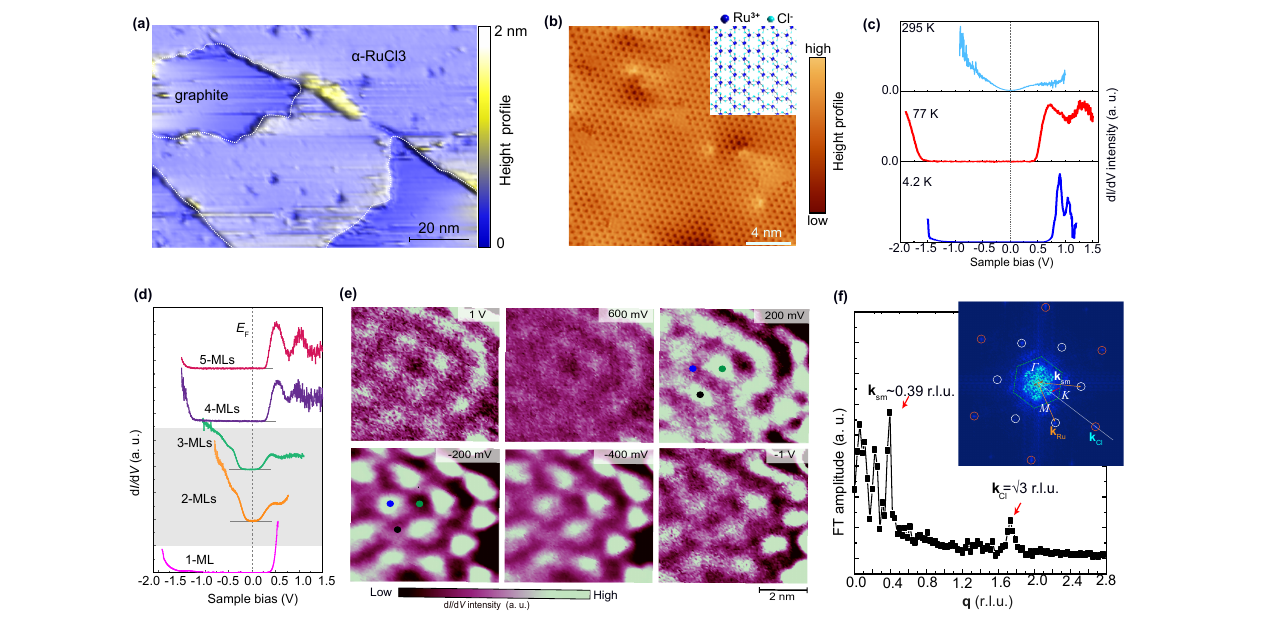}
 	\linespread{1}
 	\caption[Fig 1]{(a) STM morphology of \arucl\ flakes transferred on a graphite substrate; (b) STM morphology of the bulk \arucl\ (thickness \textgreater\ 5 nm): The lattice constructions are presented in the inset; (c)$dI/dV$ spectra were acquired at different temperatures; (d) Selected $dI/dV$ spectra taken on \arucl\ flakes with thicknesses ranging from 1-ML to 5-MLs showing two types of lineshapes, one with a large gap and the other with a reduced gap; (e) sample bias-resolved $dI/dV$ maps measured at V$_{bias}$ = 500 mV, I$_{set}$ = 1 nA on a 2-ML \arucl; (f) inset: FFT of the STM image shows the Bragg peaks of the \arucl\ lattice (white and red circles) and  peaks of the static super-modulation inside the 1st BZ (green dashed hexagon). k$_{Ru}$ and k$_{Cl}$ denote the wave vectors of ruthenium (Ru) and chlorine (Cl) lattice, respectively. k$_{sm}$ is the wave vector of the super-modulation. The line cut along the $\Gamma$-K direction shows that $k_{sm}$ is ${\sim}0.39$ r.l.u.\ (r.l.u.\ denotes reciprocal lattice unit of \arucl\ 2$\pi$/$a$, $a$ = 6 \AA). Data are taken from Ref.\ \cite{zheng_tunneling_2023} and \cite{zheng_incommensurate_2024}, © 2022, American Physical Society.}
    \label{fig:figure-IV-1}
\end{figure}

Scanning tunneling microscopy/spectroscopy (STM/STS) is a potent technique for studying the atomic-scale lattice structures and the local electron density of states (LDOS). For most Mott insulators, performing STM/STS measurements is impractical due to their insulating character. However, single layers of \arucl\ are well-suited for electron tunneling and STM/STS experiments \cite{ziatdinov_atomic-scale_2016,balgley_ultrasharp_2022}. Indeed in an early STM/STS study, Ziatdinov, \etal\ discovered the intra-unit cell symmetry breaking of charge distribution in the \arucl\ lattice at 295 K, and reported a charge-gap of approximate 0.25 eV in $dI/dV$ spectrum, which is far below the optical and ARPES experimental data \cite{ziatdinov_atomic-scale_2016}. 

By exfoliating and transferring \arucl\ thin films onto a fresh graphite surface as shown in Fig.\ \ref{fig:figure-IV-1}(a), early studies investigated the morphologies and $dI/dV$ spectra of \arucl\ as a function of film thickness and temperature \cite{zheng_tunneling_2023,zheng_insulator-metal_2024,zheng_incommensurate_2024}. It was found that when the thickness of the \arucl\ flake exceeds approximately 5 nm (hereafter referred to as thick \arucl), the STM/STS results can offer the intrinsic information of pristine \arucl.  Otherwise, for thinner flakes on graphite, the charge characteristics are somewhat influenced by the interfacial coupling between \arucl\ and the underlying graphite. The STM image in Fig.\ \ref{fig:figure-IV-1}(b), acquired at 77 K on thick \arucl, displays a clear lattice, where Ru atoms, surrounded by Cl atoms, constitute the honeycomb lattice.

These STS experiments have now unveiled the full charge gap with a magnitude of approximately 2 eV on thick \arucl\ at low temperature (T $\leq$ 77 K) \cite{zheng_tunneling_2023}. This charge gap is a correlation gap, as evidenced by the effective suppression of any trivial in-gap states. Furthermore, the edges of LHB and UHB locate approximately at -1.5 eV and around 0.5 eV, respectively, which is in close agreement with the values observed in ARPES \cite{sinn_electronic_2016,zhou_angle-resolved_2016,wang_evidence_2021,rossi_direct_2023}, IPES \cite{sinn_electronic_2016}, TPPES \cite{nevola_timescales_2021}, and fixed-node and fixed-phase diffusion Monte Carlo calculations \cite{annaberdiyev_electronic_2022}. 

Upon recording spectra at three distinct temperatures corresponding to the phases of conventional paramagnet (above 120 K), Kitaev paramagnet (ranging from 7 K to 120 K), and antiferromagnet (below 7 K), it is evident that a comparable significant Mott-gap is detected in the spectra acquired at 4.2 K and 77 K. However, the complete charge gap is almost diminished due to the emergence of in-gap states at room temperature, leading to the formation of a soft gap proximal to the FL, as shown in Fig.\ \ref{fig:figure-IV-1}(c). The room temperature $dI/dV$ spectra show a high degree of consistency with those reported in previous STM/STS studies \cite{ziatdinov_atomic-scale_2016,frick_spreading_2023}. The transition of the spectra from a substantial full gap to a reduced soft gap remains inadequately clarified in the present context. There is an ongoing discourse regarding the significant impact of spin interactions on the Mott-Hubbard framework, particularly in relation to the dynamics of charge degrees of freedom \cite{senthilCriticalFermiSurfaces2008,senthilTheoryContinuousMott2008,longDynamicChargeKondo2022}. Understanding the temperature-dependent spectra in \arucl\ may be contingent upon the interplay among itinerant Majorana spinons, emergent flux, and the charge sector\cite{yue_observation_2024}. This necessitates additional experimental and theoretical exploration. 

In addition to the temperature dependence, the robustness of the Mott-Hubbard band to device and crystal quality including strain and defects was also investigated by STM/STS. The research established that the notable Mott gap around 2 eV at 77 K stays nearly unchanged in the presence of film corrugations and lattice vacancies \cite{zheng_insulator-metal_2024}. Nevertheless, as depicted in Fig.\ \ref{fig:figure-IV-1}(d), a significant variation in the Mott gap was observed when the thin films were transferred onto a graphite surface \cite{zheng_tunneling_2023,zheng_incommensurate_2024}. An unconventional Mott transition was identified, characterized by a sharp decrease in the charge gap in few-layer \arucl\ thin films. Interestingly, an insulator-to-metal Mott transition could be even observed under slight lattice distortion in monolayer \arucl\ \cite{zheng_insulator-metal_2024}. 

This unconventional Mott transition in \arucl\ manifests where the large Mott gap of approximately 2 eV in thick \arucl\ is notably reduced to a few hundred meV on 2-4 MLs \arucl\ that are transferred onto graphite. This similar phenomenon has been documented in multiple studies. First, Zhou, \etal, documented an unconventional Mott transition in \arucl\ during ARPES measurement under \textit{in situ} rubidium doping. This transition resulted in a significant reduction in the Mott gap, which, despite substantial doping, did not completely close \cite{zhou_angle-resolved_2016}. This phenomenon was attributed to the spin fluctuations induced by the Kitaev interaction in \arucl. Koitzsch, \etal, also observed the emergence of charge disproportionation after potassium intercalation in a single crystal of \arucl\ using EELS and PES measurement techniques. Despite a substantial reduction in the charge gap, it did not fully diminish, thus preserving the inherent insulating characteristics of \arucl\ \cite{koitzsch_nearest-neighbor_2017}. Jo, \etal, investigated a decrease in the optical gap from 1.2 eV to 0.7 eV after intercalating organic cations into bulk \arucl. It was proposed that the charge doping of \arucl\ results in a distinctive density of states in the valence band, causing a shift in optical absorption without undergoing a metallic transition, even under higher doping levels \cite{jo_enhancement_2021}. And recently Rossi, \etal, identified a spectral weight transfer within the valence band of \arucl\, shifting from -1.3 eV to approximately -0.5 eV when a thin film of \arucl\ is in proximity with graphene layer, as observed through nano-ARPES \cite{rossi_direct_2023}. 

Overall, numerous experimental studies have consistently indicated that \arucl\ experiences a substantial decrease in the charge gap while retaining its insulating characteristics when exposed to charge doping through diverse methodologies. Through a thorough acquisition and analysis of the $dI/dV$ spectra and maps, the emergence of Cl 3$p$ orbital textures was observed above the LHBs, accompanied by a rapid reduction of the Mott gap in layered \arucl\ when situated in close proximity to graphite \cite{zheng_incommensurate_2024}. The Mott-Hubbard framework of \arucl\ suggests that the Mott-Hubbard bands stem from the Ru 4$d$ orbitals, whereas the majority of Cl $p$ states are conventionally believed to manifest at binding energies below -2 eV \cite{sandilands_spin-orbit_2016}. Consequently, the identification of ligand (anion) states in the Mott gap of \arucl\ lead to a significant finding:\ the aforementioned unconventional Mott transition in layered \arucl\ in close proximity to graphite is actually a transition from a Mott-insulator to a charge-transfer-insulator (CTI) \cite{zaanen_band_1985,khomskii_transition_2014,Pavarini:819465}.


The transition from Mott insulator to CTI in layered \arucl\ in the heterostructure provides the crucial insight that anion states are involved into the low-energy excitations of the strong correlation compound. The process of electron hopping in \arucl\ undergoes a fundamental change from $d^5d^5$$\rightarrow$$d^4d^6$ in a Mott insulator to $d^5L^n$$\rightarrow$$d^4L^{n+1}$ in a CTI, where $L^n$ denotes the ligand states. Early investigations of cuprate superconductors have demonstrated the significance of anion states in the internal coupling between spin, charge and lattice degrees of freedom, which ensures intriguing physics such as high-temperature superconductivity, strange metal phase, charge density wave (CDW), and other symmetry broken states \cite{yee_phase_2015,hanaguri_checkerboard_2004,da_silva_neto_ubiquitous_2014,sakurai_imaging_2011,cai_visualizing_2016,ghiringhelli_long-range_2012}. Therefore, exotic physical phenomena associated with the \arucl\ of CTI are expected. 

In recent STM/STS work \cite{zheng_incommensurate_2024}, a super-modulation of both surface DOS and lattice morphology on a 2-MLs \arucl\ at temperature of 77 K is reported, as shown in Fig.\ \ref{fig:figure-IV-1}(e). A fast Fourier transform (FFT) of the super-modulation on lattice morphology determined that the super-modulation is incommensurate, as illustrated in Figs.\ref{fig:figure-IV-1}(f). A similar incommensurate super-modulation on \arucl\ has also been reported recently in two additional separate experiments:\ One study observed quantum interference with a comparable incommensurate wave-vector on grown 1-ML \arucl\ on graphite \cite{kohsaka_imaging_2024}, while another recent study examined the surface of hetero-structure formed by a graphene layer and a \arucl\ flake. By tuning the sample bias, STM was able to see an incommensurate super-modulation occurs on the surface of underlying \arucl\ flake \cite{qiu_evidence_2024}. The wave-vector of this super-modulation under negative biases is nearly consistent across all three experimental studies. 

To interpret these findings, Kohsaka, \etal, has focused on examining the potential connection between the quantum oscillation and Majorana fermions in the KQSL candidate through detailed theoretical investigations \cite{kohsaka_imaging_2024}. The study conducted by Qiu, \etal, has attributed the super-modulation to the formation of electron and hole crystals that occur as a result of the strong electron correlations in \arucl\ \cite{qiu_evidence_2024}. Zheng, \etal, claimed that the ferroelectricity may be caused by a redistribution of electron during the transition from a Mott insulator to the CTI. As a result, the orbital polarization takes place in \arucl, leading to the generation of an incommensurate dipole order inside the electric fields created by the hetero-interfacial charges. This ultimately results in the super-modulation of both surface DOS and lattice morphology. Despite the differences in the experimental setups and also interpretations of the super-modulation in these independent researches \cite{kohsaka_imaging_2024,qiu_evidence_2024,zheng_incommensurate_2024}, the experimental finding of a super-modulation is consistent evidence for significant internal coupling between charge, spin and lattice degrees of freedom, which could potentially enable the electrical detection of fractionalized excitations in \arucl.

\section{Light-matter interactions}\label{lightmatter}

Optical probes have been an invaluable source of information in bulk \arucl, but are more challenging to apply to micron-sized nanometers-thick flakes. Several works have tackled this challenge with techniques ranging from near-field scanning probes to time-resolved pump-probe experiments.




\subsection{Spatially resolved optical probes}

As noted, charge transfer in heterostructures consisting of \arucl\ and other materials, such as graphene, is predicted to alter the electronic band structure of \arucl\ and possibly enhance the Kitaev interaction \cite{Biswas2019,Gerber2020}. To explore the electronic properties and charge dynamics of \arucl\ on its own or in heterostructures, especially in the context of quantum phenomena, the study of charge transfer at \arucl\ interfaces is crucial. This is where the role of scattering-type scanning near-field optical microscopy (s-SNOM) becomes essential. s-SNOM offers an advanced, non-invasive technique to probe electronic and optical behaviors at the nanoscale. Through its high-resolution imaging capabilities, particularly in the infrared (IR) and terahertz (THz) ranges, s-SNOM enables the investigation of charge transfer plasmon polaritons (CPPs), facilitating the visualization of charge exchange in \arucl\ heterostructures \cite{cRizzo2020-sv, Rizzo2022-lw, Vitalone2024-wz}. 

The significance of s-SNOM lies in its ability to map local optical conductivity and plasmonic excitations induced by charge transfer without requiring external probes such as electrodes. This approach offers unique advantages over traditional methods of measuring conductivity and allows for the direct visualization of charge transfer dynamics in heterostructures \cite{cRizzo2020-sv, Rizzo2022-lw, Vitalone2024-wz, Ruta2022-cq, Kim2023-do, Shen2024-sz}. The work of Rizzo, \etal\ highlights the use of s-SNOM at infrared frequencies to study the graphene/\arucl\ interface \cite{cRizzo2020-sv}, where charge transfer plasmon polaritons emerge due to the large work function disparity between graphene and \arucl\. The charge transfer across this interface, driven by work function differences, results in the formation of CPPs, allowing for the measurement of Fermi energy in graphene through nanoimaging. 

 \begin{figure}[t!]
 	\centering
 	\includegraphics[width=0.6\columnwidth]{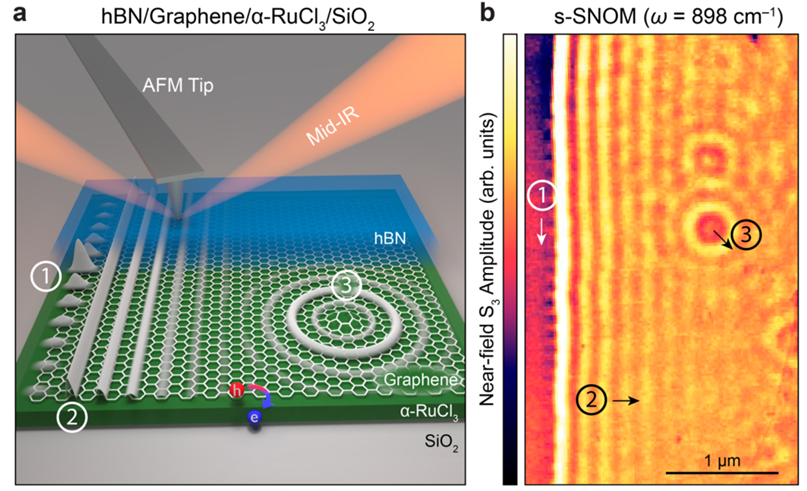}
 	\linespread{1}
 	\caption[Boyi]{Nanoimaging characterization of interlayer charge transfer in graphene/\arucl\ heterostructures using s-SNOM \cite{cRizzo2020-sv}. (a) Diagram of s-SNOM performed on hBN/graphene/\arucl/SiO$_2$. The charge transfer enables the generation of three types of plasmon features: (1) edge CPPs, (2) CPPs, and (3) circular CPPs. (b) Map of the near-field amplitude near the edge of graphene in hBN/graphene/\arucl/SiO$_2$ ($\omega$ = 898 cm$^{-1}$, $T = 60$ K) showing oscillations that are characteristic of CPPs. © 2020 American Chemical Society.} 
    \label{Boyi1}
\end{figure}

Figure \ref{Boyi1}(a) shows a schematic of s-SNOM interacting with a graphene/\arucl\ heterostructure encapsulated with hexagonal boron nitride (h-BN). In such an experiment, a metallized atomic force microscope probe tip is illuminated by a focused Mid-IR laser. The incident light field is enhanced at the tip apex and excites the plasmons \cite{Woessner2015-ap}, or CPPs in the case of graphene/\arucl\ . The CPPs of wavelength $\lambda_p$ propagate radially outward from the tip and are reflected at the sample edges or defects. The detection of the back-scattered light yields the near-field optical image of CPPs. CPP fringes with $\lambda_p$/2 spacing can be observed due to the interference of tip-launched and edge-reflected plasmons as the tip is at different locations relative to the sample edge (Fig.\ \ref{Boyi1}(b)). 

The CPP wavelength $\lambda_p$ is associated with Fermi energy E$_F$ in graphene as $\lambda_p\propto E_F^{1/2}$ \cite{LUO2013351}, enabling a direct measure of the charge transfer effect. The Fermi energy of graphene induced by charge transfer is usually found to be E$_F$=0.5-0.6 eV in the s-SNOM measurement, which agrees with theoretical predictions and other experimental observations \cite{Biswas2019, Gerber2020,zhou_evidence_2019,mashhadi_spin-split_2019, Wang2020,balgley_ultrasharp_2022,rossi_direct_2023}. This research demonstrates the capability of s-SNOM to characterize the nanoscale spatial variation of the charge transfer at the graphene/\arucl\ interface and suggests CPPs as a sensitive probe for charge transfer dynamics.

On the other hand, s-SNOM not only offers a powerful tool for visualizing charge transfer but also holds great promise for advancing our understanding of electronic and plasmonic behaviors in \arucl\ . Graphene can serve as a sensing layer and enable the study of the charge dynamics in effectively doped \arucl\ \cite{cRizzo2020-sv, Rizzo2022-lw, Vitalone2024-wz}. Future studies should include the use of cryo magneto-SNOM (m-SNOM) \cite{Dapolito2023-dr, doi:10.1126/sciadv.adp3487}, which provides access to measurements at temperatures below 5 K and magnetic fields up to 7 T. \arucl\ is suggested to be in a Kitaev paramagnetic state below $\approx$80 K and shows antiferromagnetic order below NÃ©el temperature around 7 K due to non-Kitaev interactions that can be suppressed by an in-plane magnetic field \cite{kasahara_majorana_2018}. Charge transfer from graphene to \arucl\ is expected to enhance the magnetic coupling in \arucl\ \cite{Biswas2019,Gerber2020}, potentially leading to an increase in NÃ©el temperature \cite{zhou_evidence_2019}. Cryo m-SNOM will be especially relevant for \arucl\ as the system allows us to study these temperature- and field-dependent phenomena with nanometer-scale spatial resolution. This might offer direct insights into how charge transfer mechanisms and plasmonic behavior evolve in \arucl-based heterostructures as the system approaches quantum-critical points.

Thus, incorporating cryogenic m-SNOM into the study of graphene/\arucl\ heterostructures is not just an enhancement of existing techniques, but a necessity for fully capturing the material's exotic low-temperature and high-field behaviors. This capability is essential for advancing our understanding of \arucl's quantum magnetic phases and charge transfer processes, which could have profound implications for quantum materials research and future applications in quantum information science. 

\subsection{Nature and dynamics of optical excitations}

Conventional optical spectroscopy also provides critical insights into the electronic states of \arucl. Early studies primarily focused on steady-state experiments to investigate the electronic structure and optical properties of \arucl\ by measuring the real and/or imaginary parts of the complex index of refraction as functions of light wavelength or sample temperature \cite{plumb_2014,sandilands_optical_2016,nevola_timescales_2021,b105085107,zheng_tunneling_2023}. While these studies yielded valuable information, the dynamical properties of photoexcitation have only recently been explored using time-resolved techniques. These more recent investigations suggest that the photoexcited states in \arucl\ at room temperature are predominantly governed by excitonic effects.

Photoexcitation across the gap between the upper and lower Hubbard bands of \arucl\ generates holons in the lower and doublons in the upper Hubbard bands, respectively. Due to strong correlation effects, a holon and a nearby doublon can form a tightly bound state known as a Mott-Hubbard exciton \cite{b487520}. This phenomenon is analogous to the formation of Wannier excitons in semiconductors, where photoexcited electrons and holes bind due to Coulomb interactions. Experimentally, such excitonic states have been observed in several other Mott insulators \cite{b77060501,b95235125,np191876,l103187401}.

\begin{figure} 
    \centering 
    \includegraphics[width = 0.4\textwidth]{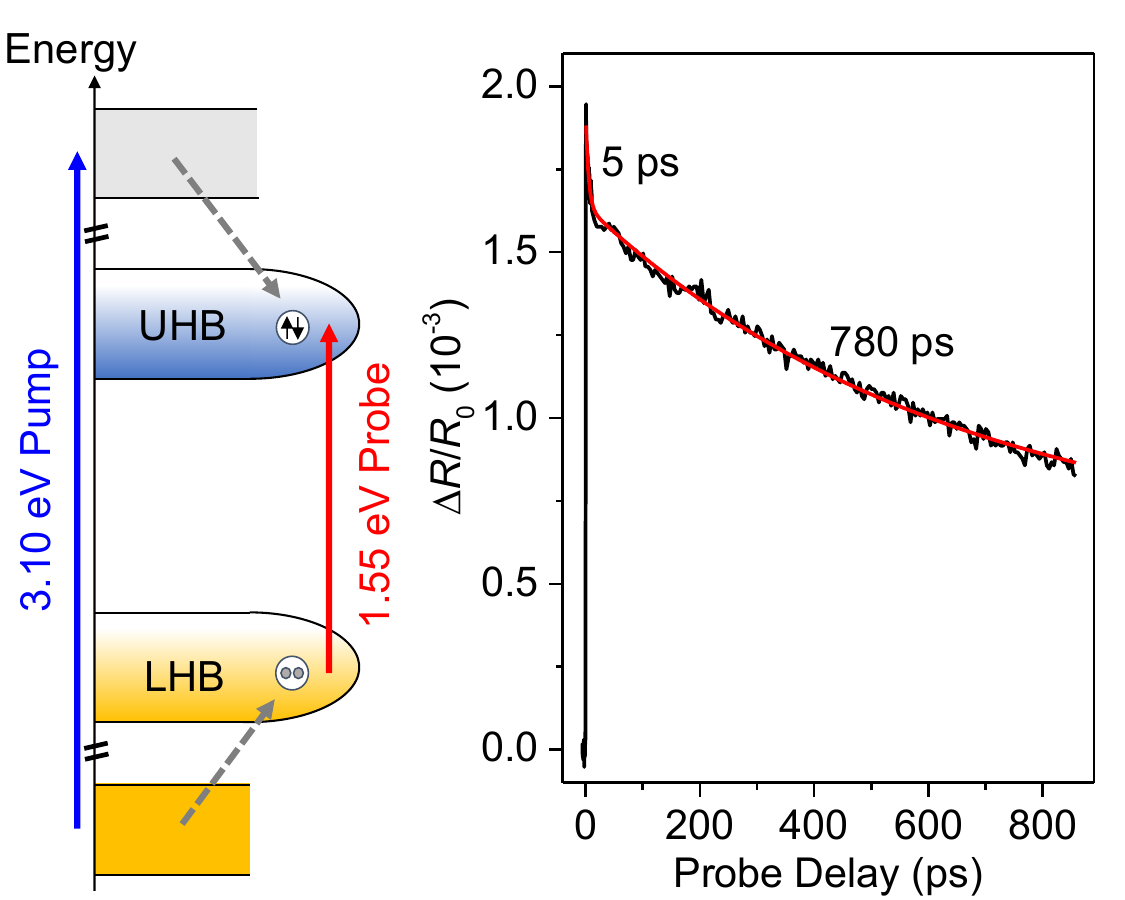}
    \caption{Differential reflectance from a 35 nm \arucl\ flake measured with 3.10 eV, 180 $\mu$J~cm$^{-2}$ pump and 1.55 eV probe pulses.}
    \label{Fig:zhao}
\end{figure}

A recent time-resolved photoemission experiment revealed features of Mott-Hubbard excitons in \arucl\ \cite{nevola_timescales_2021}. Nevola, \etal, investigated the photoexcitation dynamics by generating holon-doublon pairs with ultrashort laser pulses and time-resolving the photoemission signatures associated with holons in the upper Hubbard band and excitons. Their results indicate that excitons are formed on a time scale of 0.2 ps and can survive several hundred ps.

The photoexcitation dynamics were further investigated using optical pump-probe measurements. In this experiment, a 100 fs pump pulse with a photon energy of 3.10 eV and a fluence of 180 $\mu$J~cm$^{-2}$ excited a bulk \arucl\ flake approximately 35 nm thick. The injected peak carrier density was estimated to be about $10^{12}$~cm$^{-2}$. The pump-induced differential reflectance, defined as $\Delta R / R_0 = |R - R_0|/R_0$, where $R$ and $R_0$ are the reflectance of the sample with and without the pump, respectively, was time-resolved using a delayed probe pulse at 1.55 eV. Figure \ref{Fig:zhao} summarizes the observed signal as a function of probe delay. The decay of the signal was well-fit by a bi-exponential function, $\Delta R / R_0 = A_1 \exp(-t/\tau_1) + A_2 \exp(-t/\tau_2)$, where the deduced short time constant, $\tau_1$, of approximately 5 ps is associated with exciton formation and cooling processes. The longer time constant, $\tau_2$, of 780 ps corresponds to the lifetime of the Mott-Hubbard excitons. Further spatially resolved measurements showed that during this lifetime, the excitons diffuse over a distance on the order of 100 nm, revealing their mobile nature.


Scanning ultrafast electron microscopy (SUEM) was recently used to image the spatial-temporal dynamics of photoexcited species in a bulk \arucl\ flake \cite{choudhry2025anomalously}. SUEM is a photon-pump-electron-probe technique, where the local change in the secondary electron yield as a result of optical excitation is mapped by a scanning pulsed electron beam \cite{yang2018scanning}. In the experiment, an optical pump (2.4 eV photon energy, 150 fs pulse duration) with a range of fluences (10 $\mu$J/cm$^2$ to 400 $\mu$J/cm$^2$) was used to generate cross-gap excitations in \arucl\ at room temperature under a vacuum of 10$^{-7}$ torr and a delayed pulsed electron beam (30 keV kinetic energy, roughly 100 electrons per pulse) was used to probe the spatial distribution of the photoexcited species as a function of delay time. At lower optical fluences ($< 40 \mu$J/cm$^2$), the diffusion of Gaussian-distributed photocarriers was observed.

Interestingly, the initial regime of the fast diffusion of the hot photocarriers lasts over 1 ns, much longer than what is typically observed in conventional semiconductors \cite{choudhry2023persistent}. This can be potentially attributed to a hot photocarrier cooling bottleneck between the bands with $J_{eff}=1/2$ and $J_{eff}=3/2$. At higher optical fluences ($>$ 40 $\mu$J/cm$^2$), an intriguing central dark region was observed in the SUEM images, which grows darker with an increasing optical fluence. The suppressed SUEM contrast suggests a significant increase of the local conductivity induced by photoexcitation in \arucl\, which can be caused by a photo-induced insulator-to-metal transition as recently suggested by time-dependent density functional theory (TDDFT) simulations \cite{zhang2023ultrafast}. Notably, the transition occurs at an estimated low photocarrier density of 0.01 electrons per unit cell, suggesting the high sensitivity of the electronic structure of \arucl\ to photoexcitation.

\subsection{Floquet and cavity engineering of Kitaev magnetism}


\begin{figure} 
    \centering 
    \includegraphics[width = \textwidth]{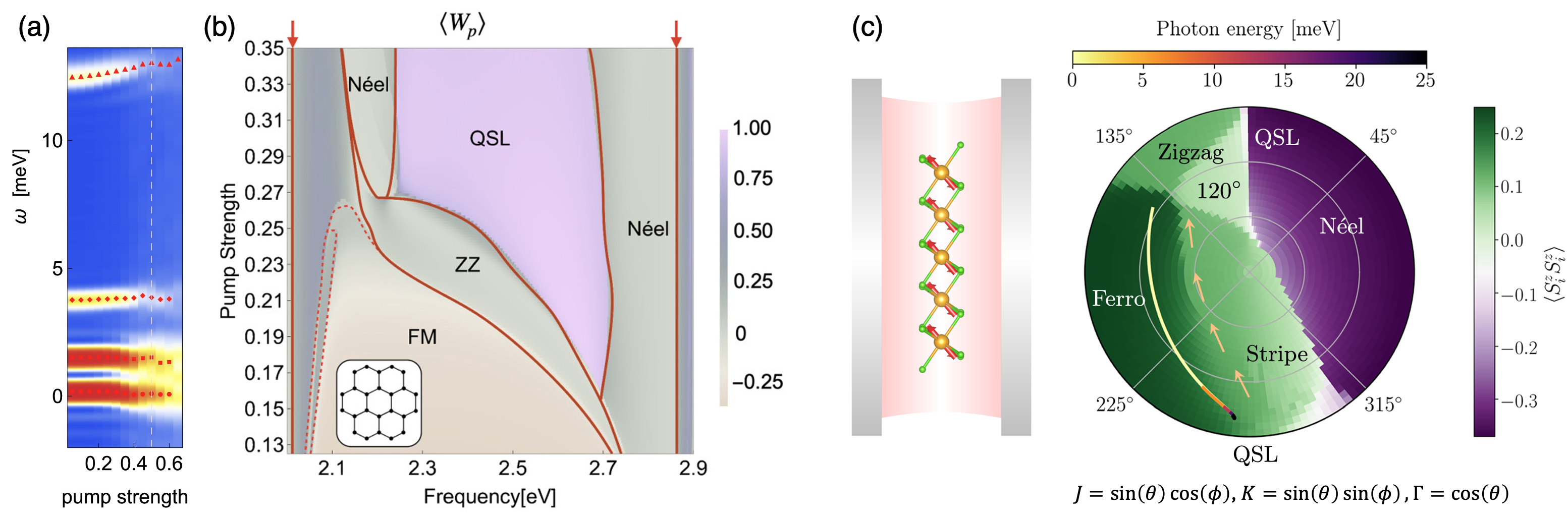}
    \caption{(a) Pump-probe coherent control of the magnetic excitation spectrum of \arucl\ irradiated with a 0.8eV Gaussian pulse and probed at the peak pump field. (b) Steady-state phase diagram of Floquet-engineered Kitaev magnetism. (c) Paths through the magnetic phase diagram traced out by \arucl\ placed inside a dark THz cavity as a function of strong light-matter coupling $g = 0 \dots 0.5$. Adapted from\ \cite{sriram2022light,bostrom2023controlling}.}
    \label{Fig:claassen}
\end{figure}

These advances in pump-probe experiments of quantum materials have opened new pathways to study and control exotic phases of matter via coherent light-matter interactions \cite{oka2019floquet,de2021colloquium,shan2021giant,mciver2020light}. Recent theoretical studies proposed \arucl\ as a particularly promising target for Floquet engineering of magnetic interactions, by virtue of strong SOC and a rich phase diagram of closely-competing orders and a proximate QSL state \cite{arakawa2021floquet,sriram2022light,kumar2022floquet,arakawa2021polarization}. In such schemes, a strong optical pump field is used to transient dress the $t_{2g}$ electronic states of Ru to favor certain superexchange pathways or to break symmetries, thereby altering the balance of competing magnetic interactions in \arucl\ for the duration of the pulse. A suitably chosen driving scheme can therefore transiently modify the equilibrium magnetic phase and its excitations, or potentially even drive the material toward the coveted Kitaev QSL phase. 

Specifically, tailored optical pulses have been predicted to selectively enhance ligand-mediated Kitaev exchange interactions while suppressing competing isotropic Heisenberg interactions. Circularly-polarized pumping can furthermore induce a ligand-mediated inverse Faraday effect \cite{sriram2022light,kumar2022floquet}, generating an effective magnetic field along the [111] crystallographic direction necessary to open a gap in the Kitaev model, potentially facilitating the stabilization of non-equilibrium spin liquid states. A key signature of such a Floquet-engineered magnetic state is the coherent \textit{transient} modification of the magnetic excitation spectrum, observable in time-resolved pump-probe experiments and depicted theoretically in Fig.\ \ref{Fig:claassen}(a) as a function of the dimensionless Floquet parameter $a_{0} e \mathcal{E}/\hbar \omega$ that parameterizes the pump strength of a monochromatic electric field $\mathcal{E}$, with $a_0$ the Ru-Ru distance. On short time scales, the light-dressed magnetic exchange interactions can in principle lead to a rich steady-state diagram of pre-thermal Floquet phases  (Fig.\ \ref{Fig:claassen}(b)) provided that heating from the driving field remains suppressed. First signatures of Floquet-engineered Kitaev magnetism have recently been reported in time-resolved resonant inelastic x-ray (RIXS) measurements of the Kitaev QSL candidate H$_3$LiIr$_2$O$_6$ subjected to a circularly-polarized 1900 nm drive \cite{kim2024signatures}.

Beyond the strong optical fields required for Floquet engineering, a complementary proposed direction is to confine an optical mode via a cavity. In such cavity quantum-electrodynamical approaches, analogous effects have been proposed to occur either at much reduced optical fields or even from vacuum fluctuations of a confined THz photon mode without external driving \cite{sentef2020quantum,schlawin2022cavity}. In \arucl, this analogously suggests two potential avenues to alter the magnetic phase. First, a dark THz single-mode cavity was proposed to stabilize ferromagnetic order from the equilibrium zigzag state purely through vacuum fluctuations at strong light-matter coupling $g$, see Fig.\ \ref{Fig:claassen}(c) \cite{bostrom2023controlling}. Second, by moderately pumping an optical cavity in the few-photon limit, an embedded \arucl\ sample can be tuned to realize the elusive KQSL phase in analogy to predictions from Floquet engineering, albeit at much reduced photon number. Such cavity-enhanced interactions between photons and charge fluctuations in the material can therefore provide a complementary route to Floquet engineering of magnetic states in \arucl\ and related materials. Moreover, quantum-optical measurement techniques of scattered photons can provide new insights into the nature of the ground state and excitations in Kitaev materials. Here, pair photon correlation measurements with frequency filtering have been proposed as a route to extract signatures of fractionalized excitations \cite{nambiar2024diagnosing}, and cavity quantum-electrodynamic modifications of the magnetic ground state at strong light-matter coupling in an optical cavity have been predicted to manifest in antibunched emitted light, which can be detected by Hanbury-Brown-Twiss measurements of the second-order photon coherence $g^{(2)}$ \cite{kass2024many}.

\section{Neutron scattering}\label{neutrons}

Neutron scattering provided some of the very first and important signatures that built confidence in the material \arucl\ as a leading Kitaev candidate. It showed that the Kitaev interaction is a dominant interaction in the spin Hamiltonian, and the material possesses the necessary inelastic signatures to host an exotic state. At the same time, it showed the effects of the symmetry-allowed non-Kitaev terms, which ultimately leads to a state which has magnons at very low energies and temperatures. We review key results in neutron scattering, and close with a focus on the layered nature of \arucl.

Neutron scattering provides one of the cleanest measurements of both the static and dynamic two-spin correlation functions in magnetic material. Neutron scattering creates spin-flip (S=1) and tracks the resultant excitations in the spin lattice.  The magnetic scattering cross section is given by the relationship:
\begin{equation}
\nonumber
\frac{d\sigma}{d\Omega dE}\propto F(q)\Big(\delta_{\alpha\beta}-\frac{q^{\alpha}q^{\beta}}{q^2}\Big)\times\sum_{\mathbf{r_i},\mathbf{r_j}}e^{i\mathbf{q}\cdot(\mathbf{r_i}-\mathbf{r_j}})\int dtdt' e^{-i\omega(t-t')}\Big(S_{\mathbf{r_i}}^{\alpha}(t)S_\mathbf{r_j}^{\beta}(t')\Big)   
\end{equation}

The term F(q) is the form factor, which in \arucl\ is the form factor of the Ru$^{3+}$ ions. The magnetic intensity of the spectrum attenuates as a function of the form factor square of Ru$^{3+}$, $|F(q)|^2$, which helps to tell apart between the magnetic parts of the spectrum from the other lattice vibrations and phonons, which rise as $|q|^2$, making NS one of the cleanest probes of magnetic excitations. Recently this function was measured in detail \cite{PhysRevB.109.104432}. The term $(\delta_\alpha\beta-q^{\alpha}q^{\beta}/q^2)$ is a selection rule dictating that only the components of the spin perpendicular to the scattering vector \textbf{q} would contribute. The rest of the equation is the Fourier transformation of the dynamic spin-spin correlation between two lattice sites i, and j, leading us to obtain the Fourier Transformation of both the static and dynamic spin structures. A sharper energy-momentum response in neutron scattering arises long-range correlations, such as magnons, while a broad excitation in conversely means the prevalence of short-range correlations. Hence a spectrum which is sharp in energy often signifies a bosonic excitation (magnons, phonons, and their hybrids). Broadening in energy can either come from nonlinear decays of the magnons, or could indicate the presence of broad fermion bands, deconfined quasiparticles \cite{Lake2005-nb}, or fractionalized excitations \cite{Knolle2019}.  

The first calculations for the neutron spectrum for the Kitaev model indicated the NS spectrum of the Kitaev model consists of gapped flux (vison) excitations and Majorana fermions \cite{PhysRevLett.112.207203, PhysRevB.92.115127}. NS for the general Kitaev model in the presence of modest non-Kitaev terms were calculated in Refs.\ \cite{Song2016,PhysRevB.107.224428}. Finite temperature calculations were included in Ref.\ \cite{yoshitake_fractional_2016}. It was shown that the only correlations in the pure Kitaev model are the self- and the nearest neighbor correlation. All other higher order correlation terms are zero, unless one accounts for non-Kitaev terms in the Hamiltonian.


\begin{figure}[t!]
 	\centering
 	\includegraphics[width=0.9\columnwidth]{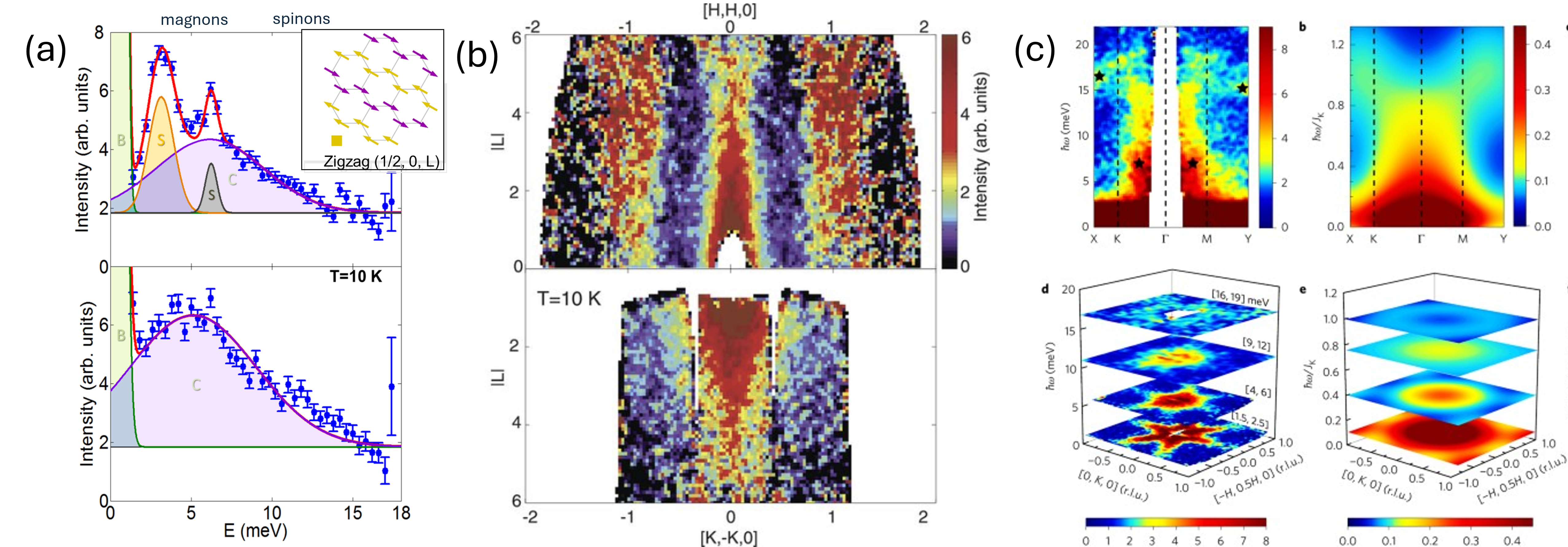}
 	\linespread{1}
 	\caption{(a) Neutron scattering on single crystals of \arucl\ show magnons arising from the zig-zag state coexisting with a spinon continuum at zero field. These magnons melt and enhances the continuum above TN. \cite{banerjee_neutron_2017}. (b) The continuum shows almost no dispersion along the out-of-plane L direction indicating that the quasiparticles are mostly constrained in one 2D plane. (c) The high energy part of the continuum matches the scattering expected from a pure ferromagnetic Kitaev model bolstering the idea that, despite sub-leading terms of the Hamiltonian renormalizing the low energy spectrum, the Kitaev interactions play a defining role \cite{do_majorana_2017}. Adapted with copyright © 2017 Springer Nature.}
    \label{BanerjeeFig3}
\end{figure}

The first elastic NS investigations of \arucl\ were reported in Ref.\ \cite{cao2016low}. The structure of the bulk \arucl\ in that experiment was hard to determine because of the prevalence of stacking faults which can occur from simply handling the vdW crystal. It was found that the structure of those \arucl\ crystals varies between the C2/m (monoclinic, 2-layer AB) and the R-3 (rhombohedral, 3-layer ABC or CBA) space groups, with a structural phase transition at 150 K \cite{ziatdinov_atomic-scale_2016}. The low temperature crystal structure (the phase below 150 K) of larger crystals of \arucl\ was determined to be R-3 from neutron diffraction studies \cite{Park2024-fz} and neutron Laue \cite{mu2022role}. NS showed that single crystals of \arucl\ can have both T$_N$=7 K and 14 K transitions coexisting in crystals with stacking faults \cite{banerjee_proximate_2016}. 

However, in thick crystals with a lower number of stacking faults, the 14 K transition is absent \cite{cao2016low, Park2024-fz}. The static spins in the low-temperature zig-zag phase were found to have spins with an in-plane spin projection perpendicular to the Ru-Ru bonds, and at an out-of-plane angle of 35 degrees from the ruthenium honeycomb plane. The direction of the tilt, whether towards the chlorine ligands (which would indicate an antiferromagnetic Kitaev-Heisenberg mode) or roughly 70 degrees away from the chlorine ligands (which would indicate a ferromagnetic Kitaev-Gamma model) was addressed subsequently using neutron diffraction \cite{Park2024-fz} and also X-Rays \cite{Sears2020-om, Kim2024-lk} and more recently, polarized neutrons \cite{braden2024directevidenceanisotropicmagnetic}, to be likely pointing away from the direction of the Ru-Cl ligands suggesting ferromagnetic Kitaev interactions with a strong Gamma term. At any rate, even in the low temperature zigzag phase, only 50\% of the total movements arise form a long-range order  indicating the presence of strong fluctuating moments at low temperatures \cite{cao2016low,Park2024-fz}---a strong indicator of the possibility of a proximate KQSL state.

Inelastic Neutron Scattering (INS) on powder samples of \arucl\ shows the presence of two distinct features in the magnetic scattering \cite{banerjee_proximate_2016}. A lower energy feature arose only below T$_N$ at $E$ $<$ 4 meV evidently from gapped magnons (with a spin gap of 2 meV at the M-point antiferromagnetic zone center \cite{banerjee_neutron_2017, ran_spin-wave_2017}. However, there was a distinct higher energy feature, with a peak at $E$ = 6 meV. This higher-energy feature was broad and increased in intensity when the long-range order vanished, and was robustly prevalent at least up to 70 K, indicating the presence of strong anisotropic excitations connected to a Kitaev exchange of roughly 6-8 meV.  

INS on single crystals of \arucl\ confirmed the co-existence of both the sharper magnons at low energies and a broad continuum of scattering close to the 2D zone center \cite{banerjee_neutron_2017}. Evidently, the low-energy magnons appearing from the zig-zag ground state gets renormalized above T$_N$ to form a bright continuum of scattering (see Fig.\ \ref{BanerjeeFig3}). The continuum of scattering has remarkable and distinct features that can arise from the presence of dominant Kitaev interactions---it is remarkably broader than the resolution, as is expected from the presence of itinerant Majorana fermions, as opposed to just magnons. The distribution of the scattering in the Brillouin zones conforms to the dominance of on-site and nearest-neighbor correlations---a defining feature of excitations from a pure Kitaev model  \cite{PhysRevLett.112.207203,yoshitake_fractional_2016,PhysRevB.96.134408}. The observations were also confirmed in a different batch of Korean-grown samples in Ref.\ \cite{do_majorana_2017}, where besides the broad excitations centered at 6 meV, a further Y-shaped higher energy excitation was observed, concomitant with a finite-temperature expansion of the ferromagnetic Kitaev model with an interaction strength of roughly $K$ = 16 meV. These experiments helped to cement the idea that, while non-Kitaev interactions are certainly present to stabilize low-energy magnons at zero field, the dominant effective interaction in \arucl\ is of Kitaev type. The description of the higher energy parts of the spectrum (above the zig-zag state magnons) could be obtained from the pure Kitaev model, which predicts the presence of thermally excited Majorana Fermions.

\begin{figure}[t!]
 	\centering
 	\includegraphics[width=0.9\columnwidth]{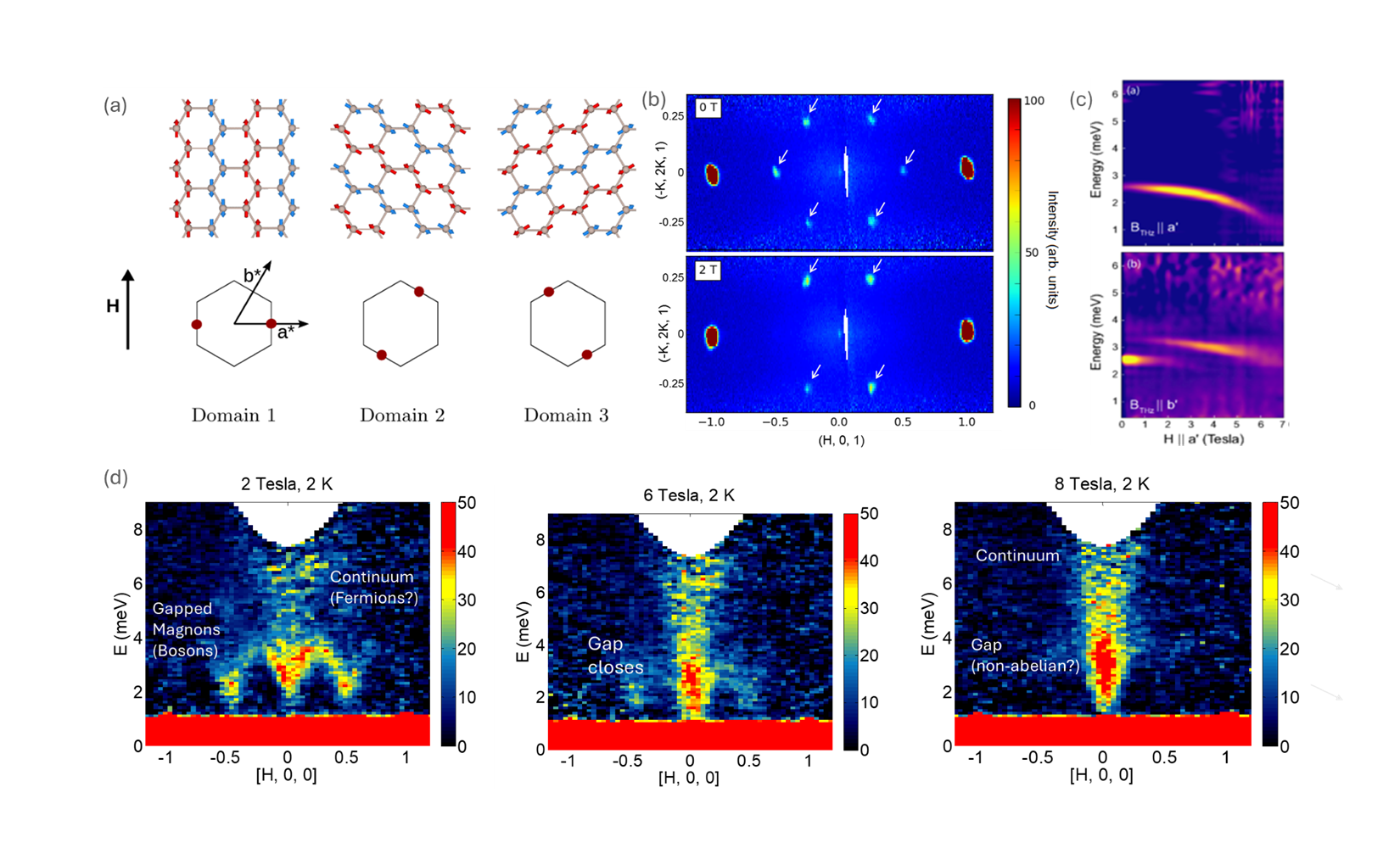}
 	\linespread{1}
 	\caption{(a) A description of the three zig-zag domains  appearing at zero field leading to 3 pairs of [1,0,1] type Bragg peaks in Fig (b) \cite{PhysRevB.98.094425}, however, by 2 T, spin reorients out of domain 1 into domains 2 and 3 leading to only 2 pairs of brighter Bragg peaks \cite{banerjee2018excitations}.  (c) Raman measurements \cite{PhysRevB.98.094425} show the evolution and the breakdown of magnons with field, (d) The evolution of the INS for field perpendicular to the Ru-Ru bonds shows that gapped magnons at zero field disappears by 6 T leaving behind a gapless continuum, which by 8 T appear to develop a new spin gap \cite{banerjee2018excitations}.}
    \label{BanerjeeFig2}
\end{figure}

The zigzag long-range order can be suppressed by applying a field of 8 T \cite{kubota_successive_2015,johnson_monoclinic_2015}. Neutron scattering has been performed with a field direction perpendicular to the Ru-Ru bonds \cite{banerjee2018excitations}. Several key features were observed in this experiment in neutron diffraction:\ six magnetic Bragg peaks are observed at six equivalent M-points at zero field \[1/2, 0, 1\], but by 2 T, due to a spin reorientation or domain rearrangement, the two Bragg peaks in the scattering plane disappeared while those out of the scattering plane become brighter. This indicated that the overall domain distribution becomes anisotropic at 2 T, a prediction confirmed in several other measurements \cite{Holleis2021-jx,idzuchi2022spinsensitivetransportspin,Wagner2022-jx,bruin_robustness_2022}. The sum-intensity of the magnetic Bragg peaks of all domains combined, however, remained constant to 4 T and steadily decreased after that (see Fig.\ \ref{BanerjeeFig2}) \cite{banerjee2018excitations} .

The precise phase transitions of \arucl\ depend on whether the field is applied perpendicular, or parallel, to the Ru-Ru bonds. Using neutron Laue diffraction, the presence of a new zigzag phase (ZZ2) was confirmed, with an effective six-layer stacking, between B$_{c1}$ = 6.0 (7.4 T) and B$_{c2}$ =7.3 T (7.8 T) when the field is applied perpendicular (parallel) to the Ru-Ru bonds, with the appearance of distinct magnetic peaks \cite{Balz2021}. These phases were later extensively studied to analyze the role of possible stacking faults in the thermal transport measurements \cite{10.1063/5.0101377,czajka_oscillations_2021}.

INS in a magnetic field in \arucl\    at intermediate fields showed the evolution of both the magnons and the broad continuum excitations as a function of magnetic field \cite{banerjee2018excitations}. At 2 T, the magnons measured in the plane of the magnetic scattering, showed an increase in intensity from the anisotropic domain redistribution. However, at 6 T, the magnons undergo a dramatic reduction in neutron intensity, accompanied by an enhancement of intensity of the broad scattering continuum. The decay of the magnons were also observed in Raman and THz spectroscopy \cite{little_antiferromagnetic_2017, PhysRevB.98.094425, PhysRevB.100.100403}. By 8 T, the magnons fully disappear leaving behind a Field Induced Quantum Disordered (FIQD) state bright scattering continuum with a seemingly small spin gap ${\sim}0.5$ meV. These results support the hope of a KQSL state in the vicinity of 8 T, with a Z2 spin gap---which can support the $1/2$-quantized thermal Hall effect predicted by Kitaev below the spin gap \cite{kim_kitaev_2015}.  

INS at higher fields, up to 13.5 T, performed in \arucl\ at FLEXX spectrometer in Berlin showed that the sharper magnon like excitations have a modulation along the vdW stacking direction clearly indicating the presence of out of plane exchange terms in the magnetic Hamiltonian \cite{balz2019finite}. These out-of-plane terms have so far been largely ignored in the theoretical predictions of the magnetic Hamiltonian from first principles \cite{Laurell2020-kk, Maksimov2020,Li2021-vk,Winter2016,Winter2017,Winter2017-dm,PhysRevB.100.144445, PhysRevB.97.134424, Suzuki2021-qm, PhysRevB.102.024415, PhysRevB.93.155143, Hou2017, PhysRevB.96.115103, PhysRevB.100.075110,ran_spin-wave_2017,PhysRevB.98.060412,PhysRevB.98.094425,PhysRevB.100.085108, PhysRevB.101.140410, Laurell2020-kk,Kaib2021,PhysRevB.102.115160,PhysRevB.101.174444,Ran2022-hl,PhysRevResearch.4.L022061,PhysRevB.105.214411,PhysRevResearch.3.033223}, including DMRG and Exact Diagonalization, which have explored possibilities of explaining the unusual spectrum both in terms of fractionalized fermionic excitations, but also non-linear multimagnon decays. The strong effect of the out-of-plane interactions was also indicated in the DFT fits to the phonon spectrum obtained in INS \cite{mu2022role}. However, the results also showed that the out-of-plane modulations diminish closer to the FIQD state perhaps indicating a diminishing contribution of the out-of-plane exchange terms in this state which could reinforce a notion of the 2D QSL state \cite{balz2019finite}. The measurement also failed to measure any magnons in the FIQD state. This was confirmed by further measurements using SIKA spectrometer \cite{Zhao2022-zu}.

Given the high saturation magnetic moment of \arucl\ ($>$ 40 T) \cite{johnson_monoclinic_2015}, \color{black} and presence of strong fluctuations to very high fields \cite{Modic2021-vb} one may doubt whether application of modest fields $\approx$10-20 T, could effectively condense all the spin excitations into magnons that could be tracked using (non-)linear spin wave theory (LSWT). It is also likely that machine-learning and AI stemming from non-linear Landau-Lifshitz-Gilbert dynamics, and quantum corrections therein, would need to play an important role in this deduction \cite{PhysRevResearch.4.L022061}. Yet, a comprehensive set of measurements showing the evolution of the magnetic continuum into sharp magnons deep into the field-induced polarized state of \arucl, as well as a comprehensive, higher-resolution measurement of the evolution of the spin-gap, which remains ambiguous from neutron measurements between various samples \cite{banerjee2018excitations,Zhao2022-zu}, as well as comprehensive polarized neutron measurements expanding on the existing measurements \cite{braden2024directevidenceanisotropicmagnetic}, could help to a complete picture of the magnetic Hamiltonian.

\subsection{Inter-layer correlations} 

There is an overall consensus from various experiments and concurrent theory that the leading order Hamiltonian of \arucl\ would certainly contain a dominant 2D Kitaev and $\Gamma$ interactions. At the same time, the overall magnetic Hamiltonian is 3D and contains sub-leading intra-layer terms. This has been proposed in several experiments such as in INS  where a clear out of plane dispersion of the magnons with a bandwidth of 0.5-1 meV was observed in the material especially in the zig-zag ordered phase at 0 T and in the partially field-polarized phase at 13.5 T \cite{balz2019finite}. The 3D nature of the long-range magnetic order was also apparent in exfoliated samples such as in the MoTe$_2$-\arucl\ inelastic electron tunneling spectroscopy experiments (see Section \ref{iets}), where a clear distinction in the spectral intensity of the magnon and spinon branches is seen in monolayer, bilayer and trilayer samples \cite{Yang2023-ke}, as well as in Raman measurements approaching monolayers \cite{Zhou2018,Du2018,Lee2021a}. The long-range orders, such as the zig-zag and the field-induced ferromagnetic orders have consistently shown out-of-plane modulation, with a change in the magnetic periodicity from a 3-layer to a 6-layer ordering in the ZZ2 phase \cite{balz2019finite}.  Consequently, there is little doubt that the magnetic long-range order is dependent on the out-of-plane interactions.

In contrast, the character of the deconfined QSL excitations high-energy continuum have consistently been two-dimensional and the QSL excitations are primarily confined within one single, or at best two, \arucl\ layers. Although some 3D interactions are natural in \arucl, the inter-plane magnetic interactions are vanishingly small $\lesssim$ 0.5 meV which is roughly one to two orders of magnitude lower than the leading order in-plane terms \cite{PhysRevB.93.155143}. Experimentally, this was first demonstrated in the out-of-plane reciprocal cuts in the neutron scattering experiments  where the high-energy part of the continuum of scattering showed rod-like dispersions, with negligible out-of-plane modulations compared to the overall broad bandwidth of these excitations \cite{banerjee_neutron_2017}. Furthermore, in a magnetic field, the measurements using triple axis spectrometer also showed that the net effect of the out-of-plane interactions tend to diminish as one approaches the field-induced QSL state at 8 T \cite{balz2019finite}. A vanishing 3D dispersion is expected from disordered magnetic layers, such as in a 2D QSL state. 

Kitaev's theory of topological protection and the non-Abelian physics and the inelastic spin excitations require a 2D Hamiltonian. The 3D nature of the Hamiltonian of \arucl---while important for explaining the long-range order---may play a negligible role for the more interesting physics of the emergent excitations at higher energies, and in the field-induced QSL phase. 

As also mentioned elsewhere, there is a degree of debate on the observation of the half-integer quantized edge modes in the thermal Hall effect, whether it is fermionic or bosonic, and the role of the out-of-plane stacking. An increasing body of work supports the out-of-plane arrangement of the RuCl$_3$ layers, and the stacking sequence can play a major role in the observation of the thermal Hall plateau \cite{PhysRevB.102.220404}. There is evidence that the observation of the thermal Hall signature is phonon mediated \cite{li_giant_2021,PhysRevLett.121.147201,Lefranois2022,Lefranois2023,Dhakal2024-np}, and the out-of-plane phonon modes play a defining role in the spin-phonon coupling, which are naturally different for differently stacked allotropes of \arucl\ \cite{mu2022role}. 

A question naturally arises:\ can one observe quantized thermal Hall effect in a monolayer, as the thermal Hall effect should be present regardless of the phonons? Or is spin-phonon coupling always so critical? Does it require several layers of \arucl\ arranged in a specific stacking sequence to stabilize specific out-of-plane phonon modes required for the observation of the thermal Hall plateau? Answering such questions would require mesoscopic thermal transport experiments in the true 2D limit and at sub-Kelvin temperatures below the phonon excitation thresholds, which would also be critical towards the future applicability of \arucl\ as a prime candidate for protected quantum electronics.

\section{Outlook and prospects for Kitaev physics in atomically thin materials}\label{outlook}

The body of work on atomically thin or layered samples of \arucl\ reviewed here represents a novel route in exploring Kitaev magnetism in two-dimensional systems, illustrating both the promise and limitations of this approach. While \arucl\ falls short of being an ideal Kitaev material due to competing magnetic interactions, key signatures of Kitaev physics---including broad magnetic continua, fractionalized excitations, and field-induced quantum disordered states---can indeed be accessed in van der Waals materials. The diverse set of experimental tools reviewed here that span charge transfer-mediated electronic transport and tunneling spectroscopy to near-field optical microscopy and heterostructure engineering, provides a comprehensive framework for future explorations beyond \arucl. 

The theoretical advances parallel to these experimental developments are equally significant, with ab initio calculations revealing how strain, charge doping, and proximity effects can tune magnetic interactions toward more ideal Kitaev limits. The discovery that \arucl\ in heterostructures may have enhanced Kitaev coupling strengths by up to 50\% while suppressing competing interactions points toward device engineering approaches that overcome the limitations of pristine \arucl. Furthermore, proposals for Floquet engineering and cavity quantum electrodynamics offer pathways to dynamically control magnetic phases, potentially accessing elusive quantum spin liquid states through external driving fields or vacuum fluctuations. The identification of moir\'e superlattices as platforms for novel spin textures and the theoretical prediction of quantum spin liquids in twisted magnetic bilayers suggest that the stacking degree of freedom in van der Waals heterostructures provides another knob for tuning toward ideal Kitaev behavior. And notably, the community interest in \arucl\ has catalyzed parallel developments in quantum simulation platforms, with researchers now exploring Kitaev and extended Kitaev models using ion-trap devices, neutral atoms, and superconducting transmons to understand transport properties and edge mode transduction both in the exactly solvable limit and beyond.

Looking forward, the works here lay the foundation for discovering and engineering superior two-dimensional Kitaev materials. The emergence of candidate systems including Cr-based compounds with antiferromagnetic Kitaev interactions, Ni-based materials on triangular lattices, and Co-containing honeycomb systems demonstrates that multiple chemical pathways to Kitaev physics exist beyond the original $d5$ transition metal framework. The experimental techniques refined on \arucl---particularly heterostructure-based transport measurements, scanning probe methods for charge-doped systems, and optical approaches for ultrafast control---will be immediately applicable to these new candidates. The insights gained regarding the crucial roles of work function engineering, strain tuning, and interlayer interactions provide design principles for creating quasi-two-dimensional devices that could finally realize the intriguing Kitaev quantum spin liquid state. Should new materials emerge that more faithfully represent the ideal Kitaev system,  the foundation laid by research on \arucl\ positions the field to rapidly characterize and develop these systems, bringing the promise of new magnetic phenomena and perhaps even the possibility of fault-tolerant topological quantum computation closer to reality.

\begin{acknowledgments}
This review is the result of a workshop held in May 2024 at Washington University in St.\ Louis on recent progress in research on 2D \arucl. We are grateful for many interesting and helpful conversations as well as research efforts in collaboration with Jiaqiang Yan, Yong P.\ Chen, Gabor Halasz, Jeroen van den Brink, Alexander Seidel, Li Yang, and Nandini Trivedi. Support for the workshop was kindly provided by the Office of Naval Research award no.\ N000142412319. CO-A acknowledges funding from the U.S.\ Department of Energy, Office of Science, Office of Basic Energy Sciences under contract DE-SC0018154 for sample fabrication, high impedance electronic transport measurements and ARPES experiments and the Cal State Long Beach and the Ohio State University Partnership for Education and Research in Topological Materials, a National Science Foundation PREM, under Grant No. 2425133 for travel. Results presented here used resources of the Advanced Light Source, which is a DOE Office of Science User Facility under contract no. DE-AC02-05CH11231. EAH acknowledges research support by the Air Force Office of Science Research under award no.\ FA9550-22-1-0340, and support by the Gordon and Betty Moore Foundation, grant doi.org/10.37807/gbmf11560. BMH acknowledges research support from the Office of Naval Research, grant number N000142112443. XHZ acknowledges research support by Open Research Fund of State Key Laboratory of Materials for Integrated Circuits (SKLJC-K2025-05). This work was further performed in part (ZN) at the Aspen Center for Physics, which is supported by National Science Foundation grant PHY-2210452. MK acknowledges support from the Deutsche Forschungsgemeinschaft (DFG, German Research Foundation) under Germany's Excellence Strategy--EXC--2111--390814868, TRR 360-492547816 and DFG grants no.\ KN1254/1-2, KN1254/2-1, the European Research Council (ERC) under the European Union's Horizon 2020 research and innovation programme (grant agreement no.\ 851161), the European Union (grant agreement No 101169765), as well as the Munich Quantum Valley, which is supported by the Bavarian state government with funds from the Hightech Agenda Bayern Plus. AB acknowledges support from the Department of Energy, Office of Science, Basic Energy Sciences Grant DE-SC0022986. AWT acknowledges support from the Dorothy Killam Fellowship. HZ acknowledges the support by the U.S. Department of Energy, Office of Basic Energy Sciences, Division of Materials Sciences and Engineering under Award DE-SC0020995 for transient absorption measurements. BL acknowledges support by the U.S. Air Force Office of Scientific Research under the award number FA9550-22-1-0468 and the U.S. Army Research Office under the award number W911NF2310188. YM acknowledges support from Japan Society for the Promotion of Science (JSPS) KAKENHI Grant (Nos. 16H02206, 19H05825, 20H00122, 
and 25H01247) and JST CREST (JP-MJCR18T2) for theoretical investigations. YK acknowledges support from the National Research Foundation of Korea (NRF) (Grant nos.\ RS-2025-00557717, RS-2023-00269616, RS-2024-00444725). The work at ASU was supported by the U.S.\ Department of Energy, Office of Science, Office of Basic Energy Sciences, Material Sciences and Engineering Division under Award Number DE-SC0025247. The work from DGIST was supported by the National Research Foundation of Korea (NRF) (Grant no.\ RS-2025-00557717, RS-2023-00274875, RS-2023-00269616)  and the Nano and Material Technology Development Program through the National Research Foundation of Korea (NRF) funded by Ministry of Science and ICT (no.\ RS-2024-00444725). We also acknowledge the partner group program of the Max Planck Society. Finally we acknowledge Washington University in St.\ Louis and the Institute of Materials Science and Engineering for supporting portions of this work.
\end{acknowledgments}

\bibliographystyle{nsf_erik}
\bibliography{rucl_clean}

\end{document}